\documentclass[rmp,aps,twocolumn,superscriptaddress,showpacs]{revtex4-1}

\usepackage{graphicx}
\usepackage{epstopdf}
\usepackage{color}
\usepackage[utf8]{inputenc}

\begin{document}

\title{The magnetic and electronic properties of Oxyselenides - influence of transition metal ions and lanthanides}

\author{C. Stock}
\affiliation{School of Physics and Astronomy, University of Edinburgh, Edinburgh EH9 3FD, UK}

\author{E. E. McCabe}
\affiliation{School of Physical Sciences, University of Kent, Canterbury, CT2 7NH, UK}

\date{\today}

\begin{abstract}

Magnetic oxyselenides have been a topic of research for several decades, firstly in the
context of photoconductivity and thermoelectricity owing to their intrinsic semiconducting
properties and ability to tune the energy gap through metal ion substitution. More
recently, interest in the oxyselenides has experienced a resurgence owing to the possible
relation to strongly correlated phenomena given the fact that many oxyselenides share
a similar structure to unconventional superconducting pnictides and chalcogenides. The
two dimensional nature of many oxyselenide systems also draws an analogy to cuprate
physics where a strong interplay between unconventional electronic phases and localised
magnetism has been studied for several decades. It is therefore timely to review the physics
of the oxyselenides in the context of the broader field of strongly correlated magnetism
and electronic phenomena. Here we review the current status and progress in this area of
research with the focus on the influence of lanthanides and transition metal ions on the
intertwined magnetic and electronic properties of oxyselenides. The emphasis of the review
is on the magnetic properties and comparisons are made with iron based pnictide and
chalcogenide systems.

\end{abstract}


\maketitle

\tableofcontents

\section{Introduction}

Materials based upon transition metal ions have consistently
been a source of interest owing to novel electronic, magnetic,
and structural properties they possess. While the description
of structural and insulating magnetic transitions has been well
understood with a robust formalism to describe such systems,
metal-insulator transitions and the new phases which exist near
these critical points are not described by any such theory. The
exploration of materials that host such transitions has led to
several notable discoveries including high temperature superconductivity
in the cuprates in 1986~\cite{Bednorz86:64} and more recently with the discovery of superconductivity in LaFeAsO$_{1-x}$F$_{x}$ in 2008~\cite{Kamihara08:130}.  This report of iron-based superconductivity lead to the discovery of many more iron-based superconducting materials which have challenged both theories and experiments, including iron arsenides, iron selenides and mixed-anion iron oxyarsenides. In this review, we discuss a class of related mixed-anion materials: oxyselenides.   Whilst this family of materials is generally not superconducting, their structural similarities with several classes of the iron-based superconductors, and the ability to tune their semiconducting nature, makes them relevant to current research in strongly correlated electron systems.

Selenium (named after the Greek, \textit{selene}, for moon due to its many similarities with tellurium which was named after the Latin, \textit{tellus}, for Earth)~\cite{Emsley}, can adopt a wide range of oxidation states from \textit{0} to \textit{+6} (in selenates) and \textit{+4} (in selenites) to \textit{-2} in selenides which are most commonly formed with the more electropositive group 1 and group 2 elements and lanthanides~\cite{GreenwoodEarnshaw}. In this review, we focus on transition metal oxyselenides and explore the role of the lanthanide and transition metal ions in the crystal structure and magnetic and electronic properties of these systems. 

The semiconducting properties of the oxyselenides and the ability to tune these properties has resulted in these materials being the topic of research for several decades.  Early interest in the oxyselenides arose in the area of photoconductivity which arguably led to the discovery of LaCuOS and related materials~\cite{Minami73:15}.  More recently, work has focussed heavily on thermoelectrics~\cite{Snyder08:7} and, again, the underlying tunable semiconducting properties resulted in many of these systems being identified as good thermoelectrics~\cite{Zhao10:97}.

Although this review focuses on a class of selenides, it is worth mentioning selenite materials such as Cu$_{2}$OSeO$_{3}$ which have received significant attention as hosts~\cite{Seki12:336,Ruff15:5} for Skyrmions, vortex-like topological spin structures~\cite{Skyrme62:31} originally discovered in MnSi~\cite{Mul09:323,Yul11:10,Seki12:336}.  The Cu$_{2}$OSeO$_{3}$ crystal structure is composed of edge- and corner-linked CuO$_{5}$ polyhedra linked by SeO$_{3}$$^{2-}$ selenite groups, with the role of the Se$^{4+}$ lone pair evident in its coordination~\cite{Meunier76:9}.   While these materials constitute an important and rapidly evolving area of physics and materials science, we will not be addressing this topic and the materials in this review which is confined to the electronic and magnetic properties of selenides relevant for the broad theme of the review paper discussing interplay between metallic and magnetic properties. 

Perhaps one of the greatest areas of interest in the search of novel electronic and magnetic properties is to study materials that are close to metal-insulator transitions.  This remains one of the least understood areas of condensed matter physics. Materials near such transition provide the possibility for the discovery of new phases of matter as illustrated by the discovery of high temperature superconductivity in the cuprates and more recently in iron based pnictides and chalcogenide systems~\cite{Stewart11:83,Johnston10:59,Paglione10:6,Wen12:74}.   Many oxyselenides contain iron on a two dimensional lattice, structurally similar to iron based superconducting systems and indeed there were suggestions that oxyselenides may represent candidate superconducting phases during early work on cuprate superconductors~\cite{Otzschi99:117}.  This analogy has led to recent investigations of magnetism and electronic phenomena in such systems.  Given the semiconducting nature and the ability to tune the electronic band gap through transition metal ion substitution, the oxyselenides provide a unique series of compounds to search for new electronic phases and critical properties.    

This review is divided into six sections including this introduction.  We first present some of the experimental techniques used to investigate the magnetic and electronic properties of the oxyselenides.  We then outline the oxyselenide structural families discussed in this review followed by a discussion of the magnetic structures and properties.  We then review the electronic properties of the oxyselenides and finish with a summary and conclusion.

\section{Experimental techniques}

In this section, we provide an outline of the various experimental techniques used to synthesise and to probe the structural, magnetic and electronic properties of the oxyselenides which are the focus of this review.  

\subsection{Synthesis routes}

Unlike the synthesis of oxides which can often be carried out in air, preparing oxyselenides requires more control to avoid oxidation of selenium or other selenide reagents to higher selenium oxidation states. Many of the synthetic routes reviewed by Sefat to prepare iron-based superconductors~\cite{Sefat13:17} are relevant to oxyselenide synthesis and we give only a brief overview here.

Polycrystalline samples of oxyselenides can be prepared by reacting stoichiometric quantities of reagents in evacuated, sealed quartz ampoules. Quartz begins to soften for temperatures $>$ 1150$^{\circ}$C and so for higher temperature reactions, a small pressure of inert gas may be required to prevent the reaction tube imploding~\cite{Altmannshofer08:634,He11:84,Hawai15:91}.  Care must be taken to ensure all reagents are thoroughly dry and that unwanted gaseous phases will not be formed during the reaction (to avoid reaction tubes exploding). We note that selenium is toxic, and is volatile above $\approx$ 640$^{\circ}$C and so low heating rates (and if necessary, a low temperature dwell) should be used to prevent a build-up of excess Se pressure and the tube exploding. Some reagents have been found to react with quartz and this can often be minimised by placing the reagents in alumina or graphite crucibles inside the quartz tubes. For stoichiometry, an ``oxygen getter" is often used (which can give more flexibility in choice of starting reagents). Typical ``getters" include Al or Zr powders but the getter should be chosen based on the relative thermodynamic stability of various oxides, for example, see the Ellingham diagram for the relevant sample cations and synthesis temperature. The metal powder reacts with oxygen to form the oxide, but the reaction may be predominantly a surface reaction and so some excess oxygen-getter may be necessary. Some cations with two oxidation states close in energy can cause synthesis problems, as highlighted by Pitcher \textit{et al.}~\cite{Pitcher09:48} for CeCuOS; some earlier reports on ``CeCuOS" had been on samples with Cu$^{+}$ vacancies and with some oxidation of Ce$^{3+}$ to Ce$^{4+}$. This is likely to occur in oxyselenide systems also (for example, slightly different unit cell parameters have been reported for Ce$_{2}$O$_{2}$Fe$_{2}$OSe$_{2}$~\cite{Ni11:83,McCabe14:90}), and quenching Ce$_{2}$O$_{2}$Fe$_{2}$OSe$_{2}$ and Ce$_{2}$O$_{2}$FeSe$_{2}$ was found to minimise any phase separation on cooling~\cite{McCabe14:90}.  Ainsworth \textit{et al.} made a careful study of the effect of excess getter and reaction temperature in the sample purity physical properties in Ce$_{2}$O$_{2}$ZnSe$_{2}$ and were able to tune the Ce oxidation state and in turn, the band gap (and colour) and unit cell volume~\cite{Ainsworth15:54}.

Post-synthesis reactions and treatments can also be used to tune and optimise properties. For example, topotactic reactions could be carried out on many of these systems, particularly the layered materials. Hyett \textit{et al.}~\cite{Hyett07:129} were able to tune the antiferromagnetic ordering temperature and the size of the ordered moment in Sr$_{4}$Mn$_{3}$O$_{7.5}$Cu$_{2}$$Q$$_{2}$ by topotactic (oxidative) fluorination or (reducing) oxygen de-intercalation reactions and the same group have used lithium exchange and reductive Li$^{+}$ insertion reactions to tune magnetic and electrochemical properties~\cite{Rutt06:5,Indris06:128}.  Ammonia intercalation into layered FeSe has also been used to tune its superconducting behaviour~\cite{Sedlamaier14:136}.  Other post-synthesis treatments to optimise microstructure can have a significant effect of properties, for example, textured samples of the thermoelectric BiCuOSe (prepared by hot-forging) doubled the carrier mobility and led to a dramatic improvement of thermoelectric behaviour~\cite{Sui13:6}.  Lower-temperature solution methods have also been used to prepare oxyselenides, including nanoplates of $Ln$$_{4}$O$_{4}$Se$_{3}$,~\cite{Gu13:135} as well as the single-step hydrothermal synthesis of polycrystalline BiCuOSe~\cite{Stampler08:47}.  Lower temperature ``metathesis" reactions have proven very successful for accessing low-temperature polymorphs and metastable phases~\cite{Martinolich14:136,Martinolich15:137,Martinolich16:28}

Characterisation of single crystal samples of the arsenide, oxyarsenide, and chalcogenide families of iron-based superconductors has given a much deeper understanding of the unusual (and often anisotropic) behaviour of these materials than could have been gained with polycrystalline samples~\cite{Nitsche10:82,Wen12:74}.  Single crystals of \textit{Ln}FeAsO materials were first grown from a NaCl/KCl flux at high pressure, but the low solubility of the flux at the reaction temperature led to very slow crystal growth~\cite{Karipinski09:469,Zhigadlo08:20,Prozorov09:11}. Similar challenges are faced when preparing single crystals of oxyselenides. Most oxyselenide single crystals have been grown from a flux, usually KCl,~\cite{Ijjaali03:176}or a eutectic mix of NaI/KI,~\cite{Peschke15:641,Nitsche14:640} but CsI~\cite{Tuxworth15:44} I$_{2}$~\cite{Meerschaut162:01,Meerschaut98:137} and Na$_{2}$Se$_{x}$~\cite{Park93:5} fluxes have also been used.

\subsection{Magnetic neutron scattering}

Neutrons are sensitive to both structural and magnetic properties making them ideal for studying the properties of magnetic oxyselenides.  Given the erratic variation of the nuclear cross section with atomic number, neutrons provide complementary structural information to that of x-ray scattering.  We focus here on recent developments in neutron inelastic scattering.  While reviews of the technique have been provided elsewhere in the context of strongly correlated electronic systems~\cite{Collins:book,Shirane:book,Kastner98:70,Birgeneau06:75}, we focus here on recent developments in neutron inelastic scattering.  Reviews on neutron diffractionm including magnetic neutron diffraction, are provided in Ref. \onlinecite{Bacon,Willis,Izyumov}.

In all neutron experiments, a neutron with a fixed incident energy is either elastically scattered off a sample or inelastically scattered either by fluctuations in the lattice (phonons) or through magnetic interactions (magnons).   In the case of unpolarised neutron scattering the scattering, cross section is a function of momentum transfer $\vec{Q}=\vec{k}_{i}-\vec{k}_{f}$ and energy transfer $\hbar \omega = E_{i}-E_{f}$.  Due to instrumentation and source qualities, neutron scattering has historically been most successful at the ``thermal" or lower energy range with typical energy transfers on the order of $\sim$ meV.  This energy scale is well matched for the study of spin interactions and low energy lattice vibrations, however the technique and selection rules associated with it are limited for studying higher energy scales on the order of $\sim$ eV.  Given that magnetism provides a probe of the underlying electronic ground state, we will focus on the magnetic contribution to the scattering cross section here.    We first outline the cross sections for the study of spin-spin correlations which is important in extracting interactions and the magnetic structure.  We then outline the cross section for single-ion excitations and the selection rules associated with these transitions.

\textit{Spin Correlations}: The differential neutron scattering cross section per element of solid angle $d\Omega$, for wavevector transfer $\vec{Q}$ and energy transfer $\hbar \omega$ is,

\begin{eqnarray}
{d^{2}\sigma \over {d \Omega d\omega}}= {(\gamma r_{0})^{2} \over 4} {k_{f} \over k_{i}} S(\vec{Q},\omega),
\end{eqnarray}

\noindent where $(\gamma r_{0})^{2} \over 4$ is 73 mbarns sr$^{-1}$.  In the dipole approximation which is valid at small momentum transfers, 

\begin{eqnarray}
S(\vec{Q},\omega)=g^{2} f^{2}(Q) \sum_{\alpha \beta} (\delta_{\alpha\beta}- \hat{Q}_{\alpha} \hat{Q}_{\beta}) S^{\alpha \beta} (\vec{Q},\omega),
\end{eqnarray}

\noindent where the $S^{\alpha \beta} (\vec{Q},\omega)$ is related to the space and time Fourier transform of the spin-correlation function,

\begin{eqnarray}
S^{\alpha \beta} (\vec{Q},\omega)={1\over {2\pi}} \sum_{ij} \exp(i \vec{Q} \cdot (\vec{R}_{i}-\vec{R}_{j})) \int dt ...\\ \nonumber
e^{i\omega t} \langle S^{\alpha}_{i}(t) S^{\beta}_{j}(t) \rangle.
\end{eqnarray}

\noindent Neutron scattering therefore provides a direct probe of spin-spin correlations and hence the coupling between them.  The geometric term in the sum ($\delta_{\alpha\beta}- \hat{Q}_{\alpha} \hat{Q}_{\beta}$) provides a selection rule that neutron scattering is only sensitive to the component of the magnetic moment perpendicular to the momentum transfer ($\vec{Q}$).  The elastic scattering ($\hbar\omega$=0) component provides information on the magnetic structure through the correlation functions above while the inelastic component ($\hbar\omega\neq$0) gives information on the coupling and energy terms of the magnetic Hamiltonian.  

The scattering intensities measured from the differential cross section can also be related to susceptibility using thermodynamic techniques.  In particular, the structure factor $S(\vec{Q},\omega)$ can be related to the imaginary part of the susceptibility ($\chi''$) using the fluctuation dissipation theorem which states,

\begin{eqnarray}
S^{\alpha \beta} (\vec{Q},\omega)={-1\over \pi} {Im(\chi^{\alpha \beta}(\vec{Q},\omega)) \over {\left[1-\exp\left(-{{\hbar \omega} \over {k_{B} T}}\right)\right]}}.
\end{eqnarray}

\noindent The usual static susceptibility can be found by taking the limit as $\vec{Q}$ and $\omega$ tend to zero.

Magnetic neutron scattering is subject to strict sum rules which allows direct information about the local spin state to be derived.  The elastic ordered magnetic moment derived from neutron scattering is equal to $gS$ where $g$ is the Lande factor and $S$ is the spin of the magnetic species.  On integrating the differential neutron scattering cross section over all momentum transfers and energy, the following relation is obtained.

\begin{eqnarray}
\int d (\hbar \omega) \int d^{3}Q S(\vec{Q},\omega)=S(S+1)
\end{eqnarray}

While the zero moment sum rule provides information on the spin value which can be related to the local crystallline electric field, the first moment sum rule can provide information regarding local spin interactions.  In particular, as shown in Ref. \onlinecite{Hohenberg74:10}, in the case of isotropic Heisenberg exchange interactions, the energy integrated $S(Q)$ takes the form,

\begin{eqnarray}
S(\vec{Q})=-{2 \over 3} {1\over {\epsilon(\vec{Q})}}\sum_{i} J_{d} \langle \vec{S}_{0}\cdot\vec{S}_{\vec{d}}\rangle [1-\cos(\vec{Q}\cdot \vec{d})]
\end{eqnarray}

\noindent where the sum (indicated by the subscript $i$) is over nearest neighbours and the bond distances $\vec{d}$.  This sum rule becomes particularly practical in the simplifying case of where a single mode dominates the inelastic spectrum, termed the single mode approximation.   In this case the full scattering structure factor can be written as $S(\vec{Q},\hbar \omega)=S(\vec{Q})\delta (\hbar\omega - \epsilon(\vec{Q}))$, where the $\delta$-function forces energy conservation.  This sum rule has in the past been applied to spin-chains but also to low-dimensional magnetic systems to extract dimensionality of the interactions~\cite{Hammar98:57,Stone01:64,Hong06:74,Stock09:103}.

Information on the intrinsic spin state can be determined through an integral of the total intensity, as discussed below in the context of the local crystal field environment.  Also, from the first moment sum rule information on the dimensionality of the spin interactions can be obtained through a means which is model independent.  This is particular important when measuring powders where directional information of the crystal is heavily smeared by the powder averaging.

In summary, neutron diffraction provides a means of deriving the magnetic structure and sum rules obtained from the dynamic structure factor (measured through the inelastic neutron cross section) provide information about the local environment and also the exchange parameters.

\textit{Single-ion excitations}:  Whilst the above discussion introduced information that can be determined from neutron scattering cross sections, measurements of single-ion excitations can also give insight into the local crystal field environment. Given the energy scale of neutron instrumentation, these types of measurements are currently best suited for rare earth or lanthanide elements where spin-orbit coupling is large and the magnetism can be understood in terms of assigning a total angular momentum $\vec{J}$ to the ion with eigenvalues of the operators $J^{2}$ and $J_{z}$ being equal to $j(j+1)$ and $m$.  The eigenstates can therefore be written in terms of these two operators as $|j,m\rangle$.  The energy scale associated with changing $j$, termed intermultiplet transitions, for the lanthanides is typically quite large and of the order of $\sim$ eV while the energy scale associated with changing $m$ is much smaller, on order $\sim$ meV (see for example the case of Pr in Refs. \onlinecite{Turberfield88:61,Taylor88:61}).   We therefore confine the discussion here to transitions where $j$ is fixed and only the eigenvalue $m$ changes.

In the dipole approximation for localised magnetic moments, the neutron scattering cross section at small momentum transfers can be written as follows,

\begin{eqnarray}
{d^{2}\sigma \over {d \Omega d\omega}}= {(\gamma r_{0})^{2} \over 4} {k_{f} \over k_{i}} f^{2}(Q) \sum_{n,m} \rho_{n} |\langle n | J_{\perp}|m\rangle|^{2} ... \\ \nonumber
\delta \left(E_{n}-E_{m} -\hbar \omega \right),
\end{eqnarray}

\noindent where $|n \rangle$, $|m \rangle$ are states belonging to a given $J$ multiplet.  The operator $J_{\perp}$ is the component of the total angular momentum operator perpendicular to the scattering vector $\vec{Q}$.  The $\delta$ function enforces energy conservation.  The effect of the form factor $f(Q)$ is to decrease the magnetic-dipole transition intensities as the momentum transfer $Q$ is increased.  At higher momentum transitions, magnetic octupole and higher-order transitions are possible.  These will not be discussed in this review.  Single-ion transitions corresponding to intramultiplet transitions of the form $|n \rangle \rightarrow |m \rangle$ are distinguished between spin-spin correlations discussed above.  We note that single-ion excitations typically lack a strong momentum dependence while spin-spin excitations typical vary rapidly with $Q$.  Dispersing single ion excitations can exist and need to be treated in terms of a multi-level spin-wave analysis.  We will not discuss these cases in this paper.  

Neutron scattering is a powerful and continuously evolving technique for the study of magnetism and electronic phenomena. Historically, it has been confined to lower energy transitions for the spin correlations and intramultiplet transitions outlined above.  One reason for this originates from kinematics and the fact that the neutron has mass meaning that high energy transfers usually correspond to large momentum transfers.  While the energy scales are typically relevant for the study of transitions of interest in condensed matter physics, one limitation is the study of local environment around $d$ transition metal ions where the crystal field excitations are close to $\sim$ eV.  In this context x-ray and optical techniques have played a pioneering role allowing high energy transitions to studied.

\subsection{X-ray and optical spectroscopy}

While efforts have been made to extend neutron scattering to higher energies approaching the $\sim$ eV energy range,~\cite{Kim11:84,Cowley13:88,Stock10:81} studying single-ion transitions in $d$ transition metal ions with neutrons is limited (owing to the energy scale and also current instrumentation) and optical and x-ray techniques are required.

One important technique for studying high energy transitions is Resonant Inelastic X-ray Scattering (RIXS).  If we consider the case where the incident photon energy is close to or above the core electron excitation threshold, the RIXS intensity can be written as follows,~\cite{Kotani01:73}

\begin{eqnarray}
F(\Omega, \omega)=\sum_{j} |\sum_{i} {{\langle j|T|i\rangle \langle i|T|g\rangle} \over {E_{g}-\Omega-E_{i}-i\Gamma_{i}}}|^{2}... \\ \nonumber
\times \delta (E_{g}+\Omega-E_{j}-\omega),
\end{eqnarray}

\noindent where the operator $T$ represents the radiative transition and $\Gamma_{i}$ represents spectral broadening.   The operator $T$ is often calculated considering the dipole approximation and therefore subject to similar selection rules stated above for neutron scattering.  The above expression for $F(\Omega, \omega)$ illustrates that RIXS is the coherent second-order process consisting of the x-ray absorption from $|g\rangle$ to $|i\rangle$ and the x-ray emission from $|i\rangle$ to $|j\rangle$.  Given the high energy scale of x-rays, this technique is particularly useful for studying electronic excitations which are typically on the energy scale of $\sim$ eV and provides a complementary technique to comparatively lower energy scale of neutrons which is primarily useful for investigating collective excitations.

\subsection{Transport and thermodynamic measurements}

Electrical resistivity and thermodynamic measurements are key tools used to characterise oxyselenides, which often have semiconducting properties that can be tuned towards metallic or insulating extremes. Metallic materials show a vanishing resistivity as the temperature is lowered and insulators show a diverging resistivity.  Semiconductors show behaviour between these and the simplest model for the electrical conductivity is an activated behaviour with $\rho=\rho_{a}\exp(E_{a}/k_{B}T)$ with $E_{a}$  being an approximate measure of the band gap.  Resistivity therefore provides a fairly simple means of characterising the electronic properties of materials.

To study the localised properties associated with magnetic moments, magnetic susceptibility provides a means of studying the real part of the magnetic susceptibility $\chi$ (defined above).  The magnetic susceptibility can be related to neutron scattering techniques via the fluctuation-dissipation theorem stated above or also through the Curie-Weiss constant derived through high temperature measurements.  Neutrons are sensitive to the local spin-spin interactions from which a set of exchange constants can be derived based on a model magnetic Hamiltonian.  Assuming Heisenberg exchange, the Curie constant ($\Theta_{CW}$) derived from susceptibility can be related to the exchange constants,

\begin{eqnarray}
\Theta_{CW}={1\over 3} S(S+1)\sum_{n}J_{n}
\end{eqnarray}

\noindent with the sum being over coupled neighbours.  The exchange constants derived from scattering experiments can therefore be cross checked against susceptibility and in particular magnetisation.

In the case of itinerant magnetism, the resistivity from the spin fluctuations can be calculated from the neutron scattering cross section using the following formula,~\cite{Moriya90:59},

\begin{eqnarray}
\rho(T) \propto T \int_{-\infty}^{\infty} {E \over T} d\left({E \over T} \right) {{e^{E/T}} \over {(e^{E/T}-1)^{2}}}\int d^{3}q \chi '' (\vec{q},E).
\label{equation_rho} \nonumber
\end{eqnarray}

\noindent Here, $\chi''$ is the spin susceptibility related to the measured neutron intensity $I(\vec{Q},E)\propto S(\vec{Q},E)={1\over \pi} [n(E)+1]\chi''(\vec{Q},E)$ with $[n(E)+1]$ being the bose factor.   The temperature dependence of the resistivity can therefore be related to the spin fluctuation in itinerant magnets.  This is discussed below in the context of comparing iron based chalcogenides and oxyselenides.

\begin{figure}[t]
\includegraphics[width=8.5cm] {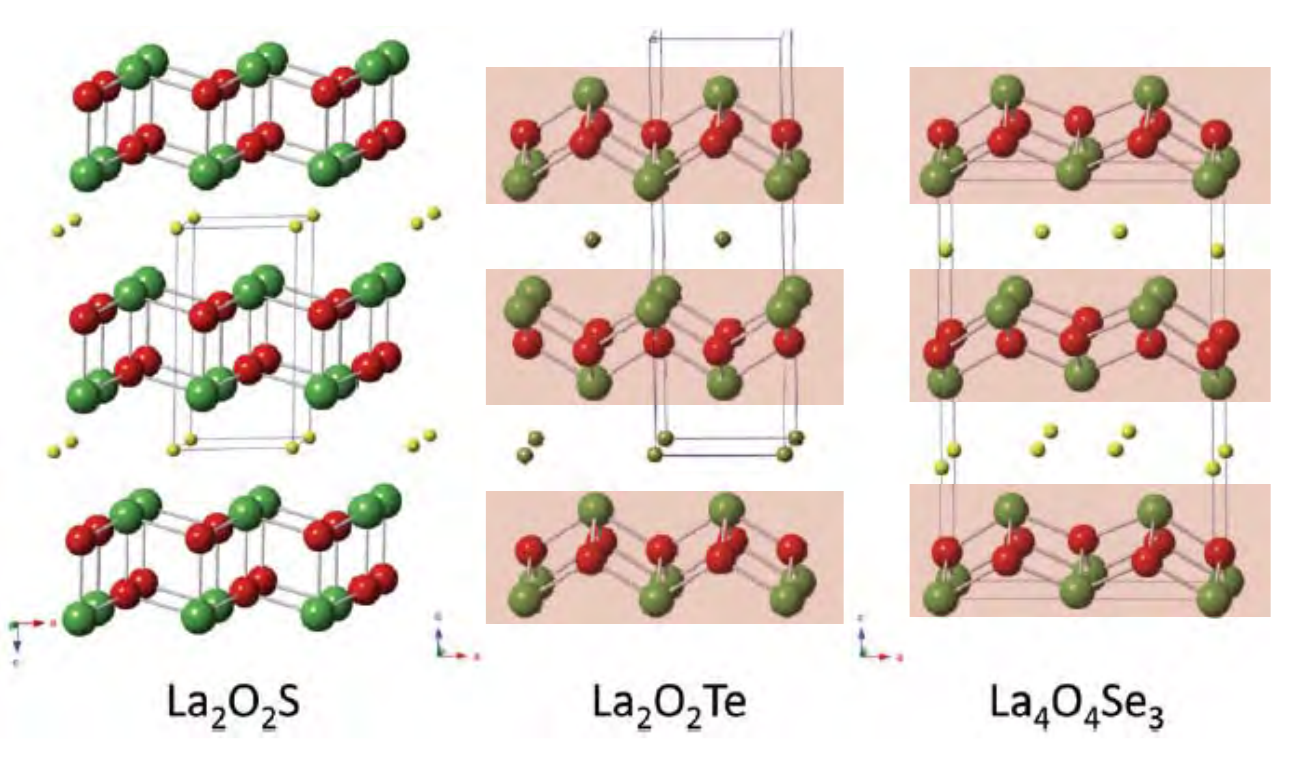}
\caption{\label{structure0}   The structures of La$_{2}$O$_{2}$S, La$_{2}$O$_{2}$Te and La$_{4}$O$_{4}$Se$_{3}$.  La=green, O=red, S/Te/Se=yellow, respectively.  This figure is reproduced from Ref. \onlinecite{Tuxworth15:44}.  Fluorite-like [La$_{2}$O$_{2}$]$^{2+}$ layers are highlighted in red.}
\end{figure}

\section{Oxyselenide structural families}

\begin{figure*}[t]
\includegraphics[width=17.7cm] {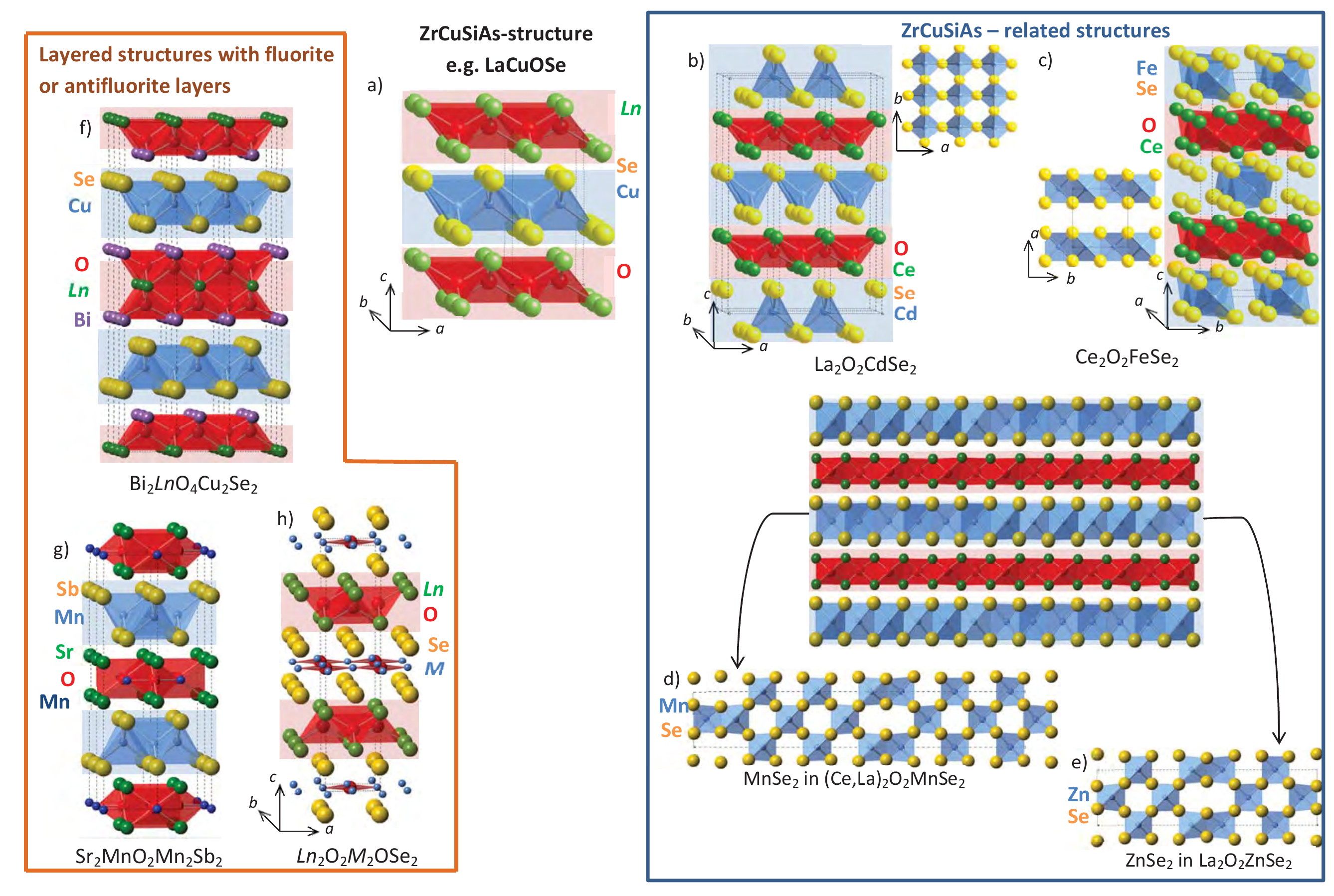}
\caption{\label{universal}  $(a)$ ZrCuSiAs structure adopted by LaCuOSe shown in centre; right hand side shows ZrCuSiAs-related structures with $(b)$ checkerboard-ordered La$_{2}$O$_{2}$CdSe$_{2}$, $(c)$ stripe-ordered Ce$_{2}$O$_{2}$FeSe$_{2}$ and $M$Se$_{2}$ layers in $(d)$ (CeLa)$_{2}$O$_{2}$MnSe$_{2}$ and $(e)$ La$_{2}$O$_{2}$ZnSe$_{2}$; left hand side of figures shows layered structure types including $(f)$ Bi$_{2}Ln$O$_{4}$Cu$_{2}$Se$_{2}$, $(g)$ Sr$_{2}$MnO$_{2}$Mn$_{2}$Sb$_{2}$, and $(h)$ $Ln$$_{2}$O$_{2}$$M$$_{2}$OSe$_{2}$.  La/Ce/Sr, Bi and $M$ ($M$=Cu, Cd, Fe, Mn, Zn), O and Se shown in green, purple, blue, red, and yellow, respectively.  Fluorite-like [$Ln_{2}$O$_{2}$]$^{2+}$ layers are highlighted in red whilst antifluorite-like [$M_{x}$Se$_{2}$] layers are highlighted in blue to emphasise the relationships between structure types. }
\end{figure*}

It is helpful to begin by considering the structural chemistry of oxyselenides and the common structural units. A previous review of layered oxychalcogenides and oxypnictides describes the structures of many transition metal oxyselenides~\cite{Clarke08:47} and allows us to highlight some structural features common to most oxyselenides:
\begin{itemize}
    \item The different sizes and characters of 1st row oxide O$^{2-}$ and 3rd row selenide Se$^{2-}$ anions usually give rise to anion-ordering. 
    \begin{itemize}
        \item The hard, polarising O$^{2-}$ anions tend to be coordinated by harder cations (often \textit{Ln}$^{3+}$ ions, as in \textit{Ln}$_{10}$OSe$_{14}$ phases~\cite{Weber12:2}) and it can be helpful to consider the parts of the structure in terms of oxide-centred tetrahedra~\cite{Krivo13:113}.
       \item ``Softer" transition metals are usually coordinated by the more covalent Se$^{2-}$ ions.
    \end{itemize}
    \item This anion-ordering often gives layered structures with quite different properties associated with the more ionic oxide and more covalent selenide layers which may be electronically isolated from one another. This anisotropy is often key to understanding the properties of the whole material.
\end{itemize}

Compared with oxides, mixed-anion systems, such as the oxyselenides discussed here, allow us to prepare materials with unusual cation coordination environments (often in low oxidation states). This, combined with their layered nature, gives them often unique properties. The structures of several oxyselenides and the relationships between them have been described in Ref. \onlinecite{Clarke08:47} and we discuss below only some of the dominant or more recent oxyselenide structure types to allow us to consider their electronic structures and properties in more depth later in this review.

\subsection{\textit{Ln} -- O -- Se phases}

The \textit{Ln} -- O -- Se phases with no transition metal provide illustrations of the structural role of the two anions of different characters.  The \textit{Ln}$_{4}$O$_{4}$Se$_{3}$ (\textit{Ln} = La - Nd, Eu - Er, Yb and Y) family adopts crystal structures composed of fluorite-like [\textit{Ln}$_{4}$O$_{4}$]$^{4+}$ oxide sheets built from edge-linked \textit{Ln}$_{4}$O tetrahedra. These sheets are separated by layers containing both  Se$^{2-}$ and diselenide  Se$_{2}$$^{2-}$ anions~\cite{Strobel08:47,Tuxworth15:44}.   Whilst the [\textit{Ln}$_{4}$O$_{4}$]$^{4+}$ sheets change little with \textit{Ln}, the arrangement of Se$^{2-}$ and Se$_{2}$$^{2-}$ anions in the interlayers varies with \textit{Ln}$^{3+}$ ionic radii~\cite{Tuxworth15:44}.   Long-range antiferromagnetic order occurs for Ln = Gd, Tb and Dy phases but has not been observed down to 1.8 K for other analogues.~\cite{Tuxworth15:44,Strobel08:47} Strobel \textit{et al.} highlight the magnetic frustration within the $Ln-O$ network as a result of arranging magnetic Ln$^{3+}$ ions on a tetrahedral motif.~\cite{Strobel08:47}  Four compositions have been reported for this family of materials including $A$$_{10}$OSe$_{14}$ ($A$=La-Nd)~\cite{Weber01:627,Wu07:222,Weber12:2}, $A$$_{2}$OSe$_{2}$ ($A$=Pr, Gd)~\cite{Tougait00:56,Weber01:627}, $A$$_{4}$O$_{4}$Se$_{3}$ ($A$=La-Nd, Sm)~\cite{Weber01:627,Dugue70:26,Strobel08:47}, and $A$$_{2}$O$_{2}$Se ($A$=La,Pr,Nd, Sm, Gd, Er, Ho, Yb)~\cite{Weber01:627,Eick60:13}.   The structures are illustrated in Fig. \ref{structure0}.

\subsection{\textit{Ln} -- O -- \textit{M} -- Se phases}
\subsubsection{ZrCuSiAs structures and related cation-ordered phases}

The fluorite-like layers of edge-linked \textit{Ln}O$_{4}$ tetrahedra are widespread in oxyselenide structural chemistry and most \textit{Ln} -- O -- \textit{M} -- Se phases contain this motif or variations of it. Perhaps the simplest structure adopted by \textit{Ln} -- O -- \textit{M} -- Se phases is the ZrCuSiAs structure,~\cite{Johnson74:11} also referred to as the 1111 structure adopted by \textit{Ln}FeAsO parent phases to the iron-based superconductors,~\cite{Kamihara08:130} and this structure type and variations upon it dominate oxyselenide chemistry. Stoichiometric oxyselenides in this family include \textit{Ln}CuOSe (\textit{Ln} = La - Sm)~\cite{Kamihara08:130,Ueda00:77,Ueda03:15,Hiramatsu04:14,Hiramatsu04:108,Hiramatsu08:20,Mizoguchi11:133,Llanos04:178,Llanos06:41} containing monovalent Cu$^{+}$ ions. These materials are particularly well known for their wide band gap, optically transparent and p-type semiconducting behaviour~\cite{Ueda00:77,Ueda03:15,Hiramatsu08:20}.  Their tetragonal crystal structure is built from fluorite-like oxide layers of edge-linked O\textit{Ln}$_{4}$ tetrahedra separated by anti-fluorite-like layers of edge-linked CuSe$_{4}$ tetrahedra (Fig. \ref{universal} $(a)$). Replacing monovalent Cu$^{+}$ ions with divalent \textit{M}$^{2+}$ ions (e.g. \textit{M} = Mn$^{2+}$, Fe$^{2+}$, Zn$^{2+}$, Cd$^{2+}$) leads to half-occupancy of \textit{M}Se$_{4}$ tetrahedra. These \textit{M}$^{2+}$ sites can be occupied in a disordered fashion (as originally reported for CeMn$_{0.5}$OSe)~\cite{Ijjaali03:176}  but this partial occupancy often gives rise to ordering of the \textit{M}$^{2+}$ ions. This cation ordering might follow a checkerboard arrangement (as in La$_{2}$O$_{2}$CdSe$_{2}$,~\cite{Hiramatsu04:14,Hiramatsu04:108}\  Fig. \ref{universal} $(b)$), a stripe arrangement (as in Ce$_{2}$O$_{2}$FeSe$_{2}$,~\cite{McCabe11:47,McCabe14:90} Fig. \ref{universal} $(c)$) or a combination of these. Intermediate structures, containing stripe- and checkerboard-ordered regions were reported for $Ln_{2}$O$_{2}$ZnSe$_{2}$~\cite{Tuxworth13:52,Ainsworth15:54} (Fig. \ref{universal} $(e)$ ) and (La,Ce)$_{2}$O$_{2}$MnSe$_{2}$ (Fig. \ref{universal} $(d)$)~\cite{Peschke15:641} Ainsworth, Wang \textit{et al.}~\cite{Ainsworth15:xx,Wang15:27} have shown recently that the ``infinitely adaptive" ordering in the La$_{2-x}$Ce$_{x}$O$_{2}$\textit{M}Se$_{2}$ solutions can be tuned between the stripe and checkerboard extremes (via incommensurately-modulated structures) by the \textit{Ln}$^{3+}$ ionic radius and Peschke and Johrendt~\cite{Peschke15:641} have extended this to the La$_{2-x}$\textit{Ln}$_{x}$O$_{2}$MnSe$_{2}$ (\textit{Ln} = Pr, Nd) solid solutions and confirmed the role of \textit{Ln}$^{3+}$ ionic radius.

\subsubsection{$\beta$-La$_{2}$O$_{2}$$M$Se$_{2}$ and related polymorphs}

\begin{figure}[t]
\includegraphics[width=9cm] {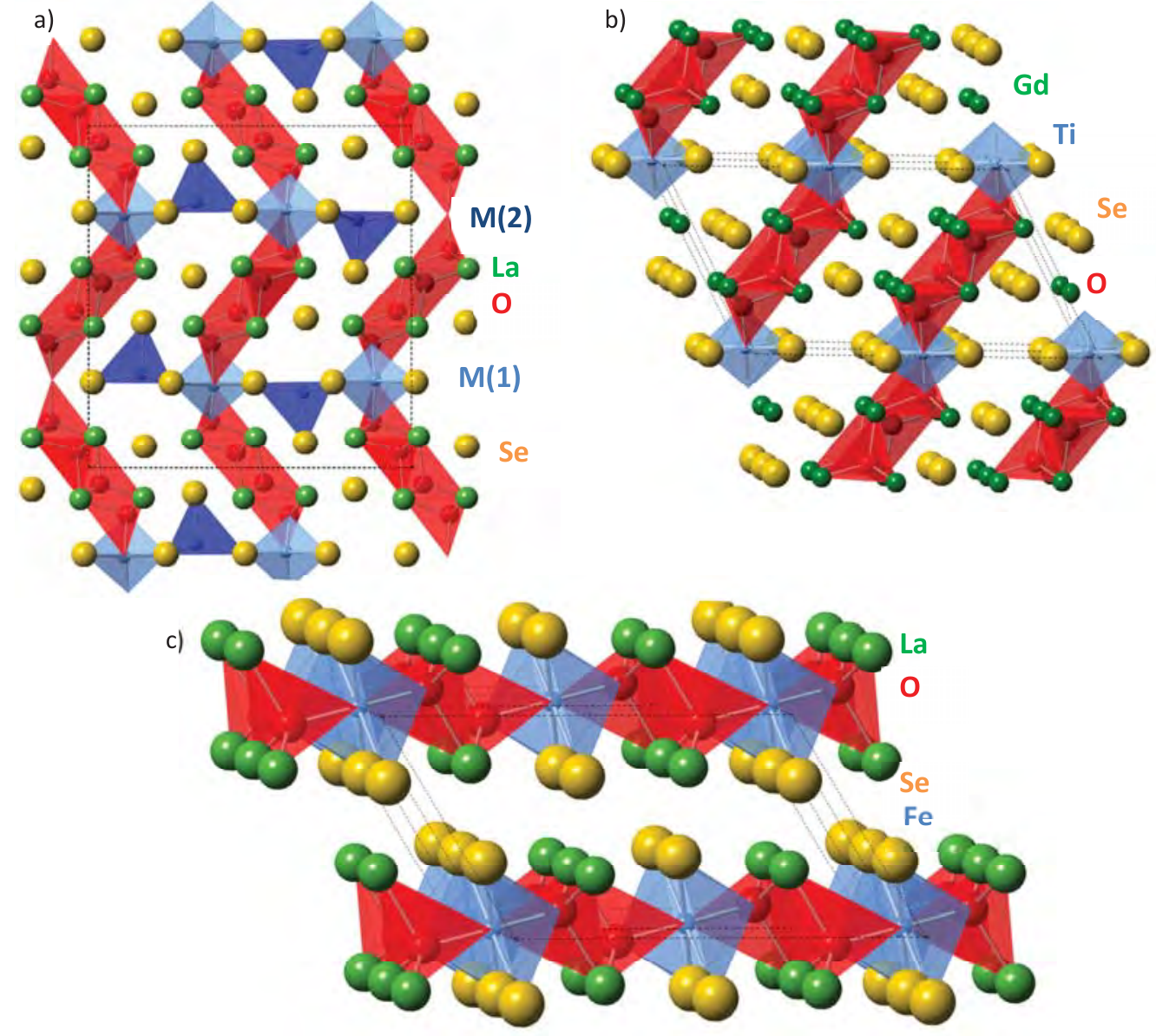}
\caption{\label{structure3}  $(a)$ $\beta$-La$_{2}$O$_{2}$$M$Se$_{2}$ structure adopted for $M$ = Mn, Fe, $(b)$ $Ln$$_{4}$O$_{4}$TiSe$_{4}$ structure and $(c)$ low temperature monoclinic polymorph of La$_{2}$O$_{2}$FeSe$_{2}$ with only pseudo-octahedral FeSe$_{4}$O$_{2}$ coordination. $Ln$, $M$ ($M$= Mn, Fe, Ti) O and Se ions are shown in green, blue, red and yellow, respectively. }
\end{figure}

Initial attempts to prepare ZrCuSiAs-related La$_{2}$O$_{2}$\textit{M}Se$_{2}$ (\textit{M} = Mn, Fe) phases lead to the discovery of a new structural family, $\beta$-La$_{2}$O$_{2}$\textit{M}Se$_{2}$. In this $\beta$-structure, the fluorite-like [\textit{Ln}$_{2}$O$_{2}$]$^{2+}$ layers are broken into ribbons and arranged in a herringbone-like fashion, separating \textit{M}Se$_{2}$ layers (Fig. \ref{structure3} $(a)$). This leaves the \textit{M}(2) site in roughly tetrahedral (\textit{M}(2)Se$_{4}$) sites, while \textit{M}(1) ions are coordinated by both O$^{2-}$ and Se$^{2-}$ anions in pseudo-octahedral coordination~\cite{McCabe10:22}.  This structure is very closely related to the \textit{Ln}$_{4}$O$_{4}$TiSe$_{4}$ structure with Ti$^{4+}$ cations occupying only the \textit{M}(1) sites (Fig. \ref{structure3} $(b)$)~\cite{Meerschaut01:162,Tuxworth14:210}.

\onlinecite{Nitsche14:640} revealed the polymorphism of the \textit{Ln}$_{2}$O$_{2}$FeSe$_{2}$ (\textit{Ln} = La, Ce) systems and were able to tune the iron coordination environment with synthesis temperature. At high temperatures, stripe-ordered ZrCuSiAs-related phases were formed with tetrahedrally-coordinated Fe$^{2+}$ ions; at intermediate temperatures the $\beta$-phases were formed with both tetragonal and pseudo-octahedral Fe$^{2+}$ coordination.  At low temperatures a new phase, with only pseudo-octahedral coordination of Fe$^{2+}$ was formed (Fig. \ref{structure3} $(c)$).

\subsubsection{ZrCuSiAs-modified structures}

\onlinecite{Clarke08:47}  describe how the ZrCuSiAs structure of $Ln$CuOSe can be modified to accommodate thicker fluorite- or antifluorite-like layers, or even additional layers.  One example of the latter is Bi$_{2}$\textit{Ln}O$_{4}$Cu$_{2}$Se$_{2}$ in which [Bi$_{2}$$Ln$O$_{4}$]$^{-}$ blocks (with $Ln^{3+}$ in square-prismatic coordination sandwiched between two Bi - O fluorite-like layers) alternate with [Cu$_{2}$Se$_{2}$]$^{+}$ sheets (Fig. \ref{universal} $(f)$), stabilising the mixed-valent Cu ions and giving metallic conductivity~\cite{Chou15:2015,Evans02:8}. Another possibility involves swapping the fluorite-like oxide layers for oxygen-deficient perovskite-like \textit{A}$_{2}$MnO$_{2}$ oxide layers to give the Sr$_{2}$MnO$_{2}$Mn$_{2}$Sb$_{2}$-type structure (Fig. \ref{universal} $(g)$),~\cite{Brechtel79:34} variations of which are adopted by several oxychalcogenides~\cite{Jin12:51,Zhu97:119,Herkelrathr08:130,Zhu97:130,Otzschi99:117,Zhao14:53,Tan14:598}.  These can also be modified to include thicker antifluorite-like Cu$_{2}$S$_{2}$ layers in the series Sr$_{2}$MnO$_{2}$Cu$_{2m-\delta}$S$_{m+1}$~\cite{Gal06:128,Barrier03:102}.

\subsubsection{$Ln$$M$OSe$_{2}$ - \textit{M}$^{3+}$ ions coordinated by selenide}

\begin{figure}[t]
\includegraphics[width=8.5cm] {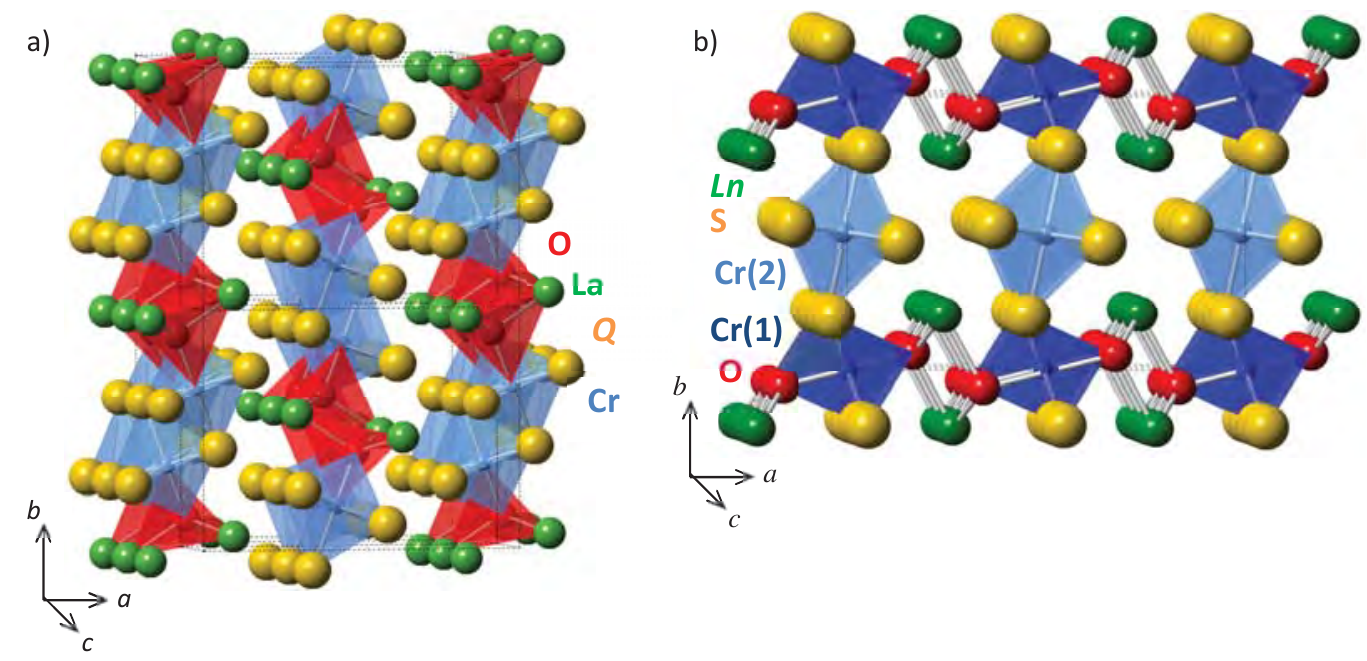}
\caption{\label{structure6}  Structure of $(a)$ LaCrO$Q$$_{2}$, $(b)$ $Ln$CrOS$_{2}$ ($Ln$ = Pr, Nd). $(c)$ $Ln$, Cr, O and Q ions are shown in green, blue, red and yellow, respectively.}
\end{figure}

Relatively few \textit{Ln} - O - \textit{M} - Se phases with \textit{M}$^{3+}$ coordinated by selenide are known (although \textit{M}$^{3+}$ oxysulfide chemistry is more extensive~\cite{Luo13:42,Ceolin76:32,Jaulmes78:34,Jaulmes82:38,Ogisu08:112,Dugue80:36,Dugue80:362,Winterberger89:79} presumably due to the redox chemistry and the challenge of maintaining the reduced Se$^{2-}$ anions in the presence of the more oxidising \textit{M}$^{3+}$ ions. This challenge has been overcome in oxyselenide systems for \textit{M}$^{3+}$ = Ga$^{3+}$ and Cr$^{3+}$. $Ln$GaOSe$_{2}$ is again built from fluorite-like [La$_{2}$O$_{2}$]$^{2+}$ oxide layers but these are separated by [Ga$_{2}$Se$_{4}$]$^{2-}$ double layers with the relatively small Ga$^{3+}$ ions in quite distorted GaSe$_{4}$ tetrahedra~\cite{Benazeth84:40}.

The crystal structures of the $Ln$CrOSe$_{2}$  phases are quite different and vary with \textit{Ln}$^{3+}$ (Fig. \ref{structure6}).  The oxyselenide and oxysulfide LaCrO\textit{Q}$_{2}$ (\textit{Q} = S, Se) are isostructural with ribbons of edge-linked pseudo-octahedral Cr$Q$$_{5}$O extending along \textit{c}, linked by fluorite-like chains ({Fig, \ref{structure3} (a))~\cite{Winterberger87:70,Dugue80:36}.  The crystal structure of CeCrOSe$_{2}$ is again different (and analogous to that adopted by several $Ln$CrOS$_{2}$ phases~\cite{Winterberger89:79,Winterberger87:70}) with the Cr-$Q$ ribbons of LaCrO\textit{Q}$_{2}$ broken to form chains of edge-linked CrSe$_{4}$O$_{2}$ pseudo-octahedra extending along \textit{c} and linked along \textit{a} by fluorite-like oxide chains. Unlike in LaCrO\textit{Q}$_{2}$, the mixed Ce - O - Cr - Se layers are also linked along \textit{b} by edge-linked CrSe$_{6}$ octahedra~\cite{Winterberger87:70}.  The structural chemistry and physical properties of these $Ln$CrOSe$_{2}$ phases have not been fully explored but the related oxysulfides show interesting magnetic behaviour with strong coupling between the $Ln^{3+}$ and Cr$^{3+}$ magnetic sublattices~\cite{Winterberger87:70,Winterberger89:79,Takano99:85,Takano02:122}.

Recently a series of insulating $A$$_{2}$O$_{2}B_{2}$Se$_{3}$ ($A$ = Sr, Ba; $B$ = Bi, Sb) oxyselenides, consisting of double-chains of edge-linked BiSe$_{4}$O square pyramids, have been reported~\cite{Panella16:28}. These materials are structurally-related to the $Ln$OBiS$_{2}$ superconducting family~\cite{Mizoguchi11:133,Yazici13:93,Lei13:52,Lin13:87} which also contain Bi-centred square-based pyramids, compatible with the ``inert pair" Bi$^{3+}$ and Sb$^{3+}$ ions. The structures adopted by these two families can be considered to be built from fluorite-related [$A$O] units (layers, ribbons or discrete units) separating [$B_{2}X_{3}$] layers: the superconducting bismuth sulfides can be written [$Ln$O]$_{2}$[Bi$_{2}$S$_{3}$]S and the $A$$_{2}$O$_{2}$$B$$_{2}$Se$_{3}$ family as [$A$O]$_{2}$[$B_{2}X_{3}$], (reflecting the additional anion needed for charge balance with the trivalent $Ln^{3+}$ ions in the $Ln$OBiS$_{2}$ superconducting family)~\cite{Panella16:28}. This results in two-dimensional Bi$X_{2}$ and fluorite-like $A$O layers in the $Ln$OBiS$_{2}$ superconducting family, whilst the $A_{2}$O$_{2}B_{2}$Se$_{3}$ family consists of quasi-one-dimensional ribbons of edge-linked BiSe$_{4}$O square pyramids, linked to fluorite-like SrO fragments by the apical Bi - O bond. This appreciation of the square pyramidal units and their connectivity with fluorite-like units opens the possibility to design and prepare further oxychalcogenides (of varying dimensionality) containing ``inert pair" ions~\cite{Panella16:28}.

\subsubsection{\textit{Ln}$_{2}$O$_{2}$$M$$_{2}$OSe$_{2}$ materials}

\begin{figure}[t]
\includegraphics[width=8.75cm] {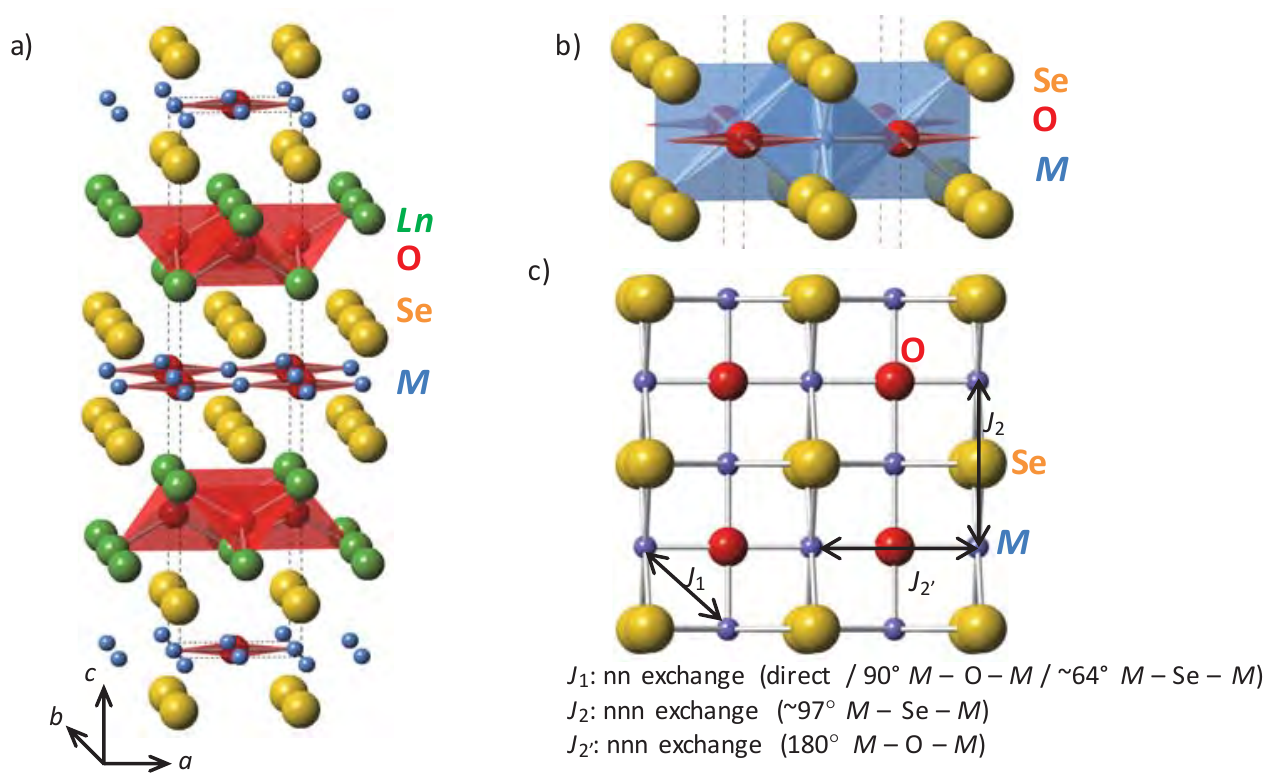}
\caption{\label{structure5}   Structure of $Ln$$_{2}$O$_{2}$$M$$_{2}$OSe$_{2}$ showing $(a)$ tetragonal unit cell, $(b)$ pseudo-octahedral FeSe$_{4}$O$_{2}$ coordination polyhedra, and $(c)$ Fe-Se-O layers illustrating exchange interactions.   $Ln$, $M$ ($M$ = Mn, Fe, Co) O and Se ions are shown in green, blue, red and yellow, respectively.}
\end{figure}

An important family of oxyselenides adopt an ``anti" form of the Sr$_{2}$MnO$_{2}$Mn$_{2}$Sb$_{2}$-type structure with cation and anion sites swapped: in the \textit{Ln}$_{2}$O$_{2}$\textit{M}$_{2}$O\textit{Q}$_{2}$ (\textit{M} = Mn, Fe, Co; \textit{Q} = S, Se), fluorite-like [\textit{Ln}$_{2}$O$_{2}$]$^{2+}$ layers and [Fe$_{2}$O]$^{2+}$ sheets (in an anti-CuO$_{2}$ type arrangement~\cite{Park93:5}) are separated along \textit{c} by layers of Se$^{2-}$ anions (Figs. \ref{universal} $(h)$ and \ref{structure5}). This tetragonal structure was first reported by Mayer \textit{et al.}~\cite{Mayer10:81} for La$_{2}$O$_{2}$Fe$_{2}$O\textit{Q}$_{2}$ (\textit{Q} = S, Se) and has since been extended to include a range of \textit{Ln}$^{3+}$ and $M^{2+}$ ions (\textit{Ln} = La, Ce, Pr, Nd, Sm; M = Mn, Fe, Co)~\cite{Free11:23,Fuwa10:132,Fuwa10:22,Kabbour08:130,Ni10:82,Ni11:83,Liu11:83,Lei12:86,Liu15:221,Popovic14:89}. 

The \textit{M}$^{2+}$ environment in the \textit{Ln}$_{2}$O$_{2}M_{2}$O\textit{Q}$_{2}$ structure is unusual with coordination by two O$^{2-}$ ions within the plane and by four Se$^{2-}$ ions above and below the plane, giving \textit{M}O$_{2}$\textit{Q}$_{4}$ pseudo-octahedral coordination (similar to that of the \textit{M}(1) sites in the $\beta$-La$_{2}$O$_{2}$\textit{M}Se$_{2}$ and \textit{Ln}$_{4}$O$_{4}$TiSe$_{4}$ structures described above). Closely related Na$_{2}$OFe$_{2}$S$_{2}$ has the same [Fe$_{2}$O]$^{2+}$ sheets but here, the fluorite-like oxide layers are replaced by layers of Na$^{+}$ ions~\cite{He11:84} and Na$_{2-x}$Cu$_{2}$Se$_{2}$Cu$_{2}$O also has analogous [Cu$_{2}$O] sheets as well as [Cu$_{2}$Se$_{2}$]$^{2-}$ layers~\cite{Park93:5}. All these systems order antiferromagnetically on cooling with N\'{e}el temperatures that can be tuned with $Ln$ ionic radius and will be discussed in more detail below. 

Free \textit{et al.}~\cite{Free11:23} considered the role on \textit{Ln}$^{3+}$ size on the range of first row transition metals that could be accommodated on the \textit{M} sites in terms of size-mismatch between [\textit{Ln}$_{2}$O$_{2}$]$^{2+}$ and [$M$$_{2}$O]$^{2+}$ layers, with $M$=Fe$^{2+}$ found to be compatible with the widest range of lanthanides.  Similar compounds were discussed by Ni \textit{et al.}~\cite{Ni11:83} for $M$=Fe$^{2+}$.  The redox chemistry of the $M$ ions (ie balancing the oxidising ability of the M$^{3+}$ ions in the presence of selenide ions) is also likely to play a role.

The [\textit{Ln}$_{2}$O$_{2}$]$^{2+}$ fluorite-like layers are analogous to those in the ZrCuSiAs structure type and are subject to similar structural distortions induced by \textit{Ln}$^{3+}$ ions.   While extra structural peaks were reported in Ref. \onlinecite{Free11:23} in temperature dependent x-ray diffraction data for $M$=Mn$^{2+}$ variants, these could not be indexed by a commensurate unit cell.  Besides this, no observable low temperature structural transitions have been reported for these compounds with the exception of $Ln$=Pr$^{3+}$~\cite{Free11:23}.  Pr$^{3+}$ (4\textit{f}$^{2}$, a non-Kramers ion) on the high symmetry (\textit{C}$_{4\textit{v}}$ or 4\textit{mm} site in the \textit{I}4/\textit{mmm} tetragonal crystal structure) tends to drive an orthorhombic distortion, lowering the symmetry of the Pr$^{3+}$ coordination environment (e.g. to \textit{C}$_{2\textit{v}}$ or 2\textit{mm} in the low temperature \textit{Immm} phase). This has been observed in the mixed-anion ZrCuSiAs-structure phases PrMnSbO~\cite{Kimber10:82} and Pr$M$AsO ($M$ = $Mn$, $Fe$),~\cite{Wildman15:54,Kimber08:78} as well as in the oxyselenides Pr$_{2}$O$_{2}$Mn$_{2}$OSe$_{2}$ and Pr$_{2}$O$_{2}$Fe$_{2}$OSe$_{2}$~\cite{Free11:23,Oogarahxx}.     For Pr$_{2}$O$_{2}$Mn$_{2}$OSe$_{2}$ a structural transition to an orthorhombic unit cell is found for temperatures below $\sim$ 50 K~\cite{Free11:23}  with $a$=4.08616(3) \AA\ and $b$=4.09417(4) \AA~\cite{Free11:23}.

\section{Magnetic Properties}

In this section we provide a review of the magnetic properties of the oxyselenides.   This section is divided into three parts discussing the local magnetism, magnetic structures, and then magnetic interactions.  We first provide a discussion of the local magnetic environment in terms of the crystalline electric field which is central to understanding the magnetic ground state in any material.  We then discuss the various magnetic structures reported in the oxyselenides.  With the recent interest in strongly correlated electron systems, the studies have primarily focussed on two dimensional variants.  We then finish with a discussion of the available work on the magnetic interactions primarily probed through neutron inelastic scattering.

\subsection{Local magnetism}

Many oxyselenides are based upon two magnetic sites with one being a $3d$ transition metal ion and the other a rare earth lanthanide site.  The local magnetism on both sites is treated differently with the rare-earth magnetism treated in terms of $j-j$ coupling and the $3d$ site being understood in terms of $L-S$ coupling.   Here we discuss the results for the two sites with a discussion of the rare earth local magnetism followed by a discussion of the local magnetism on the $3d$ transition metal ion site.

\subsubsection{Rare Earth local magnetism}

\begin{figure}[t]
\includegraphics[width=7.5cm] {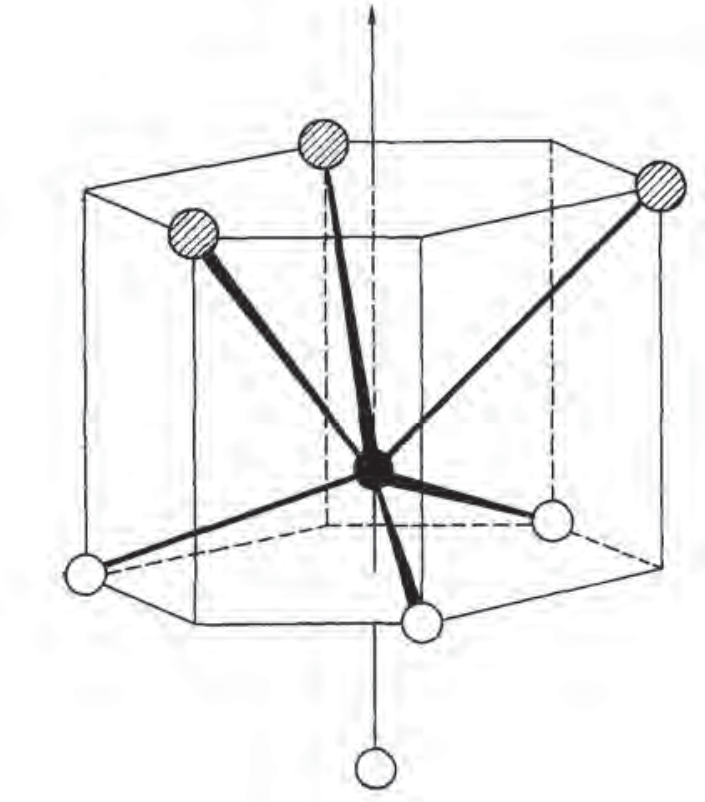}
\caption{\label{rare_cef}  The local coordination site taken from Ref. \onlinecite{Mignod74:62} which studied the local crystalline electric field in Tb$_{2}$O$_{2}$S.  Nearest neighbours of a R$^{3+}$ ion in oxysulides and oxyselenides (Tb,Dy,Ho)$_{2}$O$_{2}$(S,Se).  The empty and cross-hatched circles are oxygen and sulfur/selenium, respectively.}
\end{figure}

Oxyselenides containing only rare earth magnetic ions were a topic of interest in the 1970's.  Rare earth local magnetism is treated in terms of $j-j$ coupling where the spin-orbit coupling is much larger than the local crystalline electric field.  Typical energy scales for the crystalline electric field are $\sim$ meV while the spin-orbit coupling is on order of $\sim$ 1 eV.  In the $j-j$ coupling scheme, the spin-orbit Hamiltonian is therefore diagonalised first with the crystalline electric field treated as a perturbation on the ground state.  Following the formalism in terms of Steven's operators, the single ion Hamiltonian ($H_{CEF}$) can be written as,

\begin{eqnarray}
H_{CEF}=\sum_{l,m}V^{m}_{l}\theta_{l}O^{m}_{l}
\end{eqnarray}

\noindent where $V^{m}_{l}$ are the adjustable crystal field parameters, $O^{m}_{l}$ the Stevens operators (which are functions of the total angular momentum operator $J$, with $J^{2}|j,m\rangle=j(j+1)|j,m\rangle$ and $J_{z}|j,m\rangle=m|j,m\rangle$), and $\theta_{l}$ are the multiplicative factors that depend on the free ion level and are tabulated~\cite{Hutchings65:45}.  An important result in understanding crystalline electric fields is Kramer's theorem which states that the crystal electric field scheme of ions with half-integer $j$ is made up of doublets.  This is always the case for such ions and the degeneracy can only be broken with a field which breaks time reversal symmetry, such as a magnetic field but not an electric.

\begin{figure}[t]
\includegraphics[width=8.5cm] {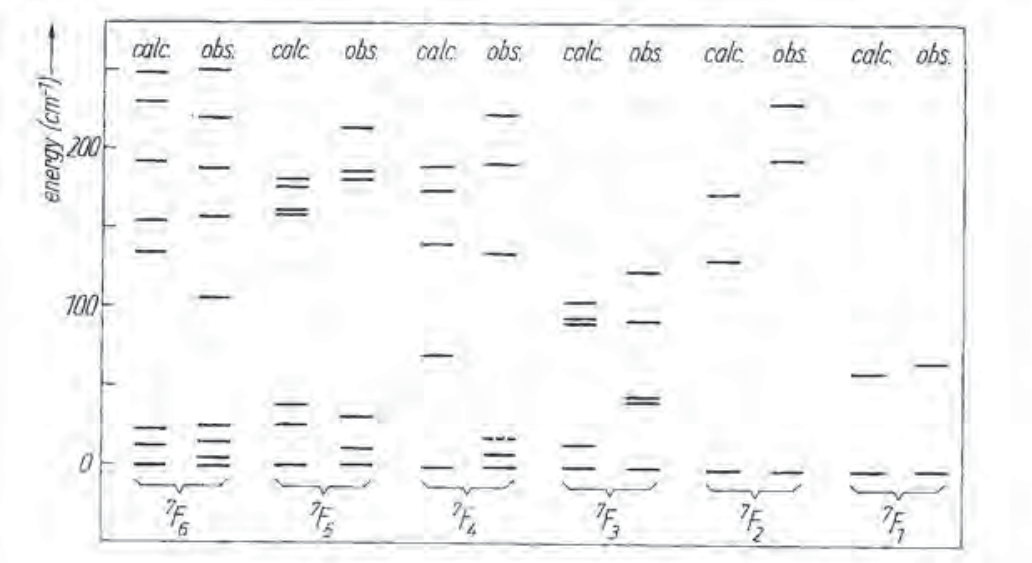}
\caption{\label{rare_scheme}  Crystal field scheme determined using Stevens operators and also measured with optical spectroscopy.  The figure is taken from Ref. \onlinecite{Mignod74:62} and shows a comparison between the observed and calculated levels of $^{7}F_{J}$ manifold.}
\end{figure}

Some of the first rare earth oxyselenides studied were (Tb,Dy,Ho)$_{2}$O$_{2}$(S,Se) and the local coordination geometry is shown in Fig. \ref{rare_cef} for  Tb$_{2}$O$_{2}$S~\cite{Abbas74:14,Abbas73:12}.  The case of Tb$_{2}$O$_{2}$(S,Se)~\cite{Abbas73:12,Mignod74:62} is particularly interesting as the magnetic structure involves a large canting of the Tb$^{3+}$ moment which is not expected given the symmetry of the lattice.  It was therefore concluded in this system that the ground state could not be considered as a doublet, but as a set of doublets with the energy scale between the low-energy crystal fields to be small and of order $\leq$1-2 meV.  These were studied using optical spectroscopy and compared against predictions from the susceptibility and the magnetic structure with reasonable agreement being obtained (shown in Fig. \ref{rare_scheme}).  The magnetic structure of Dy$_{2}$O$_{2}$(S,Se) was found to be more unaxial consistent with expectations from the crystal symmetry.

These early studies of rare earth oxyselenides and sulfides illustrate a common theme reflected in studies on oxypnictides and also two dimensional oxyselenides that these systems display well defined crystal field excitations and levels from the rare earth site.  Generally these excitations do not disperse substantially indicating weak coupling between the rare earth sites.  The local nature of these excitations has been used to probe crystalline electric field symmetry in rare earth oxypnictides.~\cite{Gore11:83,Xiao13:88}  The well defined nature of the crystal field excitations is also indicative of their localised nature in contrast to expectations for itinerant systems which would not show well defined excitations in energy.

\subsubsection{Transition metal ion local magnetism}

\begin{figure}[t]
\includegraphics[width=7.5cm] {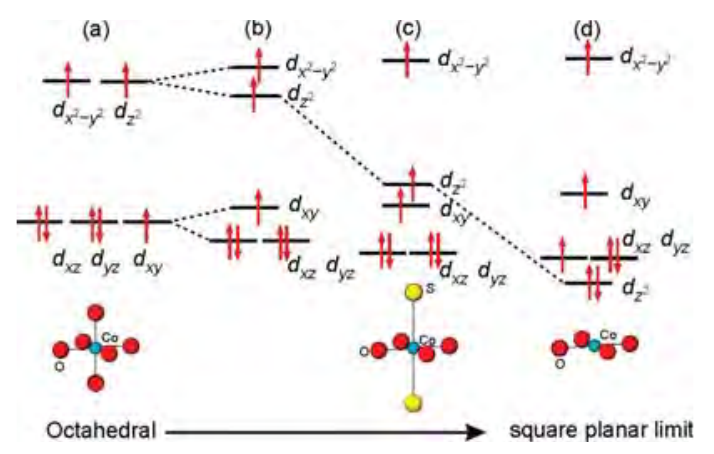}
\caption{\label{tm_cef}   Schematic illustrating the $d$ orbital splitting for Co$^{2+}$ with the crystal field being octahedral $(a)$ to square planar $(d)$ via a slightly tetragonally elongated octahedron $(b)$ and the highly tetragonally elongated octahedron $(b)$. The figure is taken from Ref. \onlinecite{Smura11:133} in the context of (Sr,Ba)$_{2}$CoO$_{2}$Cu$_{2}$S$_{2}$.}
\end{figure}

\begin{figure}[t]
\includegraphics[width=7.5cm] {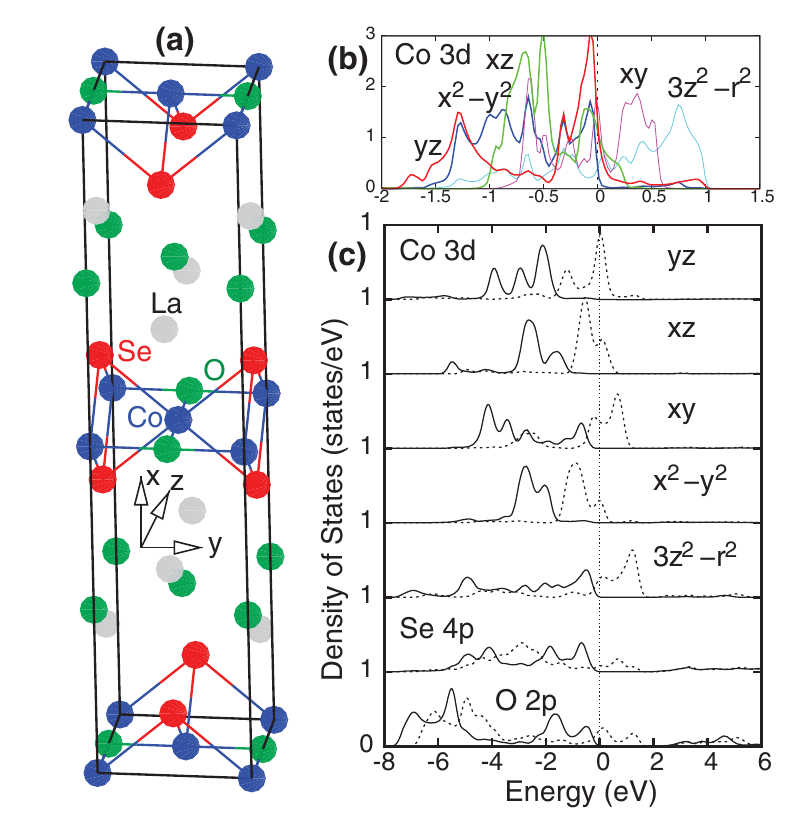}
\caption{\label{cef}   The crystal field scheme and local crystalline electric field taken from Ref. \onlinecite{Wu10:82}. $(a)$ shows the crystal structure of La$_{2}$O$_{2}$Co$_{2}$OSe$_{2}$ having Co$_{2}$Se$_{2}$O layers.  $(b)$ Orbitally resolved Co 3d density of states for the nonmagnetic state taken from the first principle calculations in Ref. \onlinecite{Wu10:82}.  The low-lying $t_{2g}$-like orbitals ($x^{2}-y^{2}$, $yz$, and $xz$) and the higher $e_{g}$ ($xy$ and $3z^{2}-r^{2}$) are split.  $(c)$ Density of states of the high-spin state.}
\end{figure}

\begin{figure}[t]
\includegraphics[width=7cm] {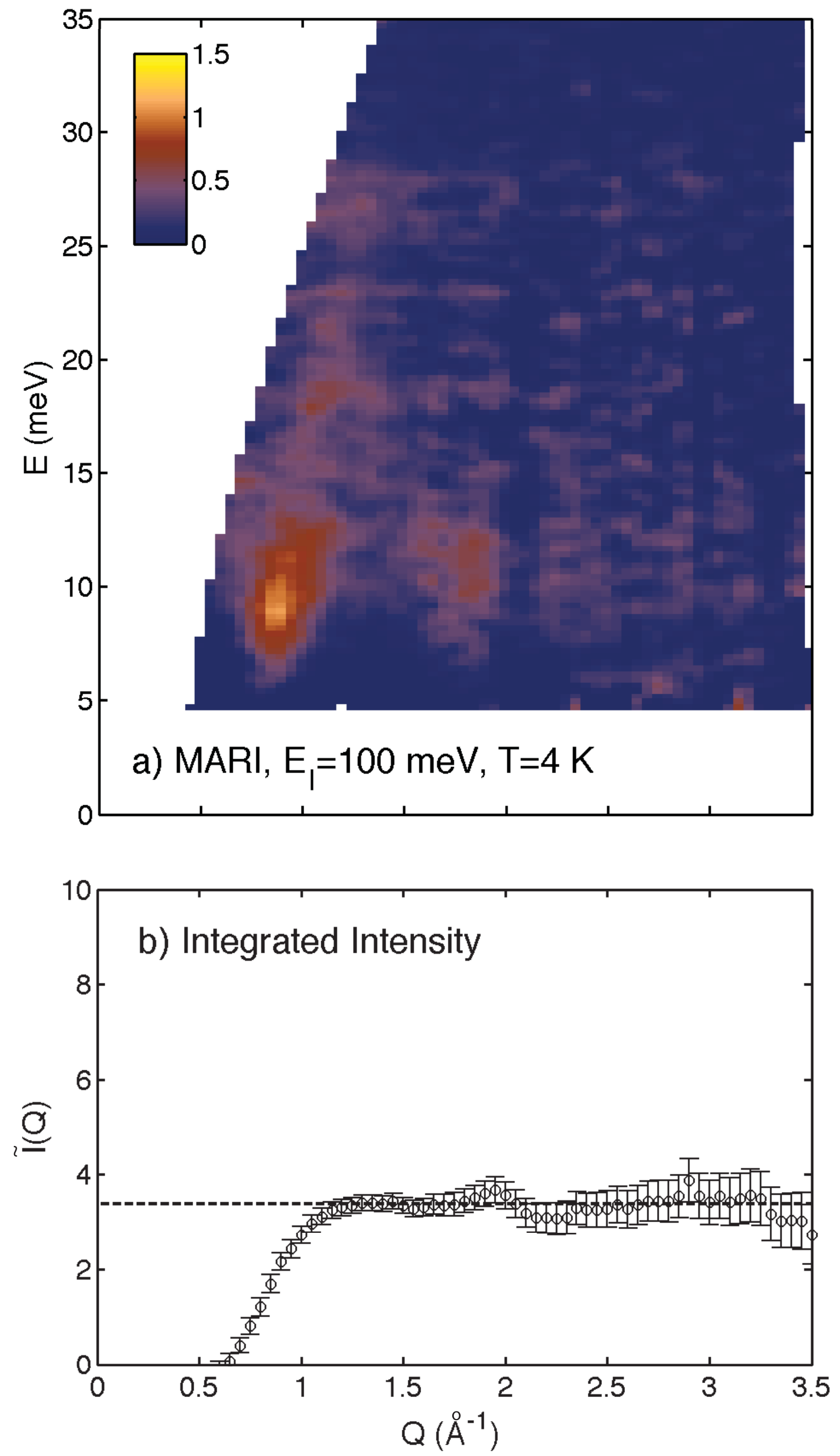}
\caption{\label{moment}   The zeroeth moment sum rule applied to powders of La$_{2}$O$_{2}$Fe$_{2}$OSe$_{2}$ from McCabe \textit{et al.}~\cite{McCabe10:150}  $a)$ illustrates the low temperature powder averaged spectrum taken using the MARI spectrometer with E$_{i}$=100 meV.  $b)$ illustrates the averaged integrated spectral weight as a function of $Q$ which can be related to the zeroeth moment sum rule.  The dashed line is predicted value for $S$=2.}
\end{figure}

The basic building block of the magnetic properties of the oxyselenides is the local crystalline electric field environment surrounding the transition metal ion (examples include Fe$^{2+}$ or Co$^{2+}$).   As outlined elsewhere~\cite{Griffiths:book,Abragam:book,Ballhausen:book,McClure59:9}, there are two competing interactions in understanding the local single-ion magnetic properties - the crystal field splitting which splits the degeneracy of the $d$ orbital levels and the Hund's energy scale which characterises the energy barrier for allowing double occupancy of the orbitals. The competition between the two energy scales is described through Tanabe-Sugano diagrams which show the energy of electronic states relative to the ground electronic state as a function of crystal field strength.

The competition between these two energy scales can be seen to cause an uncertainty in the value of $S$ for certain $d$ transition metal ions and hence an ambiguity in the sum rules discussed above in the context of neutron scattering and also the magnetic structure.  For example, an octahedral coordination environment will split the $d$ orbitals into a lower energy triplet, referred to as the $|t_{2g}\rangle$ orbitals and a higher energy doublet, the $|e_{2g}\rangle$ orbitals. Conversely, a tetrahedral coordination environment will give a much smaller splitting and a lower energy doublet ($|e\rangle$ orbitals) and a higher energy triplet ($|t_{2}\rangle$ orbitals). For a tetrahedrally coordinated Fe$^{2+}$, there are two scenarios for populating the five 3d orbitals with 6 electrons: for weak or intermediate crystal fields, the Hund's energy scale dominates and the ``high spin" configuration is found, with $e_{g}^{3} t_{2}^{3}$ ($S=2$); for larger crystal fields, the crystal field splitting dominates and the ``low spin" configuration is found, with $e_{g}^{4} t_{2}^{2}$ ($S=1$). These scenarios give different  $S$ and also different possible magnetic Hamiltonians with the later introducing an orbital degree of freedom.    This distinction between weak/intermediate and strong crystal field limits has been suggested to give rise to a spin state transitions or additional orbital terms in the magnetic Hamiltonian~\cite{Kruger09:79,Gretarsson13:110,Stock14:90}. 

In Figs. \ref{cef} and \ref{tm_cef}, we consider the example of Co$^{2+}$ in the two dimensional La$_{2}$O$_{2}$Co$_{2}$OSe$_{2}$ with its pseudo-octahedral CoO$_{2}$Se$_{4}$ coordination (taken from Refs. \onlinecite{Wu10:82,Smura11:133}).  Figure \ref{cef} shows the crystal structure highlighting the local environment around the Co$^{2+}$ site.  Figure \ref{tm_cef} illustrates the crystal field scheme, assuming a point charge distribution, given different environments with $(a)$ illustrating a perfect octahedron and a highly distorted case in $(d)$.  First principle calculations in panel $(b)$ also resolve the different orbital contributions with the relative energies consistent with the local $\textit{D}_{2\textit{h}}$ (\textit{mmm}) Co$^{2+}$ site symmetry. The deviation from a perfect octahedral environment has direct implications for the case of Fe$^{2+}$ oxyselenides and their magnetic excitation spectrum.  As noted in \onlinecite{Dai05:44}, high spin d$^{6}$ ions in an undistorted octahedral environment cannot have uniaxial magnetic properties.  The above discussion does not solve the ambiguity over which energy scale dominates and hence - whether we should expect high spin or low spin Fe$^{2+}$ ions.  To address this we consider evidence from magnetic diffraction and sum rules from neutron inelastic scattering for the case of Fe$^{2+}$ oxyselenides.  

Fig. \ref{moment} shows an analysis of the zeroeth moment sum rule in a powder sample of La$_{2}$O$_{2}$Fe$_{2}$OSe$_{2}$ from inelastic neutron scattering experiments (from the supplementary information in Ref. \onlinecite{McCabe10:150}).  Panel $a)$ shows the low temperature powder averaged spectrum and panel $b)$ shows a plot of the average spectral weight as a function of $Q$,

\begin{eqnarray}
\tilde{I}(Q)={{\int_{-\infty}^{+\infty} dE \int_{0}^{Q} d^{3}q S(\vec{q},E)} \over {\int_{0}^{Q} d^{3}q}} = S(S+1).
\label{zeroeth_sum}
\end{eqnarray}

\noindent The dashed line in Fig. \ref{moment} $(b)$ is the theoretical value for $S$=2.  The good agreement between the powder average spectral weight and the zeroeth sum rule value for $S$=2 implies that Fe$^{2+}$ has 4 unpaired electrons, consistent with a high spin configuration for the pseudo-octahedral environment in La$_{2}$O$_{2}$Fe$_{2}$OSe$_{2}$. 

Evidence supporting the high spin state $S=2$ (weak-intermediate crystalline electric field) is also found from neutron diffraction.   Table \ref{iron_moments} shows the refined ordered magnetic moments for the oxyselenides where magnetic diffraction and complete refinement have been performed.  From diffraction theory, the ordered moment should be equal to $gS$ and it can be seen that all of the ordered moments are $\sim$ 3 $\mu_{B}$.  Taking the Lande factor $g$=2, these results are inconsistent with low spin state  (strong crystal field limit) yet clearly closer to $S=2$.  The values are very close to the reported value of 3.3 $\mu_{B}$ for FeO further corroborating the fact that Fe$^{2+}$ is in a high spin state in these compounds~\cite{Roth58:110}.  Given the total moment sum rule applied to the neutron inelastic scattering spectrum and a summary of the ordered moments from neutron diffraction and magnetic refinement, we conclude the local crystalline electric field environment surround the transition metal ion in the oxyselenides is in a weak-intermediate crystal field limit.

\begin{table*}[ht]
\caption{Ordered Magnetic moments in several Fe$^{2+}$ oxyselenides and selenides (Note that T refers to a tetrahedral coordination environment; ps-O refers to pseudo-octahedral coordination environment.)}
\centering
\begin{tabular} {c c c c c}
\hline\hline
Compound & $\mu$ ($\mu_{B}$) & Coordination environment & Site symmetry & Reference\\
\hline\hline
BaFe$_{2}$Se$_{3}$ & 2.80(8) & T & C$_{1}$ (1) & ~\cite{Caron11:84} \\
\hline
Sr$_{2}$F$_{2}$Fe$_{2}$OSe$_{2}$ & 3.3(1) & ps-O & D$_{2h}$ $(mmm)$ & ~\cite{Zhao13:87} \\
La$_{2}$O$_{2}$Fe$_{2}$OSe$_{2}$ & 3.50(5) & ps-O & D$_{2h}$ $(mmm)$ & ~\cite{McCabe10:150} \\
Ce$_{2}$O$_{2}$Fe$_{2}$OSe$_{2}$ & 3.33(3) & ps-O & D$_{2h}$ $(mmm)$ & ~\cite{McCabe11:47} \\
Ce$_{2}$O$_{2}$FeSe$_{2}$ & 3.14(8) & T & D$_{2}$ (222) & ~\cite{McCabe10:150} \\
CaOFeS & 2.59(3) & T & C$_{3\nu}$ $(3m)$ & ~\cite{Jin15:91} \\
\hline \hline
\label{iron_moments}
\end{tabular}
\end{table*}

The high spin $S$=2 nature of the local magnetism in iron oxyselenides is a distinguishing point over their iron arsenide, pnictide, and chalcogenide counterparts~\cite{Yin11:10}.  For example, in Fe$_{1+x}$Te the momentum and energy integrated spectral weight over the spectrum up to $\sim$ 150 meV only yields a value consistent with a value of $S$ slightly larger than $S$=1, and well below the value expected for high spin $S$=2~\cite{Stock14:90}.  Consistent with this general statement, magnetic diffraction studies of FeAs only find an ordered moment 0.5 $\pm$ 0.05 $\mu_{B}$~\cite{Rodriguez11:83}, a value much less than the ordered moment expected for $S$=2.  Indeed, as tabulated in Table 10 in Ref. \onlinecite{Johnston10:59}, the magnetic ordered moment in the iron based pnictides and chalcogenides is universally less than $\sim$ 1 $\mu_{B}$.   

\subsection{Magnetic structures}

Having discussed the local magnetic properties of oxyselenides, we now discuss the magnetic structure which is sensitive to exchange interactions.  The magnetic structures of the oxyselenides have been investigated for a wide range of materials containing both a series of lanthanides and also various $d$ transition metal ions (particularly Mn$^{2+}$, Fe$^{2+}$, and Co$^{2+}$).  Because of the strong cross section for magnetic moments and the ability to study spin correlations, magnetic neutron scattering has played a central role in these studies.   Given the interest in iron based systems and the structural similarities with iron based superconductors, there has been a number of studies of two-dimensional variants iron based oxyselenides.  We first briefly outline currently available results in oxyselenides that only host a lanthanide magnetic ion and then discuss oxyselenides where both $d$ transition metal ions and lanthanides are present.  The structure of this section largely follows the outline presented above classifying the structural types.

\subsection{\textit{Ln} -- O -- Se phases}

The magnetic structural properties of $Ln$-O-Se phases where no $d$ transition metal ion is present have not been as fully investigated as transition metal oxyselenide systems.  Early work on Tb$_{2}$O$_{2}$(S,Se) and Dy$_{2}$O$_{2}$(S,Se) described above motivated crystal field work to understand the unusual canting measured in the Tb variant~\cite{Mignod74:62,Abbas73:12,Abbas74:14}. For $A_{4}$O$_{4}$Se$_{3}$ compounds long-range antiferromagnetic order occurs for \textit{Ln} = Gd, Tb and Dy phases but has not been observed down to 1.8 K for other analogues~\cite{Strobel08:47,Tuxworth15:44}.  Interestingly, these later compounds have been suggested to show geometric frustration based on the lack of obvious magnetic order and the fact that spins on the [O$M_{4}$]$^{10+}$ tetrahedron cannot all minimise the magnetic Hamiltonian in analogy to rare earth pyrochlore lattices~\cite{Gardner10:82}.  

\subsection{\textit{Ln} -- O -- \textit{M} -- Se phases}

\subsubsection{$Ln$CrOS$_{2}$ (oxysulfides)}

Although most work on $Ln$CrO$Q$$_{2}$ systems has been carried out on the oxysulfides, the magnetic interactions present and the resulting magnetic structures give some insight into the structurally related oxyselenide materials discussed below.  For completeness we therefore provide an overview of the results in this section and then return to our review of oxyselenides.  As described above, the crystal structure of LaCrO$Q$$_{2}$ is composed of double chains of edge-linked Cr$Q$$_{5}$O pseudo-octahedra with intrachain Cr - Cr distances of $\sim$ 3.4 \AA\ and $\sim$ 3.7 \AA, and Cr - $Q$ - Cr angles of $\sim$ 90$^{\circ}$ and $\sim$ 100$^{\circ}$.  Susceptibility measurements indicate that Cr$^{3+}$ ions order ferromagnetically (T$_{\mathrm{C}}$ = 35 K (LaCrOS$_{2}$); T$_{\mathrm{C}}$ = 51 K (LaCrOSe$_{2}$))~\cite{Winterberger87:70} and recent magnetization and specific heat studies suggest that LaCrOS$_{2}$ can be described in terms of Ising chains with relatively weak ferromagnetic interchain coupling~\cite{Takano02:122,Takano99:85}.

For smaller lanthanides $Ln$$^{3+}$ = Pr, Nd, a slightly different structure is formed with single chains of edge-linked CrS$_{6}$ and CrO$_{2}$S$_{4}$ pseudo-octahedra with only the longer intrachain Cr - Cr distance of $\sim$ 3.7 \AA.  $Ln$CrOS$_{2}$ ($Ln$ = Pr, Nd) phases order antiferromagnetically (T$_{\mathrm{N}}$ = 83 K for PrCrOS$_{2}$, T$_{\mathrm{N}}$ = 72 K for NdCrOS$_{2}$) with ferromagnetic coupling within chains (presumably via $\sim$ 100 $^{\circ}$ Cr - S - Cr exchange), but antiferromagnetic coupling between chains (via $\sim$ 125$^{\circ}$ Cr - S - Cr exchange).   Moments are close to the [010] direction. $Ln^{3+}$ moments order at fairly high temperatures in both systems suggesting that there is significant interaction between the magnetic ordering on the Cr$^{3+}$ sublattice and Ln$^{3+}$ moments~\cite{Winterberger87:70,Winterberger89:79}.  To the best of our knowledge, the magnetic behaviour of analogous oxyselenide systems has not been fully investigated.

\subsubsection{ZrCuSiAs structures}

\begin{figure}[t]
\includegraphics[width=8.5cm] {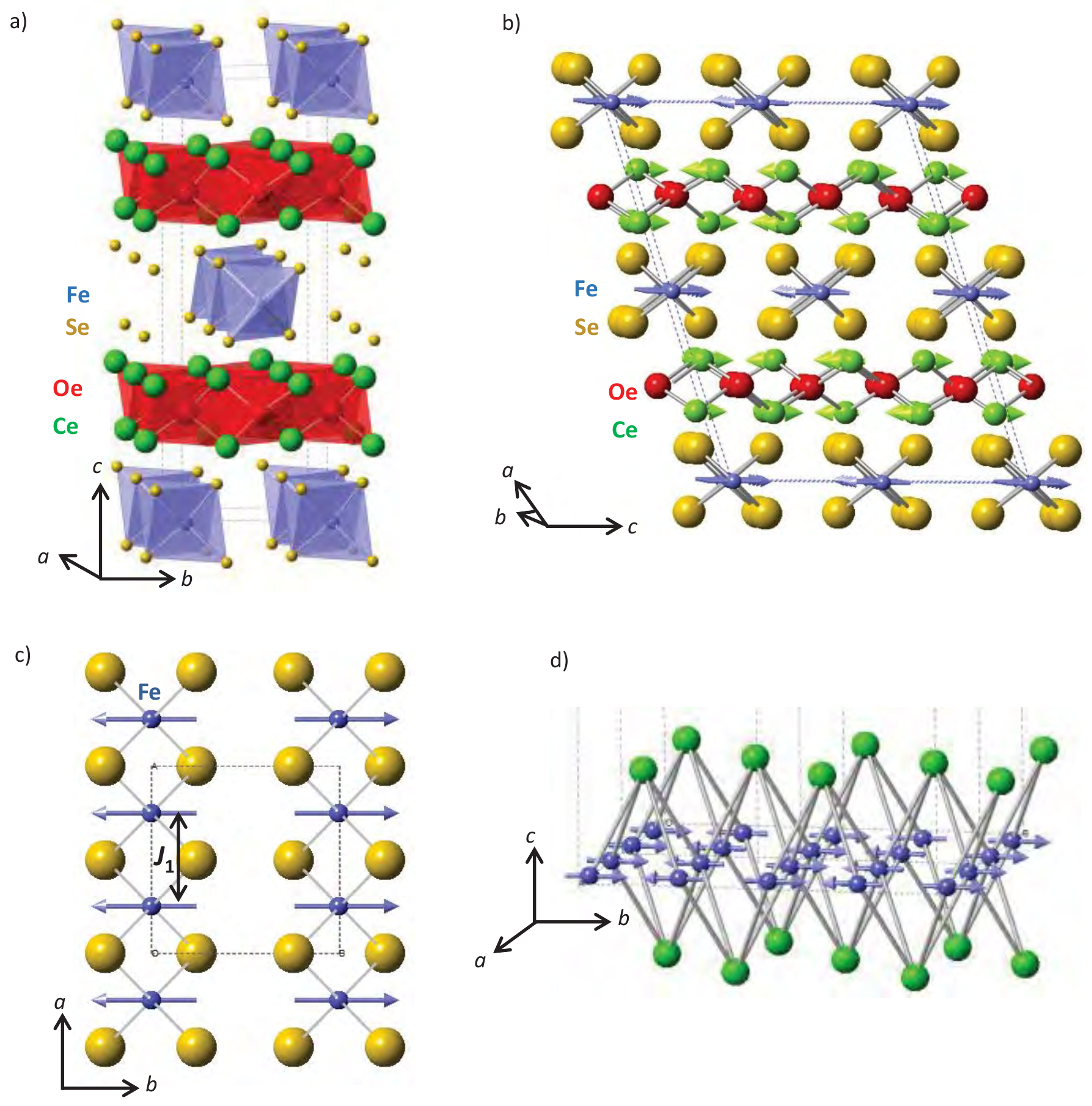}
\caption{\label{chain} The magnetic structure taken from Ref. \onlinecite{McCabe14:90}.  $(a)$ and $(b)$ show the orthorhombic and monoclinic magnetic unit cells of Ce$_{2}$O$_{2}$FeSe$_{2}$ (Ce-green, Fe=blue, O=red, and Se=yellow spheres). $(c)$ and $(d)$ displays isolated sheets of the FeSe$_{4}$ tetrahedra.}
\end{figure}

${\bf{Ce_{2}O_{2}FeSe_{2}}}$-  Ce$_{2}$O$_{2}$FeSe$_{2}$ orders antiferromagnetically at T$_{\mathrm{N}}$=171 K as seen by a local maximum in susceptibility~\cite{McCabe11:47,McCabe14:90}.   Neutron powder diffraction experiments indicate that the magnetic structure of Ce$_{2}$O$_{2}$FeSe$_{2}$ (Fig. \ref{chain}) consists of ferromagnetic chains of edge-linked FeSe$_{4}$ tetrahedra (Fe - Fe nearest neighbour distance $\sim$ 2.84 \AA, Fe - Se - Fe angle $\sim$ 71$^{\circ}$) with antiferromagnetic coupling between chains.  These exchange interactions inferred from the magnetic structure are consistent with Goodenough rules and bond angle analysis compiled experimentally from the cuprates~\cite{Shimizu03:68,Mizuno98:57}.  Fe$^{2+}$ moments are consistent with high spin d$^{6}$ Fe$^{2+}$ ions (3.14(8) $\mu_{B}$ at 4 K) and are oriented along [010]~\cite{McCabe11:47}.  The Fe$^{2+}$ magnetic structure is illustrated in Fig. \ref{chain} $(c)$ and $(d)$.  A discussion of the magnitude of the exchange interaction is discussed below in the context of neutron spectroscopy measurements.

A change in relative intensity of magnetic Bragg reflections in neutron powder data is observed for Ce$_{2}$O$_{2}$FeSe$_{2}$~\cite{McCabe14:90,Wang15:27} on cooling, similar to observations made for similar experiment in $Ln$CrOS$_{2}$ (see above), consistent with two components contributing to the low-temperature magnetically-ordered state.  An ordered moment of 1.14(1) $\mu_{B}$ on the cerium site was also found to be required in the magnetic powder refinement.  Interestingly, the onset of magnetic order for the rare earth cerium was found in Ref. \onlinecite{McCabe14:90} to be coincident with the iron ordering and onset at high temperatures.  The coupling between the rare earth cerium site and the Fe$^{2+}$ lattice will be discussed below in the context of the localised crystal field excitations of cerium.   The combined iron and cerium magnetic structures are shown in Fig. \ref{chain} $(b)$.  

\begin{figure*}[t]
\includegraphics[width=15cm] {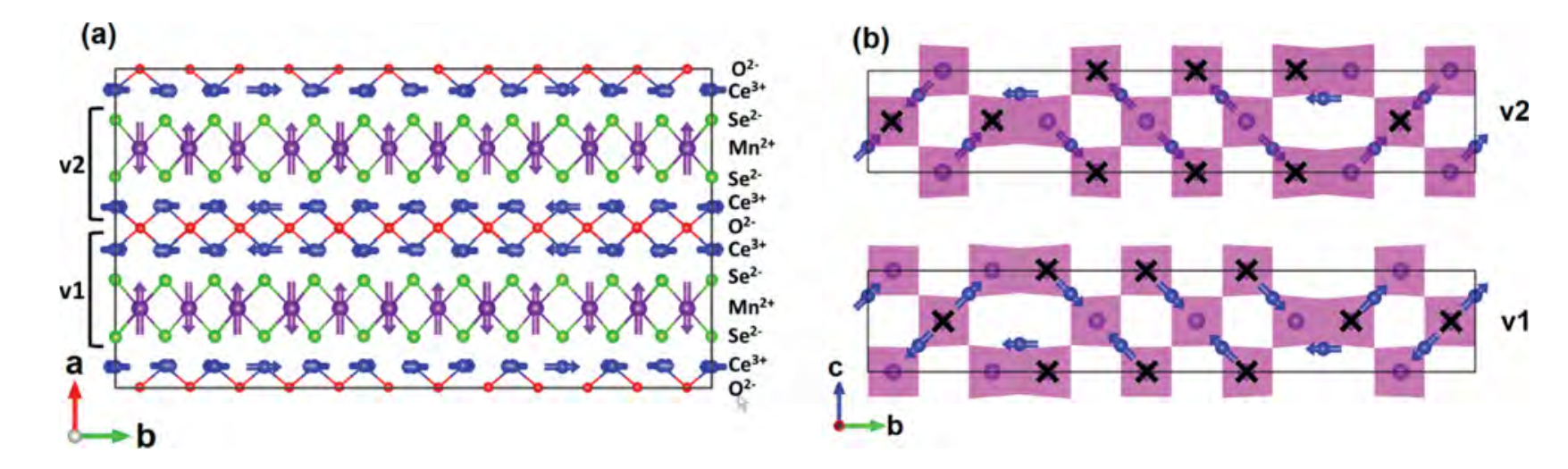}
\caption{\label{mn_struc_chain} The magnetic structure of (Ce$_{0.78}$La$_{0.22}$)$_{2}$O$_{2}$MnSe$_{2}$ at 30 K.  $(a)$ shows the view along the $c$ axis and $(b)$ illustrates the structure along the $a$ axis.  The figure is taken from Ref. \onlinecite{Wang15:27}.}
\end{figure*}

${\bf{(Ce,La)_{2}O_{2}MnSe_{2}}}$-  (Ce,La)$_{2}$O$_{2}$MnSe$_{2}$ orders antiferromagnetically at T$_{\mathrm{N}}$=150 K and the magnetic structure has been investigated using neutron powder diffraction in Ref. \onlinecite{Wang15:27}.  (Ce,La)$_{2}$O$_{2}$MnSe$_{2}$ has both corner- and edge-linked MnSe$_{4}$ tetrahedra and, in contrast to the case of Ce$_{2}$O$_{2}$FeSe$_{2}$, both nearest neighbour and next nearest neighbour exchange interactions are antiferromagnetic (Mn - Mn nearest neighbour distance $\sim$ 3.18 \AA\ and Mn - Se - Mn angles of $\sim$ 75$^{\circ}$ and $\sim$ 100$^{\circ}$ for nearest and next nearest neighbour interactions). Based on magnetic neutron diffraction, the Mn$^{2+}$ moments (4.12(1) $\mu_{B}$ at 30 K) are oriented perpendicular to the layers as shown in Fig. \ref{mn_struc_chain}.  The ordered moment direction is also perpendicular to the Fe$^{2+}$ moment direction in $R_{2}$O$_{2}$Fe$_{2}$OSe$_{2}$ discussed above.  Above T$_{\mathrm{N}}$, momentum broadened peaks in the neutron response indicate the presence of short-range correlations and are possibly the reason for the transition being less obvious in the susceptibility than the Fe$^{2+}$ counterpart discussed above.  Such short range correlations were not observed in Ce$_{2}$O$_{2}$FeSe$_{2}$.

A similar variation in relative intensity of magnetic Bragg reflections on cooling is also observed for (Ce,La)$_{2}$O$_{2}$MnSe$_{2}$ and the onset of Ce$^{3+}$ ordering is $\sim$100 K. Ce$^{3+}$ moments are oriented in-plane, parallel to the nearest Mn - Mn vector with ordered moments of 0.85(1) $\mu_{B}$~\cite{Wang15:27} at 30 K and the magnetic structure is illustrated in Fig. \ref{mn_struc_chain}.  Unlike the case of the pnictide CeOMnAs~\cite{Tsukamoto11:80,Corkett14:54,Zhang15:91}, no evidence of a reorientation of the Mn$^{2+}$ is observed at low temperatures when cerium magnetic order is present.

\subsubsection{$Ln$$_{2}$O$_{2}$$M$$_{2}$OSe$_{2}$ phases with $M$=Fe$^{2+}$}

${\bf{Nd_{2}O_{2}Fe_{2}OSe_{2}}}$-  The magnetic structure in Nd$_{2}$O$_{2}$Fe$_{2}$OSe$_{2}$ was initially investigated using Mossbauer spectroscopy and compared with first principles calculations in Ref. \onlinecite{Fuwa10:22}.  Magnetic ordering was observed at T$_{\mathrm{N}}$=90 K and the principal axis of the electric field gradient tensor was measured to be parallel to the Fe-O bond and the change in the electric field quadrupole splitting was found to be temperature independent indicating antiferromagnetic order.  The principle axis of the electric field gradient is shown in Fig. \ref{fuwa_structure} $(a)$ taken from Ref. \onlinecite{Fuwa10:22}. As noted in \onlinecite{Fuwa10:22}, high-spin Fe$^{2+}$ is expected to have an anisotropy in this crystalline electric field environment due to the possibility of unquenched orbital magnetic moments and also the deviation from a perfect octahedral environment described above.   Because of this magnetic anisotropy, combined with the tetragonal structural symmetry, it was suggested that the directions between nearest neighbour Fe$^{2+}$ moments along the Fe-O bond axes are perpendicular to each other with the moments oriented in the $a-b$ plane (see Fig. \ref{fuwa_structure} $(c)$ and $(d)$ for examples).  

The exchange paths are illustrated in Fig. \ref{structure5} $(c)$ where three different exchange interactions are defined.  Given that antiferromagnetic nearest neighbour interactions are frustrated, the next nearest neighbour interactions are expected to be key in determining the magnetic structure.  There are two such interactions with one mediated by a 97$^{\circ}$ bond through selenium and a second through an oxygen.  Based on Goodenough rules the 180$^{\circ}$ Fe-O-Fe interaction is expected to be antiferromagnetic.  Therefore, the next nearest neighbour interaction through selenium is expected to be central and the two possible magnetic structures are displayed in Fig. \ref{fuwa_structure} for $(c)$ antiferromagnetic interactions and $(d)$ ferromagnetic interactions.  

These two possibilities have very different consequences for neutron diffraction studies of the magnetic structure.  For antiferromagnetic alignment through the Fe-Se-Fe (Fig. \ref{fuwa_structure} $(c)$), a single propagation wave vector of $({1\over 2}, {1 \over 2})$ would be needed while for ferromagnetic Fe-Se-Fe (Fig. \ref{fuwa_structure} $(d)$), two propagation vectors of  $(0, {1 \over 2})$  and  $({1\over 2},0)$ would be required.  This is now discussed in the the other Fe$^{2+}$ compounds where magnetic neutron diffraction studies have been completed.

\begin{figure}[t]
\includegraphics[width=9.5cm] {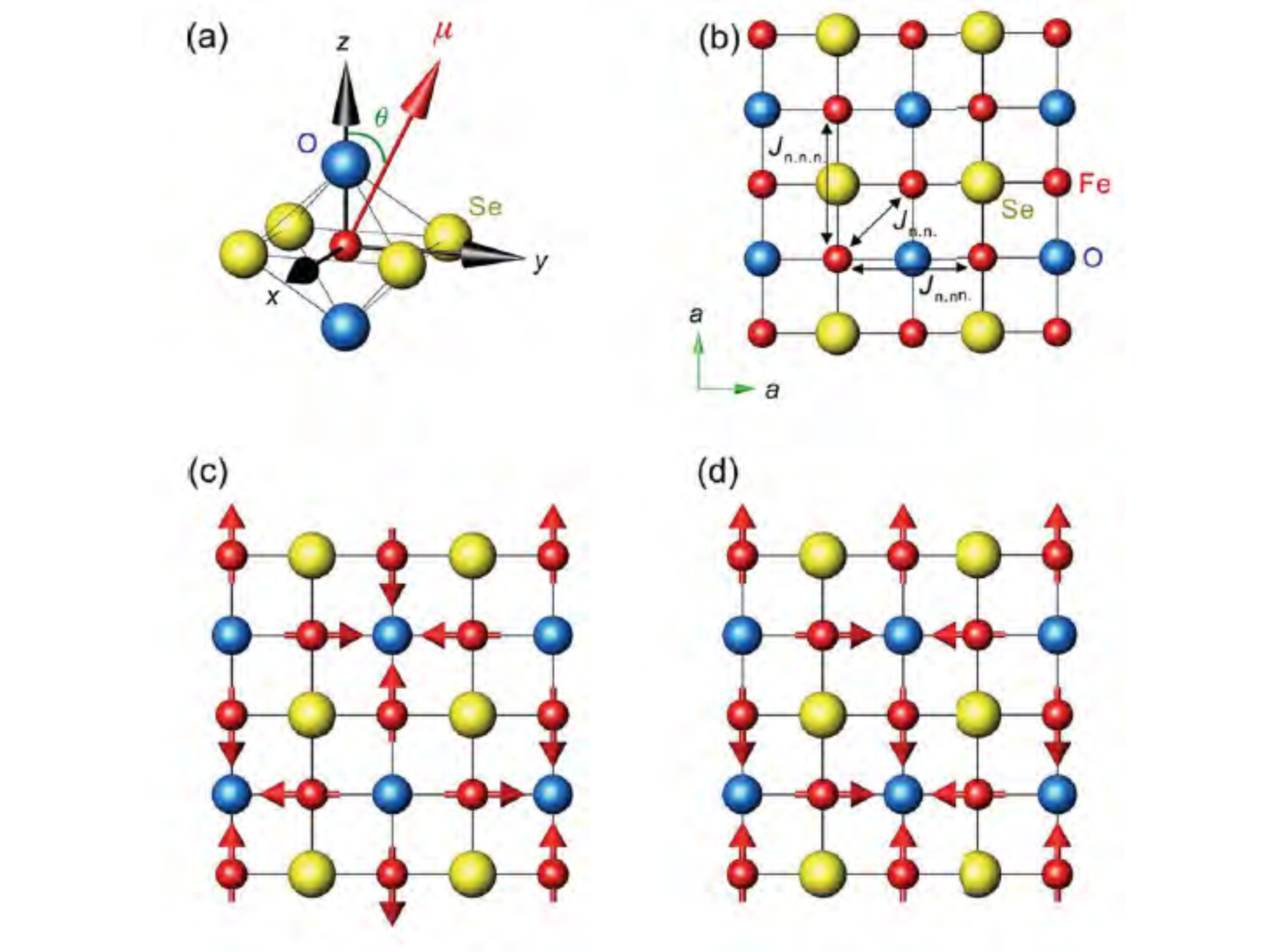}
\caption{\label{fuwa_structure}   Proposed magnetic structures for Nd$_{2}$O$_{2}$Fe$_{2}$OSe$_{2}$ taken from Fuwa $\textit{et al.}$~\cite{Fuwa10:22} $(a)$ shows the principle axis of the electric field gradient(EFG) tensor.  $(b)$ illustrates the magnetic exchange pathways. $(c)-(d)$ show the magnetic structures are proposed based upon an analysis of the electric field gradient for Fe-O-Fe antiferromagnetic and Fe-O-Fe ferromagnetic respectively.  Note that in $(c)$ the next nearest neighbour interaction mediated by selenium is antiferromagnetic while in $(d)$ the interaction is ferromagnetic.}
\end{figure}

${\bf{(Sr,Ba)_{2}F_{2}Fe_{2}O(Se,S)_{2}}}$ {\bf{and}} ${\bf{La_{2}O_{2}Fe_{2}OSe_{2}}}$- Susceptibility data collected for the oxide-fluoride-chalcogenides (Sr,Ba)$_{2}$F$_{2}$Fe$_{2}$O(Se,S)$_{2}$~\cite{Kabbour08:130} indicate similar magnetic behaviour to the oxyselenides $Ln$$_{2}$O$_{2}$Fe$_{2}$OSe$_{2}$ (including Nd$_{2}$O$_{2}$Fe$_{2}$OSe$_{2}$ described above) with T$_{\mathrm{N}}$ $\sim$ 90 K and significant deviation from Curie-Weiss behaviour at temperatures above T$_{\mathrm{N}}$, consistent with short-range magnetic correlations.~\cite{Mayer10:81} Kabbour \textit{et al.} were the first to investigate the magnetic order using neutron powder diffraction in this class of materials. Their work, using reactor neutron powder diffraction data for Ba$_{2}$F$_{2}$Fe$_{2}$OSe$_{2}$, suggested that the magnetic structure was incommensurate with modulation wavevector $\vec{q}$ = (0.42, 0.00, 0.00), see Fig. \ref{2d_thermo}.~\cite{Kabbour08:130} 

\begin{figure}[t]
\includegraphics[width=8.5cm] {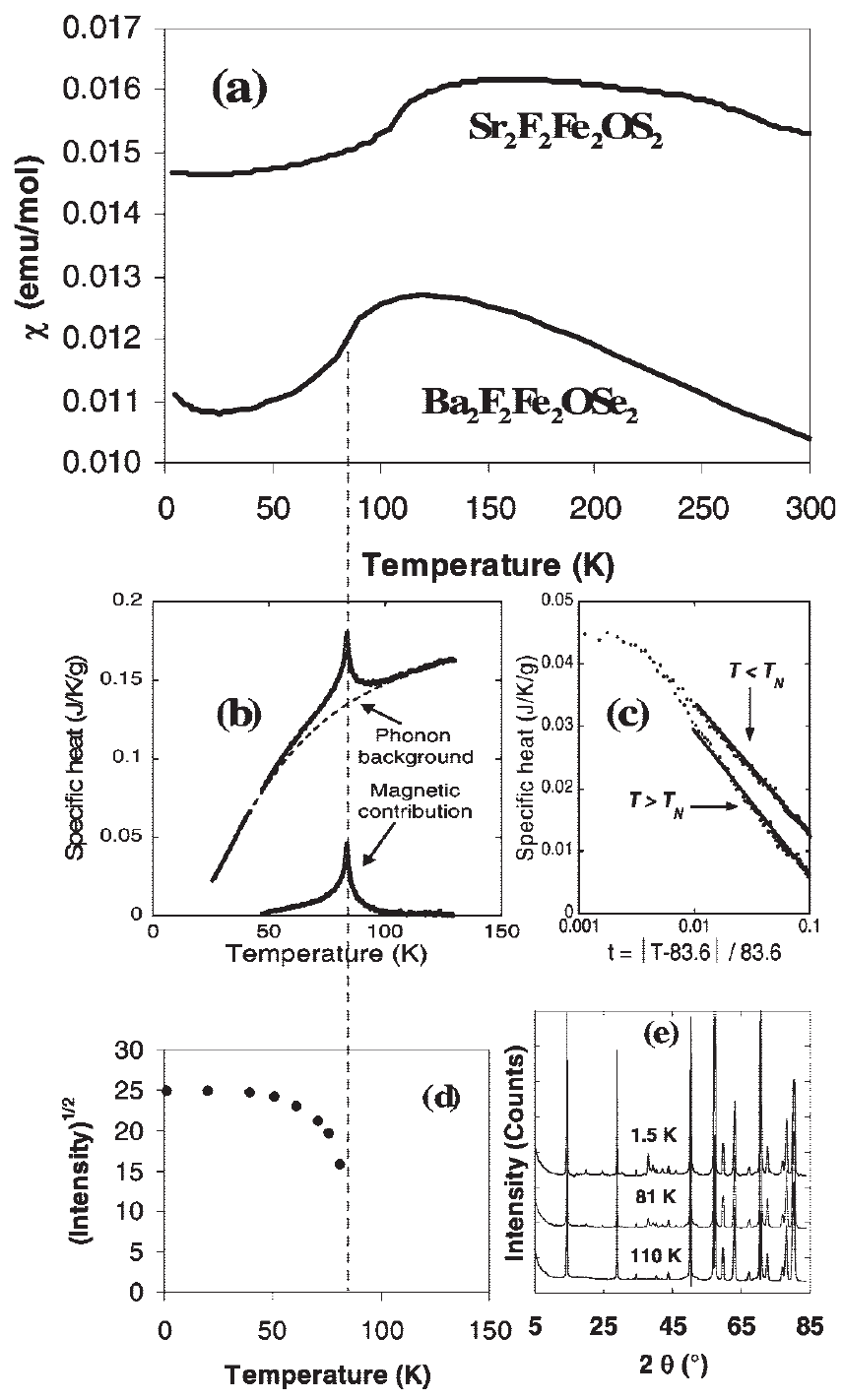}
\caption{\label{2d_thermo}  Thermodynamic data and diffraction results on Ba$_{2}$F$_{2}$Fe$_{2}$OSe$_{2}$ and Sr$_{2}$F$_{2}$Fe$_{2}$OS$_{2}$ taken from Ref. \onlinecite{Kabbour08:130}.  $(a)$ is the magnetic susceptibility and $(b)$ is the specific heat for Ba$_{2}$F$_{2}$Fe$_{2}$OSe$_{2}$ with $(c)$ showing a logarithmic plotting showing the divergence near the Neel temperature. $(d)$ is the evolution of the magnetic Bragg peak intensity with temperature taken from neutron diffraction.  $(e)$ example neutron diffraction taken on Ba$_{2}$F$_{2}$Fe$_{2}$OSe$_{2}$ collected above and below the magnetic transition.  Magnetic Bragg peaks are observable particularly at scattering angles 2$\theta$ $\sim$ 25$^{\circ}$.}
\end{figure}

High resolution neutron powder diffraction data were collected for La$_{2}$O$_{2}$Fe$_{2}$OSe$_{2}$ by Free and Evans and analysis of these data first suggested a single-$k$ vector model with Fe$^{2+}$ moments oriented in the $a-b$ plane with collinear spins (Fig. \ref{mstruct_compare} $(a)$).~\cite{Free10:81} This collinear model is consistent with the $\vec{q}$ = (${1\over 2}$, 0, ${1\over 2}$) propagation vector observed from these neutron powder diffraction data and is very similar to the magnetic structure adopted by Fe$_{1+x}$Te (with the same modulation vector). However, in Fe$_{1+x}$Te for small values of $x$, this magnetic transition is accompanied by a tetragonal - monoclinic structural transition~\cite{Bao09:102,Rodriguez11:83,Rodriguez13:88,Turner09:80,Koz13:88}. There is no evidence for such a distortion in La$_{2}$O$_{2}$Fe$_{2}$OSe$_{2}$ from these high resolution neutron powder diffraction data, although there may be some disorder of the O(2) sites within the Fe$_{2}$O layers. In La$_{2}$O$_{2}$Fe$_{2}$OSe$_{2}$, this collinear model has $J_{1}$, $J_{2}$ and $J_{2'}$ interactions all partially frustrated and so is difficult to justify on energy grounds and seems surprising given the symmetry of the nuclear structure.

Zhao \textit{et al.} found that high resolution neutron powder diffraction data for Sr$_{2}$F$_{2}$Fe$_{2}$OS$_{2}$ were consistent with a commensurate magnetic structure~\cite{Zhao13:87} (in contrast to earlier work on Ba$_{2}$F$_{2}$Fe$_{2}$OSe$_{2}$ mentioned above~\cite{Kabbour08:130}).  Zhao \textit{et al.} were able to show that the $2-k$ model proposed by Fuwa \textit{et al.} (with ferromagnetic Fe - Se - Fe exchange, Fig. \ref{fuwa_structure}) and Free and Evans' collinear model (Fig. \ref{mstruct_compare} $(a)$) are indistinguishable using neutron powder diffraction data and that the $2-k$ model is more appropriate for Sr$_{2}$F$_{2}$Fe$_{2}$OS$_{2}$~\cite{Zhao13:87}. They observed significant anisotropic broadening of magnetic Bragg reflections (which was fitted by a Warren-like lineshape), consistent with shorter range correlations along $c$. The in-plane correlations were found to be resolution-limited, indicating an in-plane correlation length $>$ 300 \AA\ and out-of-plane correlations with $\xi_{c}$= 17(3) \AA\ gave a good fit to the data~\cite{Zhao13:87}.

\begin{figure*}[t]
\includegraphics[width=15cm] {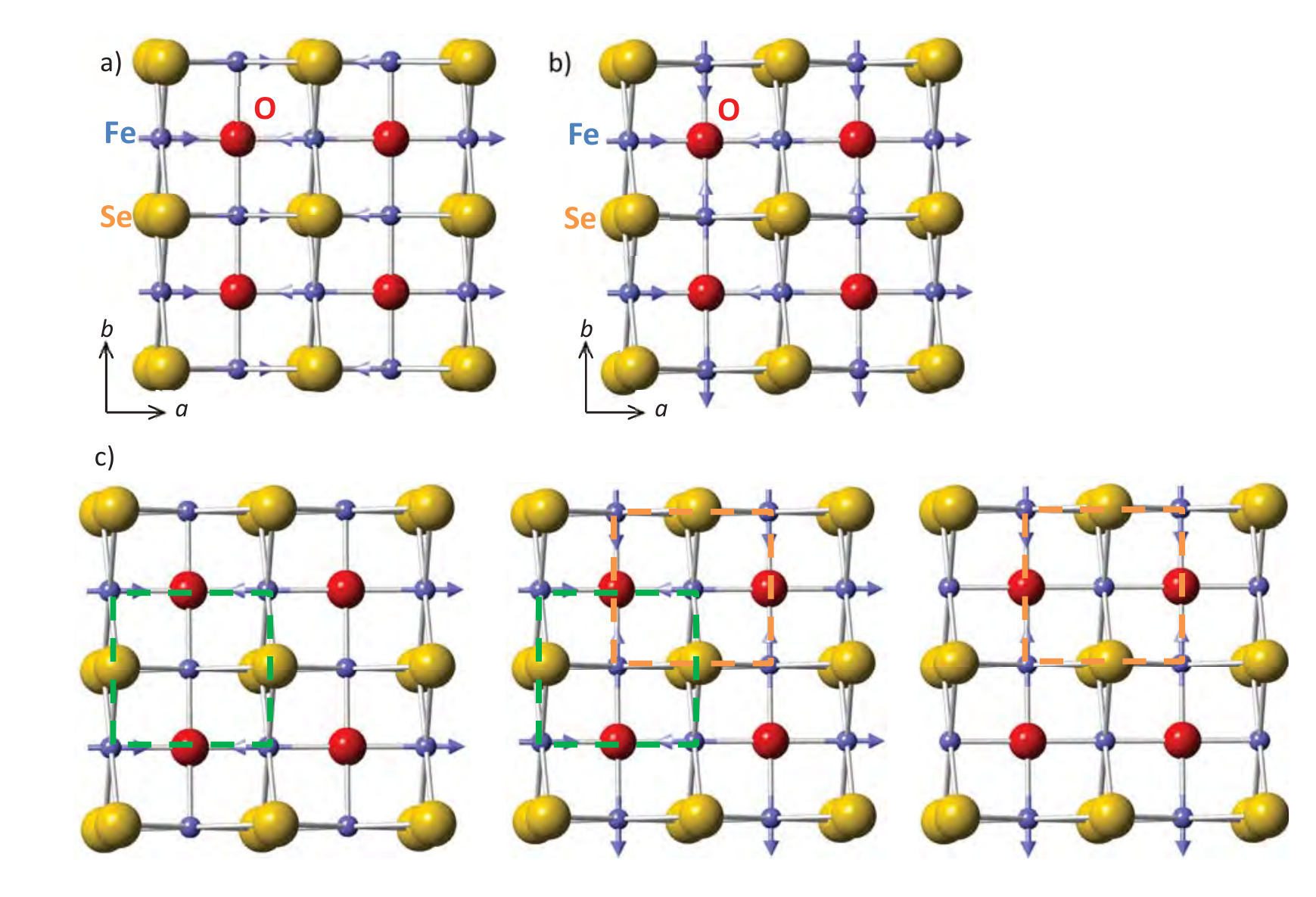}
\caption{\label{mstruct_compare} Collinear model showing in-plane spin arrange of the $(a)$ full collinear magnetic structure; $(b)$ 2-k model and $(c)$ illutrating the in-plane spin arrangement viewed as two interpenetrating square lattices.}
\end{figure*}

At about the same time, the magnetic structure of La$_{2}$O$_{2}$Fe$_{2}$OSe$_{2}$ was investigated using high-flux (lower resolution) neutron powder diffraction data.  As observed for Sr$_{2}$F$_{2}$Fe$_{2}$OS$_{2}$, the collinear and $2-k$ models gave equivalent fits to the low temperature neutron powder diffraction data for La$_{2}$O$_{2}$Fe$_{2}$OSe$_{2}$, and similar anisotropic broadening of magnetic Bragg reflections (Fig. \ref{aniso_line})~\cite{McCabe10:150} was also observed. These were fitted using a model to simulate stacking faults~\cite{Her07:63} in the magnetic structure. The high-resolution neutron powder diffraction data and x-ray powder diffraction data collected for La$_{2}$O$_{2}$Fe$_{2}$OSe$_{2}$ revealed no defects in the nuclear crystal structure, indicating that these stacking faults exist only in the magnetic ordering and that La$_{2}$O$_{2}$Fe$_{2}$OSe$_{2}$ and Sr$_{2}$F$_{2}$Fe$_{2}$OS$_{2}$ have similar magnetic microstructures~\cite{McCabe10:150}. Neutron powder diffraction data were also collected for La$_{2}$O$_{2}$Fe$_{2}$OSe$_{2}$ on cooling through the magnetic phase transition and showed that the onset of magnetic order had 2D-Ising like character (discussed further below, Fig. \ref{order_param})~\cite{McCabe10:150}, consistent with the results from Mossbauer spectroscopy studies of Nd$_{2}$O$_{2}$Fe$_{2}$OSe$_{2}$~\cite{Fuwa10:22} described above and with a predominantly two-dimensional character to the magnetism (rather than resulting from nearer one-dimensional chains).  A Warren-like peak was observed in neutron powder diffraction data for La$_{2}$O$_{2}$Fe$_{2}$OSe$_{2}$ in a narrow temperature range ($\sim$13 K) above T$_{\mathrm{N}}$, indicating short-range magnetic ordering within the Fe$_{2}$O layers immediately above the three-dimensional ordering temperature. This is on contrast to La$_{2}$O$_{2}$Mn$_{2}$OSe$_{2}$ for which a similar Warren peak is observed up to 140 K above T$_{\mathrm{N}}$~\cite{Ni10:82}. This indicates a low degree of frustration in the La$_{2}$O$_{2}$Fe$_{2}$OSe$_{2}$ magnetic structure, consistent with the $2-k$ model and not the collinear model. This conclusion was consistent with neutron inelastic scattering data: simulated inelastic spectra for various models were obtained and were comparable with experimentally observed spectra (Figure 25) but only the exchange constants determined for the $2-k$ model are consistent with the magnetism observed.~\cite{Kabbour08:130,McCabe10:150,Fuwa10:22}

Most recently, Gunther \textit{et al.} have investigated the magnetic ordering in La$_{2}$O$_{2}$Fe$_{2}$OSe$_{2}$ using the local probes of Mossbauer spectroscopy and muon spin rotation~\cite{Gunther14:90}.  These Mossbauer studies are consistent with those described above for Nd$_{2}$O$_{2}$Fe$_{2}$OSe$_{2}$,~\cite{Fuwa10:22} with the Fe$^{2+}$ moments directed along the Fe - O bond axes~\cite{Gunther14:90}. Their recent muon spin rotation experiments confirmed the static, long-range magnetic order on Fe$^{2+}$ sites (consistent with the stacking faults suggested by neutron powder diffraction work described above~\cite{Zhao13:87,McCabe10:150}, rather than slow magnetic dynamics contributing to the unusual peak shapes) but also revealed a dynamic component due to muons affected by a large hyperfine coupling constant~\cite{Gunther14:90}.

\begin{figure}[t]
\includegraphics[width=8.5cm] {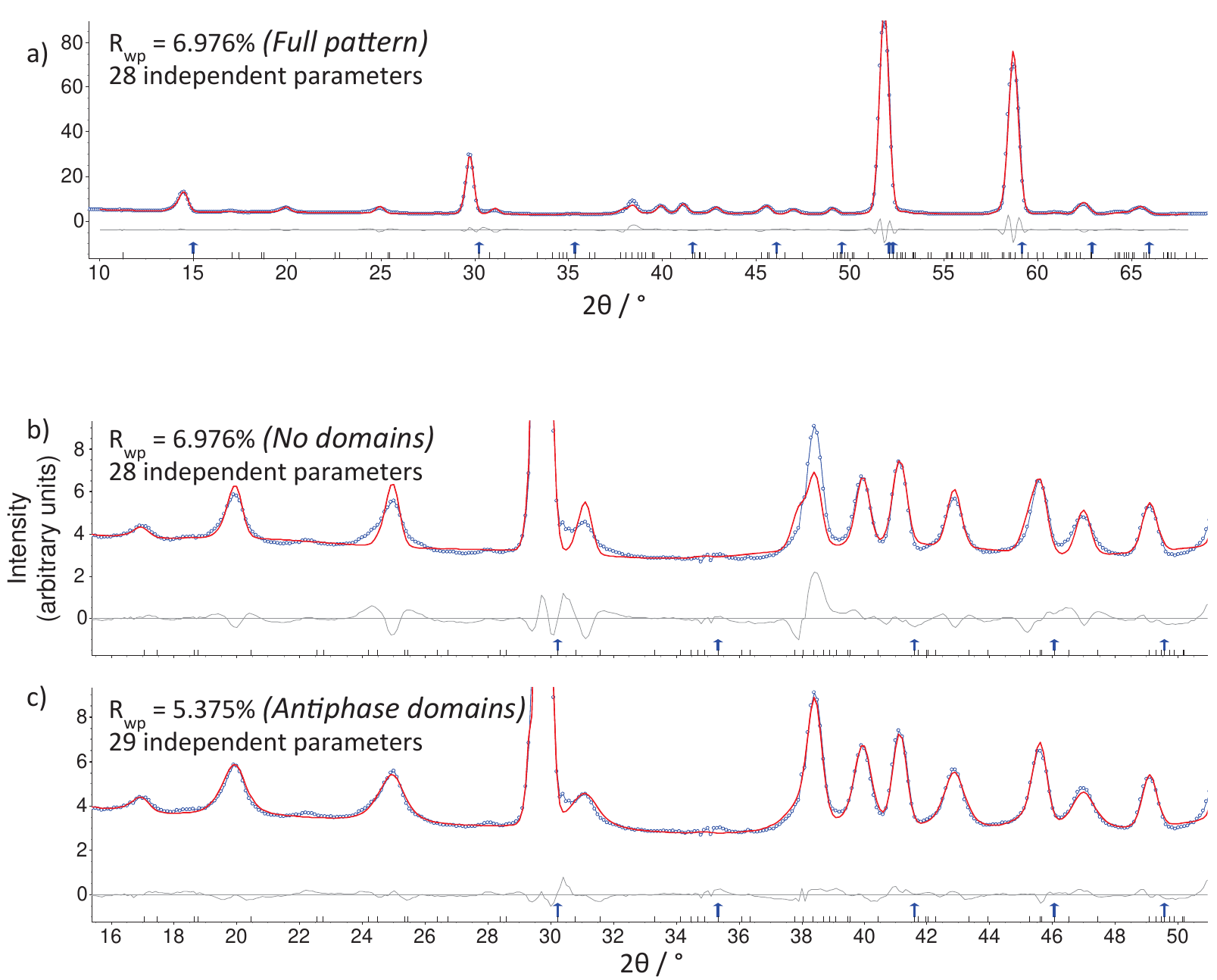}
\caption{\label{aniso_line} Neutron diffraction data on La$_{2}$O$_{2}$Fe$_{2}$OSe$_{2}$ taken from Ref. \onlinecite{McCabe10:150}.  $(a)$ shows Rietveld refinments with the $2-k$ model showing both nuclear (blue arrows) and magnetic (black tick marks) phases.  $(b)$ shows the refinement with \textit{same} peak shape for both nuclear and magnetic phases.  $(c)$ shows refinement including antiphase boundaries in the magnetic phase.  Note the increase in the quality of the fit with the inclusion of just one additional free parameter.  Note that the tick marks do not include a refined zero offset of $\sim$ 0.4$^{\circ}$.}
\end{figure}

${\bf{R_{2}O_{2}Fe_{2}OSe_{2}}}$ ($R=Ce, Pr, Nd$, and $Sm$)-  Free \textit{et al.} explored the compositional flexibility of the $Ln$$_{2}$O$_{2}$$M$$_{2}$OSe$_{2}$ structure and found that $M$ = Fe is compatible with the widest range of $Ln^{3+}$ radii in the fluorite-like layers from $Ln$ = La (eight-coordinate ionic radius 1.16 \AA) to $Ln$ = Sm (1.079 \AA)~\cite{Free11:23}. This series allows us to investigate the role of the Ln$^{3+}$ ion in the magnetic behaviour of these materials.

\begin{table*}[ht]
\caption{Summary of key structural and magnetic ordering behaviour in Fe$_{2}$O materials; unit cell parameters and Fe-O bond lengths are from references listed and T$_{\mathrm{N}}$ is from diffraction results except for Na$_{2}$Fe$_{2}$OSe$_{2}$ and Sm$_{2}$O$_{2}$Fe$_{2}$OSe$_{2}$ for which T$_{\mathrm{N}}$ is from mangetic susceptibility measurements.  $^{**}$ indicates T$_{\mathrm{N}}$ extracted from heat capacity measurements.}
\centering
\begin{tabular} {c c c c c c c}
\hline\hline
Compound & Ln$^{3+}$ ionic radius (\AA) & $a$ (\AA) (295-300 K) & d$_{Fe-O}$ (\AA) & T$_{\mathrm{N}}$ (K) & $\mu$ ($\mu_{B}$) & Ref. \\
\hline\hline
Na$_{2}$Fe$_{2}$OSe$_{2}$              & -         & 4.107(8)        & 2.054(8)         & 75            & -                & \onlinecite{He11:84}  \\
La$_{2}$O$_{2}$Fe$_{2}$OSe$_{2}$ & 1.16    & 4.084466(9)  & 2.042233(9)   & 89.50(3)  & 3.50(5)       & \onlinecite{Free10:81}  \\
 &  &  &  &  &  & ~\onlinecite{McCabe10:150}  \\
Ce$_{2}$O$_{2}$Fe$_{2}$OSe$_{2}$ & 1.143  & 4.06134(5)   & 2.03067(5)     & 92.3(2)  & 3.32(1)       & \onlinecite{McCabe14:90}  \\
Pr$_{2}$O$_{2}$Fe$_{2}$OSe$_{2}$ & 1.126  & 4.0447(1)      & 2.0224(1)       & 92.09(2); 88.6$^{**}$   & 3.36(2)       & \onlinecite{Oogarahxx} \\
 &  &  &  &  &  & ~\onlinecite{Ni11:83}  \\
Sm$_{2}$O$_{2}$Fe$_{2}$OSe$_{2}$ & 1.079  & 3.9976(1)      & 1.9988(1)       & 85.3  & -       &  \onlinecite{Ni11:83}  \\
\hline \hline
\label{fe_moments}
\end{tabular}
\end{table*}

The first point to note is that there is no change in the ordered magnetic structure as the $Ln^{3+}$ ion is changed~\cite{Free11:23,McCabe14:90,Ni11:83}. This suggests that despite the decrease in unit cell volume (and particularly the contraction within the ab plane) as Ln$^{3+}$ radius decreases, the $2-k$ magnetic order is robust in terms of the effects of chemical pressure~\cite{McCabe14:90}. With decreasing $Ln^{3+}$ radius, there is a slight increase in T$_{\mathrm{N}}$ (Table \ref{fe_moments}). This is presumably due to increased overlap of orbitals involved in magnetic exchange interactions as the unit cell (and therefore Fe - O bond lengths) decrease. 

Ordering of Nd$^{3+}$ moments has not been observed for Nd$_{2}$O$_{2}$Fe$_{2}$OSe$_{2}$~\cite{McCabe14:90}.  However, Ce$^{3+}$ moments are thought to order below 16 K in Ce$_{2}$O$_{2}$Fe$_{2}$OSe$_{2}$, with a similar in-plane arrangement to the Fe moments but with some out-of-plane canting, perhaps indicating some coupling between Fe$^{2+}$ and Ce$^{3+}$ sublattices~\cite{McCabe14:90}. The behaviour of the Pr$_{2}$O$_{2}$$M$$_{2}$OSe$_{2}$ analogues is unusual and the low temperature tetragonal - orthorhombic distortion has been discussed above. Analogous distortions in related PrMnSbO~\cite{Kimber10:82}, PrMnAsO~\cite{Wildman15:54} and PrFeAsO~\cite{Kimber08:78} are accompanied by long-range ordering of Pr$^{3+}$ moments. Ni \textit{et al.} observed a peak in heat capacity data at $\sim$ 23 K for Pr$_{2}$O$_{2}$Fe$_{2}$OSe$_{2}$ and slight changes in neutron powder diffraction data at low temperature suggesting that Pr$^{3+}$ moments may order but this is not yet fully understood~\cite{Ni11:83}. Ordering of Sm$^{3+}$ moments is also thought to occur below 6 K~\cite{Ni11:83}.

\subsubsection{$Ln$$_{2}$O$_{2}$$M$$_{2}$OSe$_{2}$ phases with $M$=Co$^{2+}$}

\begin{figure}[t]
\includegraphics[width=8.5cm] {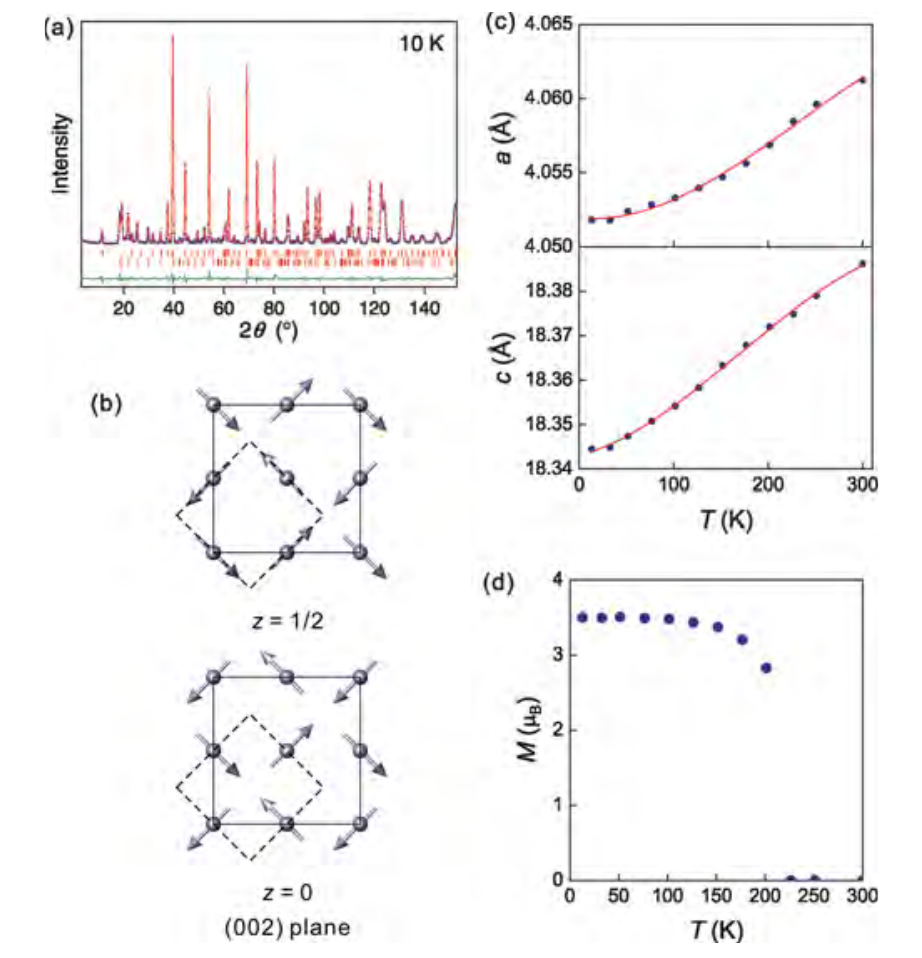}
\caption{\label{cobalt_diffract} Neutron diffraction and magnetic structure for La$_{2}$O$_{2}$Co$_{2}$OSe$_{2}$ reported in Ref. \onlinecite{Fuwa10:132}. $(a)$ neutron diffraction patter taken at 10 K.  The upper and lower sets of vertical markers in the pattern are the calculated nuclear and magnetic peak positions, respectively. $(b)$ shows possible magnetic structure model for the cobalt moments.  $(c)$ is the temperature dependence of the lattice constants.  $(d)$ is a plot of the ordered magnetic moment as a function of temperature.}
\end{figure}

${\bf{La_{2}O_{2}Co_{2}OSe_{2}}}$- The magnetic structure for La$_{2}$O$_{2}$Co$_{2}$OSe$_{2}$ has been reported in Ref. \onlinecite{Fuwa10:132} with order parameter shown in Fig. \ref{cobalt_diffract} (T$_{\mathrm{N}}$ =217 K).   The magnetic structure for La$_{2}$O$_{2}$Co$_{2}$OSe$_{2}$ is different from Fe$^{2+}$ analogue discussed above in that it is determined by a single propagation wave vector of $({1\over2}, {1\over2},0)$.   As noted in Ref. \onlinecite{Free11:23}, an ambiguity exists in the magnetic structure from powder diffraction data with magnetic moments directed either along or perpendicular to the Co-O bonds.  The two possible structures noted in \onlinecite{Free11:23,Fuwa10:132} are both consistent with the next nearest neighbour Co-Se-Co (bond angle $\sim$ 99$^{\circ}$) and Co-O-Co (bond angle 180$^{\circ}$) interactions being antiferromagnetic.  This differs from La$_{2}$O$_{2}$Fe$_{2}$OSe$_{2}$ where, as noted above, the Fe-Se-Fe interaction is ferromagnetic.  

While momentum broadened scattering in the diffraction data was observed in the range of 225-250 K, sharp Bragg peaks were present at lower temperatures.   The high temperature momentum broadened scattering was considered to originate from magnetic diffuse scattering.  The momentum broadened peaks at high temperatures occur over a similar temperature range where strong deviation from Curie-Weiss behaviour is observed in susceptibility~\cite{Fuwa10:150}. Below T$_{\mathrm{N}}$, no anisotropic lineshape to the magnetic Bragg peaks was reported in contrast to analogous La$_{2}$O$_{2}$Fe$_{2}$OSe$_{2}$ and Sr$_{2}$F$_{2}$Fe$_{2}$OS$_{2}$.

The magnetic moment for Co$^{2+}$ was measured to be 3.53 $\pm$ 0.01 $\mu_{B}$ in Ref. \onlinecite{Fuwa10:132} and 3.29(3) $\mu_{B}$ in Ref. \onlinecite{Free11:23}.  It was noted in Ref. \onlinecite{Fuwa10:132} that this was much larger than the theoretical first principle calculations that predicted 2.70 $\mu_{B}$~\cite{Wu10:82}  leading to the suggestion of a possible orbital contribution to the magnetic moment.  

\subsubsection{$Ln$$_{2}$O$_{2}$$M$$_{2}$OSe$_{2}$ phases with $M$=Mn$^{2+}$}

${\bf{La_{2}O_{2}Mn_{2}OSe_{2}}}$-  The magnetic and structural properties of La$_{2}$O$_{2}$Mn$_{2}$OSe$_{2}$ were investigated and reported in \onlinecite{Ni10:82,Free11:23}.  The nuclear and magnetic structure (from Ref. \onlinecite{Ni10:82}) refined from neutron powder data is shown in Fig. \ref{mn_struct} with long-range magnetic order being reported below T$_{\mathrm{N}}$=163 K (Ref. \onlinecite{Ni10:82}) and 168.1 K (Ref. \onlinecite{Free11:23})  with  magnetic propagation vector $\vec{q}$=(0,0,0).  The magnetic structure of La$_{2}$O$_{2}$Mn$_{2}$OSe$_{2}$ is very different to those of the Fe$^{2+}$ and the Co$^{2+}$ analogues: the Mn$^{2+}$ spins are oriented perpendicular to the Mn$_{2}$O (along $c$) planes with different relative spin arrangements.  The magnetic structure of La$_{2}$O$_{2}$Mn$_{2}$OSe$_{2}$ does have strong similarities to that reported in PrOMnSb~\cite{Kimber10:82}.

The magnetic moment was reported to be 4.147$\pm$ 0.028 $\mu_{B}$ (Ref. \onlinecite{Ni10:82}) and 4.5 $\pm$ 0.3 $\mu_{B}$ (Ref. \onlinecite{Free11:23}).    This value is in excellent agreement with the magnetic moments reported in BaMn$_{2}$P$_{2}$ (4.2(1) $\mu_{B}$) and BaMn$_{2}$As$_{2}$ (3.88(4) $\mu_{B}$)~\cite{Brock94:113,Singh09:79}.  They are also in agreement with MnO (4.892 $\mu_{B}$ at 10 K) and Mn$_{2}$SiSe$_{4}$ (4.36 $\mu_{B}$ at 2 K)~\cite{Bodenane96:164,Bonfante72:10}.

\begin{figure}[t]
\includegraphics[width=8.5cm] {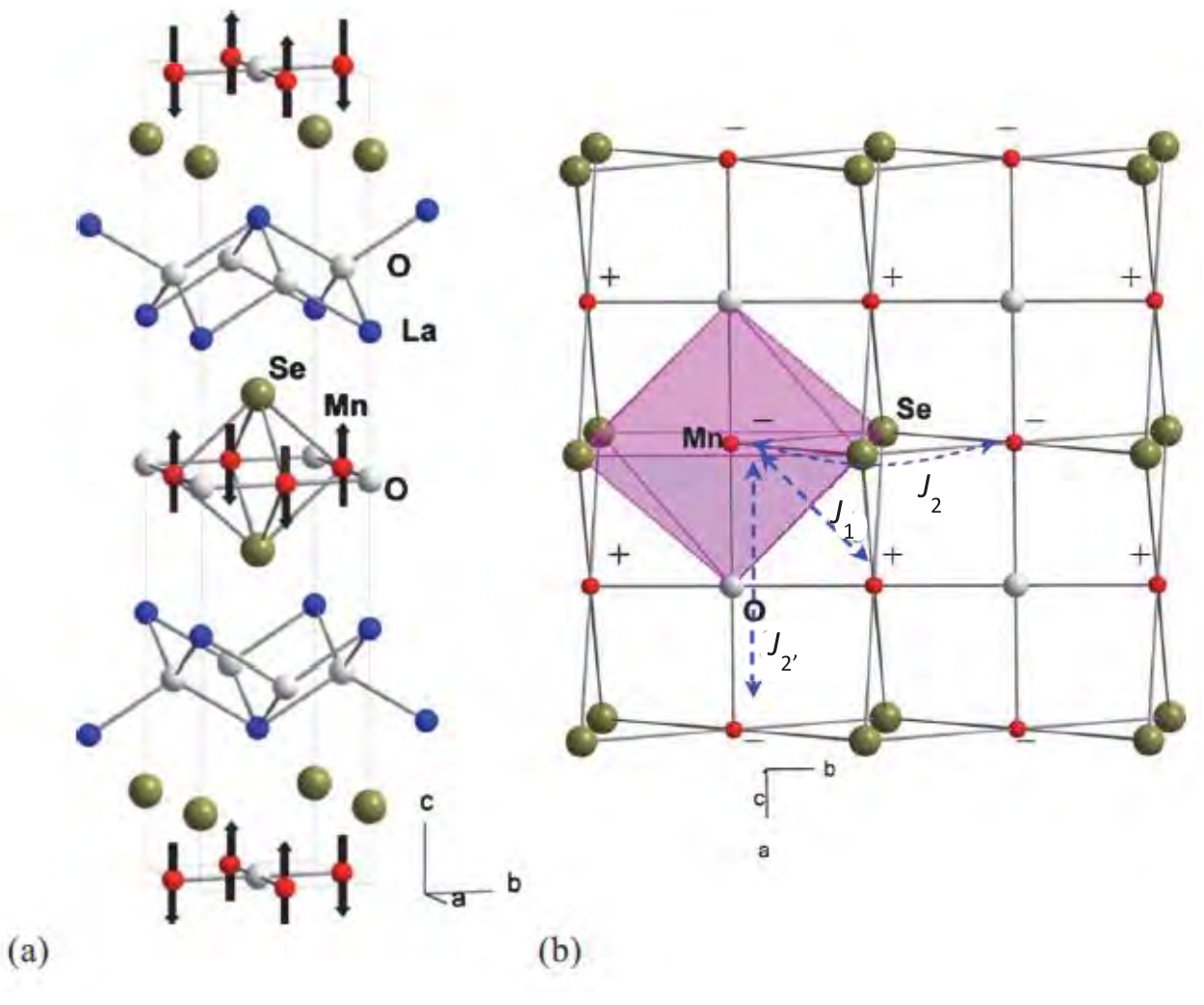}
\caption{\label{mn_struct} The magnetic structure of La$_{2}$O$_{2}$Mn$_{2}$OSe$_{2}$ taken from Ref. \onlinecite{Ni10:82}.  $(a)$ illustrates the chemical and magnetic structure and $(b)$ shows a view of a single [Mn$_{2}$OSe$_{2}$]$^{2-}$ with the three exchange interactions highlighted.  The labelling of the exchange interactions have been edited to be consistent with the text of this review.}
\end{figure}

Similar to the case of the iron based oxyselenides discussed above, there are three magnetic interactions between Mn$^{2+}$ ions that need to be considered.  These include the superexchange interaction through the 180$^{\circ}$ Mn-O-Mn pathway, the superexchange interaction through the $\sim$ 95$^{\circ}$ Mn-Se-Mn channel, and the interaction between the nearest Mn$^{2+}$ ions.  These are denoted as $J_{1}$, $J_{2}$, and $J_{2'}$ in Fig. \ref{mn_struct}.   Based on Goodenough rules, the 180$^{\circ}$ Mn-O-Mn exchange is expected to be antiferromagnetic while the $\sim$ 95$^{\circ}$ Mn-Se-Mn coupling is expected to be ferromagnetic.    Unlike La$_{2}$O$_{2}$Fe$_{2}$OSe$_{2}$ discussed above, the nearest-neighbour antiferromagnetic $J_{1}$ exhange interactions dominate in La$_{2}$O$_{2}$Mn$_{2}$OSe$_{2}$, resulting in a G-type antiferromagnetic structure with nearest-neighbour Mn$^{2+}$ spins antiparallel and next-nearest-neighbour spins parallel.~\cite{Ni10:82}  This leaves the 180$^{\circ}$ Mn-O-Mn $J_{2'}$ interactions frustrated.

\begin{table*}[ht]
\caption{Summary of key structural and magnetic ordering behaviour in Mn$_{2}$O materials; unit cell parameters and Mn-O bond lengths are from references listed and T$_{\mathrm{N}}$ is from diffraction results.  Data taken from Ref. \onlinecite{Free11:23}. }
\centering
\begin{tabular} {c c c c c c}
\hline\hline
Compound & Ln$^{3+}$ ionic radius (\AA) & $a$ (\AA) (295-300 K) & d$_{Mn-O}$ (\AA) & T$_{\mathrm{N}}$ (K) & $\mu$ ($\mu_{B}$) \\
\hline\hline
La$_{2}$O$_{2}$Mn$_{2}$OSe$_{2}$ & 1.16    & 4.138921(4)  & 2.06435(1)   & 168.1(4)  & 4.5(2) \\     
Ce$_{2}$O$_{2}$Mn$_{2}$OSe$_{2}$ & 1.143  & 4.11304(2)   & 2.05124(1)     & 174.1(2)  & 4.8(3) \\       
Pr$_{2}$O$_{2}$Mn$_{2}$OSe$_{2}$ & 1.126  & 4.09739(2)      & 2.04308(1)      & 180.3(4)   & 4.5(1) \\       
\hline \hline
\label{mn_moments}
\end{tabular}
\end{table*}

Susceptibility measurements on La$_{2}$O$_{2}$Mn$_{2}$OSe$_{2}$ suggest~\cite{Ni10:82,Free11:23}. Indications of high-temperature short-range correlations are indicated by a deviation from Curie Weiss behaviour.  At low temperatures the published data shows several features and a marked difference between field and zero field cooled responses. However, as noted in Ref. \onlinecite{Free11:23}, it is likely that the features are due to trace quantities of Mn$_{3}$O$_{4}$ which has an ordering temperature of T$_{\mathrm{N}}$=42 K~\cite{Regmi09:321,Seo04:43}.  Features near 140 K in the magnetic susceptibility can also be explained by trance amounts of LaMnO$_{3}$~\cite{Ghivelder99:60,Huang97:55}.

Evidence for short range, two-dimensional magnetic order in La$_{2}$O$_{2}$Mn$_{2}$OSe$_{2}$ was observed from neutron powder diffraction, where a Warren peak, indicative of two dimensional ordering~\cite{Ni10:82}, was observed over a wide temperature range. This peak diminished on cooling and was unobservable at 6 K. At 100 K, the out-of-plane correlation length was extracted to be 60 \AA~\cite{Ni10:82}. Broadening of magnetic Bragg reflections (as observed for $Ln_{2}$O$_{2}$Fe$_{2}$OSe$_{2}$ ($Ln$ = La, Ce, Nd)~\cite{McCabe10:150,McCabe14:90} and Sr$_{2}$F$_{2}$Fe$_{2}$OS$_{2}$ phases~\cite{Zhao13:87}) is not observed for La$_{2}$O$_{2}$Mn$_{2}$OSe$_{2}$~\cite{Ni10:82,Free11:23}. This suggests more similar magnetic correlation lengths within, and perpendicular to the $M_{2}$O planes in the three-dimensional magnetic structures of Mn$^{2+}$ and Co$^{2+}$ materials~\cite{Ni10:82,Free11:23,Fuwa10:150,Fuwa10:132},  compared with the Fe$^{2+}$ analogues~\cite{McCabe10:150,McCabe14:90,Zhao13:87}. 

${\bf{R_{2}O_{2}Mn_{2}OSe_{2}}}$ ($R=Ce, Pr, Nd,$ and $Sm$) - The magnetic properties of two dimensional Mn$^{2+}$ based oxyselenides have been investigated for a series of lanthanides and reported in Ref. \onlinecite{Free11:23}.  There is no reported changed in the magnetic structure with $Ln^{3+}$ ion and the results reported in Table \ref{mn_moments} show no observable change of the ordered magnetic moment with rare earth substitution.  However, a decrease in Mn$^{2+}$ T$_{\mathrm{N}}$ in proportion to the rare earth radius is observed.

To allow a direct comparison with the Fe$^{2+}$ materials discussed above, bond distance information is reproduced from Ref. \onlinecite{Free11:23} in Table. \ref{mn_moments}.  The manganese ordering temperature for two dimensional oxyslenides increases with decreasing rare-earth radius and has been associated with increased overlap of the orbitals involved in the magnetic coupling.  Finally, no evidence for rare earth ordering in either Pr$_{2}$O$_{2}$Mn$_{2}$OSe$_{2}$ or Ce$_{2}$O$_{2}$Mn$_{2}$OSe$_{2}$ has been reported. 

\subsubsection{Mixed Fe/Mn-based oxyselenides}

${\bf{Nd_{2}O_{2}(Fe_{1-x}Mn_{x})_{2}OSe_{2}}}$ - Several groups have investigated the magnetic behaviour of compositions within the $Ln_{2}$O$_{2}$Fe$_{2-x}$Mn$_{x}$OSe$_{2}$ solid solution and magnetic susceptibility measurements suggest some ferromagnetic behaviour on cooling~\cite{Lei12:86,Liu15:618,Land13:25}. We note that this has also been observed for powder samples of La$_{2}$O$_{2}$Mn$_{2}$OSe$_{2}$~\cite{Ni10:82,Free11:23} whilst single crystal studies suggested only antiferromagnetic ordering on cooling~\cite{Liu11:83}. Landsgesell \textit{et al.} have investigated the magnetic ordering in La$_{2}$O$_{2}$MnFeOSe$_{2}$ with a disordered arrangement of Mn$^{2+}$ and Fe$^{2+}$ ions within the $M_{2}$O layers. Their results from neutron powder diffraction experiments indicate that La$_{2}$O$_{2}$MnFeOSe$_{2}$ orders with $\vec{k}$ = (0 0 0) (as for La$_{2}$O$_{2}$Mn$_{2}$OSe$_{2}$) but with moments within the ab plane, although the exact spin arrangement has yet to be confirmed~\cite{Land13:25}.

\begin{figure}[t]
\includegraphics[width=8.5cm] {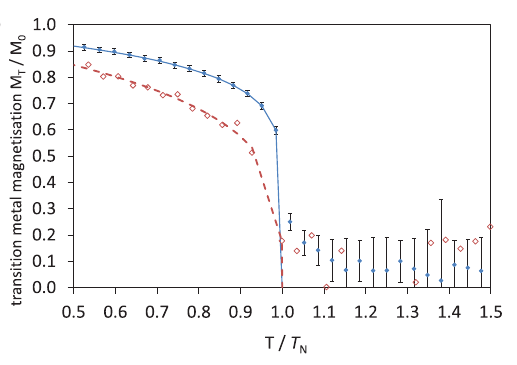}
\caption{\label{order_param} The magnetic order parameters for La$_{2}$O$_{2}$Mn$_{2}$OSe$_{2}$ (blue) and La$_{2}$O$_{2}$Fe$_{2}$OSe$_{2}$ (red) from Ref. \onlinecite{McCabe10:150}.  The data was obtained from the magnetic intensities derived from magnetic neutron powder diffraction.}
\end{figure}

\subsubsection{Critical scattering}

The temperature dependence of the magnetic order parameter can provide helpful information on the universality class and also the dimensionality of the magnetism~\cite{Collins:book}.  The critical properties classify the phase transition and allow commonalities to be established with other systems.  To this end, the magnetic order parameter has been studied in a number of materials and relies on the fitting of a critical exponent to the magnetisation $M(T)=M_{0} (1-{T\over{T_{N}}})^\beta$.    A summary of results obtained from several groups using neutron diffraction to measure the magnetisation is presented in Table \ref{exponents}. 

\begin{table}[ht]
\caption{Critical exponents for the magnetic order parameter extracted from magnetic neutron diffraction}
\centering
\begin{tabular} {c c c}
\hline\hline
Compound & $\beta$ & Reference \\
\hline\hline
La$_{2}$O$_{2}$Mn$_{2}$OSe$_{2}$ & 0.24(7) &  \onlinecite{Free11:23} \\
Ce$_{2}$O$_{2}$Mn$_{2}$OSe$_{2}$ & 0.29(7) &  \onlinecite{Free11:23} \\
Pr$_{2}$O$_{2}$Mn$_{2}$OSe$_{2}$ & 0.27(4) &  \onlinecite{Free11:23} \\
\hline\\
Ce$_{2}$O$_{2}$FeSe$_{2}$ & 0.28(1) &  \onlinecite{McCabe14:90} \\
\hline \\
La$_{2}$O$_{2}$Fe$_{2}$OSe$_{2}$ & 0.122(1) &  \onlinecite{McCabe10:150} \\
Ce$_{2}$O$_{2}$Fe$_{2}$OSe$_{2}$ & 0.11(1) &  \onlinecite{McCabe10:150} \\
Ba$_{2}$F$_{2}$Fe$_{2}$OSe$_{2}$ & 0.118 &  \onlinecite{Kabbour08:130} \\
Sr$_{2}$F$_{2}$Fe$_{2}$OS$_{2}$ & 0.15 &  \onlinecite{Kabbour08:130} \\
\hline \hline
\label{exponents}
\end{tabular}
\end{table}

The critical exponents can be seen to fall into two broad categories with the manganese variants having exponents $\sim$ 0.2-0.3 and the iron based two dimensional oxyselenides have exponents $\sim$ 0.1.   The critical exponents for the 2D Ising universality class is $\beta$= 0.125 and 2D XY is 0.13.  3D Ising has an exponent of 0.326 and 3D Heisenberg is 0.36~\cite{Collins:book}.  The two dimensional Fe$^{2+}$ based oxyselenides clearly display 2D character and the exponents are similar to $Re$FeAsO materials studied in Ref. \onlinecite{Wilson10:81} with $\beta \sim 0.125$, close to the ideal 2D Ising universality class.  The exponents are also similar to Fe$_{1+x}$Te which, for large values of interstitial iron, displays $\beta$=0.15~\cite{Rodriguez13:88}  and FeAs with $\beta$=0.16(2)~\cite{Rodriguez11:83}. The manganese compounds and structurally one dimensional Ce$_{2}$O$_{2}$FeSe$_{2}$ are closer to the 3D limit and are similar to to the critical exponents in (Ba,Sr)Fe$_{2}$As$_{2}$~\cite{Wilson09:79,Christianson09:203} and also more recently extracted in SrMn$_{2}$As$_{2}$~\cite{Das16:xx}.

\onlinecite{Chen09:80} has considered the dimensionality of the order parameters in two dimensional pnictide iron based systems in terms of coupling to an orbital degree of freedom.  This orbital order parameter possess 2D-Ising character and results from a spin-orbital Hamiltonian.  This idea is broadly consistent with the response discussed above in the case of the magnetic oxyselenides.  Fe$^{2+}$ based materials with 6 $d$ electrons potentially have an orbital degree of freedom as discussed above while Mn$^{2+}$ has only 5 $d$ electrons and therefore no orbital component in the weak/intermediate crystal field limit.  Based on the orbital model proposed in Ref. \onlinecite{Chen09:80}, it is therefore expected that Fe$^{2+}$ would display a stronger 2D-Ising character as illustrated in the data in Table \ref{exponents}.

\subsubsection{Summary of the magnetic structure variation with transition metal ion}

\begin{figure}[t]
\includegraphics[width=8.5cm] {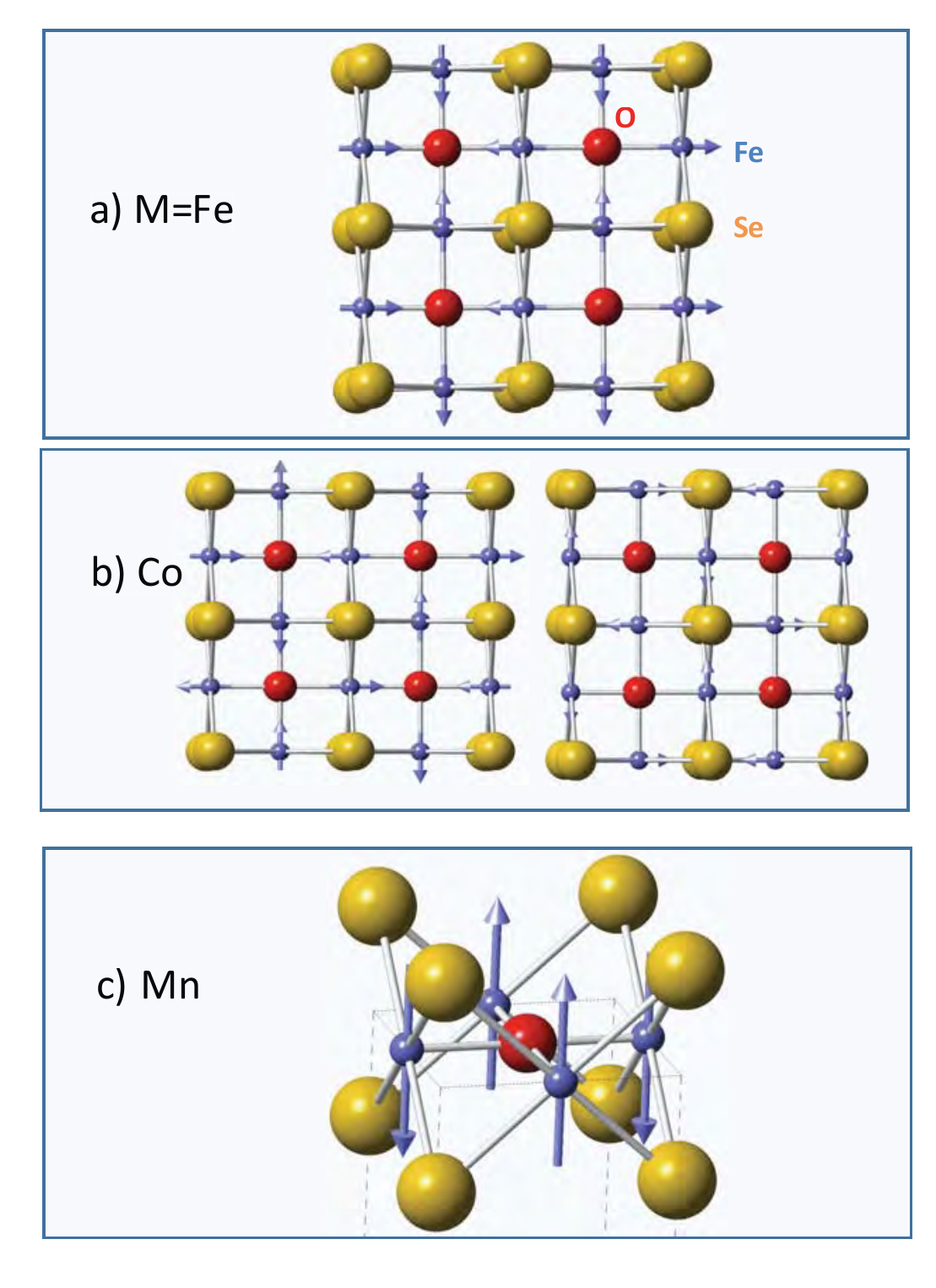}
\caption{\label{mag_summary} A summary of the reported magnetic structure for the $Ln_{2}$O$_{2}M_{2}$OSe$_{2}$ series of compounds with the transition metal ion M= $(a)$ Fe$^{2+}$, $(b)$ Co$^{2+}$ (note the ambiguity of magnetic structures and the two possible results) and $(c)$ Mn$^{2+}$.}
\end{figure}

While the magnetic structures of the $Ln$$_{2}$O$_{2}M_{2}$OSe$_{2}$ do not vary drastically with changing $Ln$ ion, there are dramatic changes with transition metal ion ($M$) substitution. While we have discussed these structures in depth above, here we summarise the three magnetic structures observed for $M$=Fe$^{2+}$, Co$^{2+}$, and Mn$^{2+}$ in Fig. \ref{mag_summary}.  It should also be noted that there is a large difference in the critical properties between $M$=Fe$^{2+}$ and Mn$^{2+}$ as noted above.  

The magnetic structures adopted by the $Ln_{2}$O$_{2}M_{2}$OSe$_{2}$ materials are governed by the exchange interactions within the $M$$_{2}$O planes.  We now discuss the magnetic interactions governing these structures discussed above.

\subsection{Magnetic interactions}

The sign and relative strength of the magnetic interactions can be postulated based upon magnetic diffraction data establishing the magnetic structure of the material.  These can also be compared with Goodenough rules and expectations on systematically characterised systems like the cuprates.  A definitive measure of the coupling strength is obtained through spectroscopy and neutron inelastic spectroscopy is ideal given the sensitivity to magnetic moments and the energy scale that is sensitive to.  However, few low-energy spectroscopy measurements have been reported with the focus on the iron based two dimensional oxyselenides given their possible interesting Mott insulating behaviour and also the close analogy to the iron based pnictide and chalcogenide superconductors.  In this section we discuss the magnetic interactions with first a review of neutron inelastic work on two dimensional Fe-based oxyselenides and then a short summary of the results found for other materials discussed above in the context of the static structure.

A summary of the magnetic interactions is presented in Fig. \ref{structure5} $(c)$.   The nearest neighbour $M$ - $M$ distance within the $M_{2}$O planes ranges from $\sim$ 3.02 \AA\ (in Ba$_{2}$F$_{2}$OMn$_{2}$Se$_{2}$) to $\sim$ 2.83 \AA (in Sm$_{2}$O$_{2}$Fe$_{2}$OSe$_{2}$).  The nearest neighbour $J_{1}$ exchange could be direct, or could proceed via 90$^{\circ}$ $M$ - O - $M$ superexchange or $\sim$ 64$^{\circ}$ $M$ - Se - $M$ superexchange, and is expected to be antiferromagnetic  for $M$ = Mn, Fe and Co~\cite{Kabbour08:130,Zhu10:104,Zhao13:87,Koo12:324,Wu10:82,Wang10:132}. The next nearest neighbour $J_{2'}$ exchange interaction (180$^{\circ}$ superexchange via M - O - M) is also expected to be antiferromagnetic from Goodenough-Kanamori rules for $M$=Mn, Fe and Co.  The next nearest neighbour $J_{2}$ exchange ($\sim$ 97$^{\circ}$ $M$ - Se - $M$ superexchange) is expected to be ferromagnetic for $M$ = Fe~\cite{Kabbour08:130,Zhu10:104,Zhao13:87} and Co~\cite{Wu10:82,Wang10:132}, despite the apparent antiferromagnetic interaction from the magnetic structure.   

The case of $M$=Mn is different for the $J_{2}$ exchange ($\sim$ 97$^{\circ}$ M - Se - M superexchange) with density functional calculations differing in terms of the sign of $J_{2}$~\cite{Liu11:83,Koo12:324}.  Antiferromagnetic ordering of (Ce,La)$_{2}$O$_{2}$MnSe$_{2}$ described above (with M - Se - M angles of $\sim$ 75$^{\circ}$ and $\sim$ 100$^{\circ}$) might suggest antiferromagnetic $J_{2}$ in La$_{2}$O$_{2}$Mn$_{2}$OSe$_{2}$.  A key factor for the relative strength of the three exchange interactions in Fig. \ref{structure5} $(c)$ with M is the electronegativity: for M = Mn, nearest neighbour $J_{1}$ interactions dominate and $J_{2'}$ interactions are frustrated while for the more electronegative M = Co, $J_{2'}$ interactions dominate at the expense of nearest neighbour $J_{1}$ interactions. 

\subsubsection{La$_{2}$O$_{2}$Fe$_{2}$OSe$_{2}$ and Ce$_{2}$O$_{2}$FeSe$_{2}$}

Given the semiconducting nature of the oxyselenides, a localised model for the spin interactions is appropriate and therefore the dominant term in the magnetic Hamiltonian that needs to be considered is $H=J\sum_{i,j}\vec{S}_{i}\cdot\vec{S}_{j}$, where the sum is performed over nearest neighbours.  This model is much more applicable to the oxyselenides over the cuprates or iron based superconductors which derive from metallic ground states and hence display strong evidence of coupling between electronic and magnetic moments.   This is particularly evident in the high energy neutron scattering response in the cuprates~\cite{Stock10:82,Stock07:75} and iron based systems~\cite{Stock14:90}.

One of the key questions that arise from the magnetic diffraction data is how to stabilise the orthogonal $2-k$ magnetic structure reported for La$_{2}$O$_{2}$Fe$_{2}$OSe$_{2}$.  As noted in Ref. \onlinecite{McCabe10:150}, second order terms in the spin involving either antisymmetric  (such as Dzyaloshinskii-Moriya interactions $\sim$ $\vec{D}\cdot (\vec{S}_{i} \times \vec{S}_{j})$) or symmetric (such as Heisenberg) interactions are not able to stabilise the $2-k$ structure for these tetragonal crystal structures.  

Other terms that may be relevant to the magnetic Hamiltonian and discussed in the literature include the biquadratic spin-spin interactions.  These terms have the form $H_{1}=-K\sum_{i,j}(\vec{S}_{i}\cdot\vec{S}_{j})^{2}$ and are required to understand~\cite{Wysocki11:7,Stanek11:84,Yu12:86} spin excitations in the pncitides near the Brillouin zone boundary~\cite{Harriger11:84,Zhao09:5}.    Without consideration of this term in the magnetic Hamiltonian, anisotropic exchange terms need to be considered which are difficult to reconcile given the tetragonal nuclear structure.  Studies of the magnetic structure are not able to uniquely determine which term is present in the Hamiltonian and therefore neutron inelastic scattering is required to obtain a better understanding of the magnetic interactions. 

Due to lack of large single crystals, complete neutron scattering data is currently quite sparse for the oxyselenide materials.  While powders provide limited information and, most importantly, are not able to determine the dispersion near the zone boundary, they can provide helpful information on the integrated intensity (a rough measure of the overall exchange constant), anisotropy gap (directly related to the local crystalline electric field environment), and also the integrated spectral weight which helps in understanding the spin state.

\begin{figure}[t]
\includegraphics[width=8.7cm] {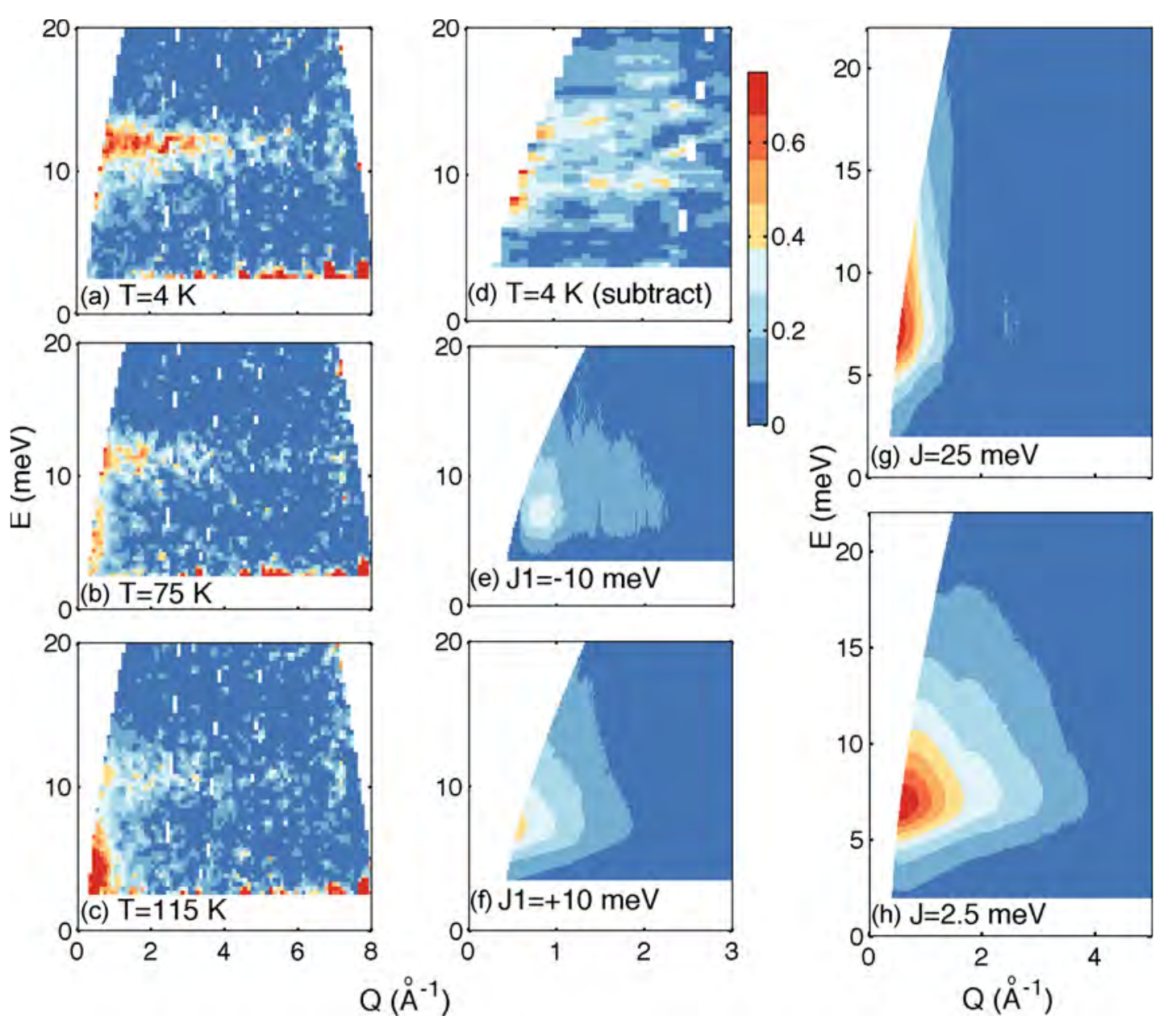}
\caption{\label{ce_ins}   Neutron inelastic data and calculations using the first-moment sum rule combined with the single mode analysis on Ce$_{2}$O$_{2}$FeSe$_{2}$ .  The figure is taken from Ref. \onlinecite{McCabe14:90}.  $(a-c)$ shows temperature dependent data illustrating a cerium crystal field peak near $\sim$ 12 meV and low-energy ferromagnetic fluctuations near $Q$=0.  $(d)$ shows an estimate of the iron contribution subtracting off the cerium crystal field using high-angle data.  $(e-h)$ show calculations from which is concluded that the Fe-Fe exchange is ferromagnetic and $\sim$ 25 meV in magnitude.  Note the sign convention that negative (-) is antiferromagnetic and positive (+) is ferromagnetic.}
\end{figure}

Inelastic data on powders has been obtained on powders of La$_{2}$O$_{2}$Fe$_{2}$OSe$_{2}$ and Ce$_{2}$O$_{2}$FeSe$_{2}$ allowing the magnetic interactions to be investigated with increasing levels of structural complexity.   We first outline the results for the structurally one-dimensional Ce$_{2}$O$_{2}$FeSe$_{2}$ and then discuss this in the context of two-dimensional La$_{2}$O$_{2}$Fe$_{2}$OSe$_{2}$.

The powder averaged magnetic excitations for a sample of Ce$_{2}$O$_{2}$FeSe$_{2}$ is shown in Fig. \ref{ce_ins} with data taken from the MARI direct geometry spectrometer at ISIS.  Ce$_{2}$O$_{2}$FeSe$_{2}$ has two magnetic sites (Ce$^{3+}$ and Fe$^{2+}$) which complicates the neutron excitation spectrum as it consists of both crystal field excitations from the Ce$^{3+}$ sites and also collective excitations for $S$=2 Fe$^{2+}$ moments.    However, the crystal field excitations are momentum independent while the collective excitations of Fe$^{2+}$ moments are comparatively localised in momentum. This difference was used in Ref. \onlinecite{McCabe14:90} to subtract the crystal field contribution from the powder average neutron inelastic spectrum in Ce$_{2}$O$_{2}$FeSe$_{2}$.  The results are shown in Fig. \ref{ce_ins} where the remaining spectral weight after subtraction is concentrated near $Q$=0, indicating ferromagnetic interactions consistent with neutron magnetic diffraction.  

The coupling between the rare earth site and the iron site were also studied by investigating the response of the Ce$^{3+}$ crystal field excitations to Neel ordering on the iron site.  Ce$^{3+}$ can be assigned a $J={5\over2}$ and the crystal field scheme consists of three doublets.  Kramers theorem ensures that the degeneracy of these doublets is not split unless there is a field which breaks time reversal symmetry such as a magnetic field.  Crystalline electric fields will not split the doublet degeneracy alone.   Since no splitting of the Ce$^{3+}$ crystal field doublets was observed at low temperatures, it was concluded in Ref. \onlinecite{McCabe14:90} that the coupling between the iron and rare earth sites is weak.  This is in contrast to rare earth substituted pnictides where a strong coupling is observed between the iron and rare earth sites as demonstrated by a splitting of the crystal field doublets at temperatures below the iron ordering~\cite{Chi08:101}.  High resolution neutron spectroscopy studies even observe a dispersion of the crystal field excitations implying coupling between the rare-earth sites~\cite{Li14:9}.  No such effects have been reported in rare earth substituted oxyselenides.

An estimate of the Fe-Se-Fe exchange constant was obtained by modelling the powder average neutron spectrum using the first moment sum rule combined with the single mode approximation outlined above.  Given the kinematics of neutron spectroscopy which masks lower momentum transfers at higher energies, the errorbar on this analysis is large however an estimate can be obtained.  As shown in Fig. \ref{ce_ins} $J$=-25 meV provides a reasonable description of the results.   Note in Ref. \onlinecite{McCabe14:90}, positive (+) exchange indicates ferromagnetic exchange while usually (-) is often taken and is the convention used here.  

\begin{figure}[t]
\includegraphics[width=8.7cm] {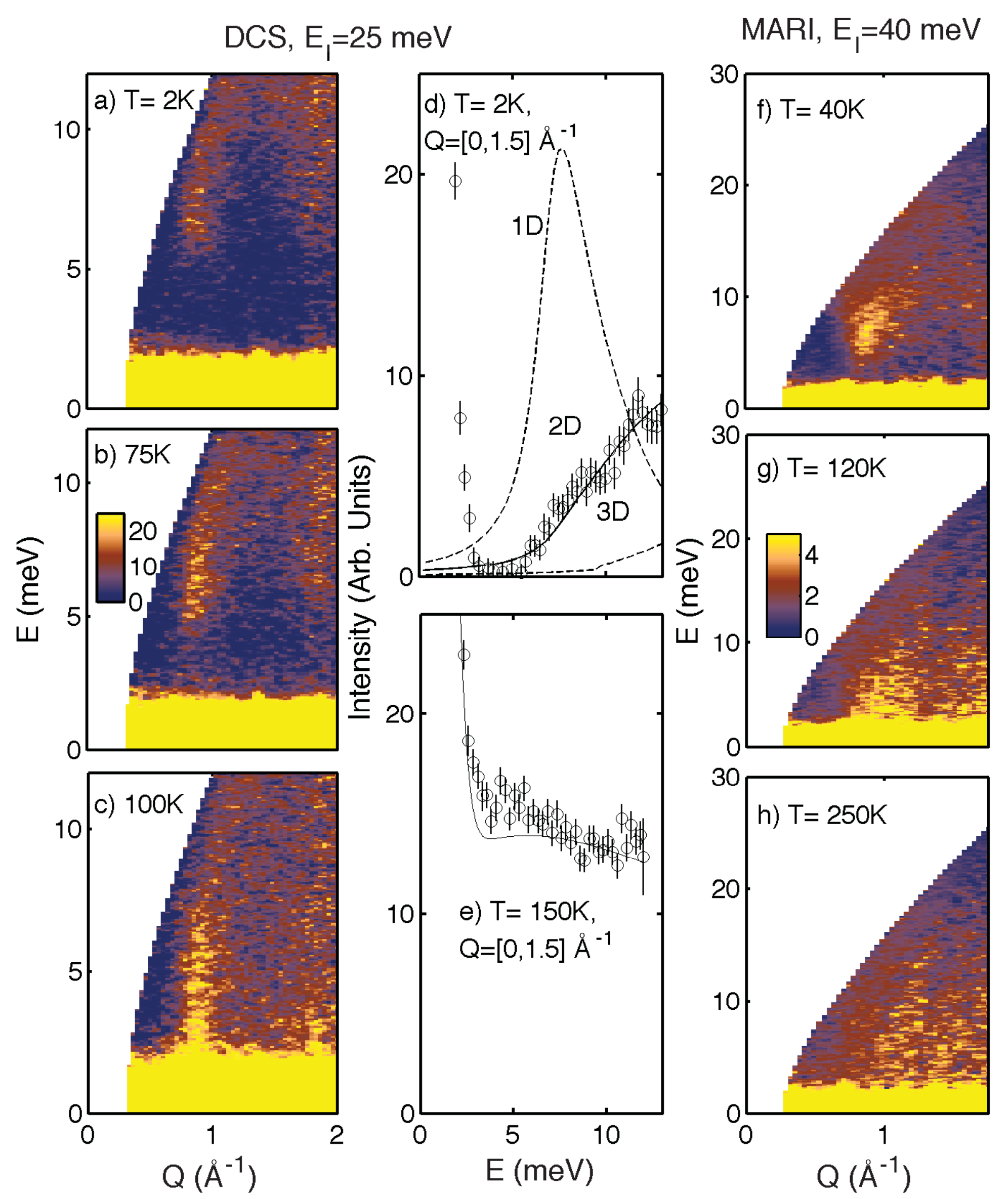}
\caption{\label{T_depend}   $(a)-(c)$ Powder averaged spectra for La$_{2}$O$_{2}$Fe$_{2}$OSe$_{2}$ measured on DCS. $(d)$ Momentum-integrated energy scan at 2 K (upper) and 150 K (lower); the curves are calculations using a single-mode analysis with a 1D model, a 2D model, and a 3D model. $(f)-(h)$ Plots of the powder-averaged temperature spectra taken on the MARI spectrometer.  The figure is taken from McCabe \textit{et al.}~\cite{McCabe10:150}}
\end{figure}

Having analysed the chain compound and established the ferromagnetic exchange, we now discuss work done on two dimensional La$_{2}$O$_{2}$Fe$_{2}$OSe$_{2}$.   A summary of the magnetic interactions and the bond angles governing them in this compound is illustrated in Fig. \ref{structure5} $(c)$. The powder average inelastic spectrum for La$_{2}$O$_{2}$Fe$_{2}$OSe$_{2}$ is illustrated in Fig. \ref{T_depend} at several temperatures.  Panels $(a)$-$(c)$ illustrate the low-energy part of the magnetic structure showing a gap of $\sim$ 6 meV which softens with increasing temperature.   The presence of a gap is also confirmed by NMR studies in Ref. \onlinecite{Gunther14:90} where the relaxation rate was fitted to $(1/T_{1})\propto T^{2}e^{-\Delta/T}$ and a gap value of $\Delta$=55 K was extracted which is close to that measured with neutron inelastic scattering. 

It is interesting to compare the magnetic anisotropy gap and its temperature dependence to Fe$_{1+x}$Te.   The magnetic anisotropy is very similar to that measured in single crystals of Fe$_{1+x}$Te where commensurate $\vec{q}= ({1\over 2},0,{1\over 2})$ is observed~\cite{Stock11:84}.  As shown in Ref. \onlinecite{Rodriguez13:88}, Fe$_{1+x}$Te for small values of $x$ undergoes a metal to ``semi metal" transition characterised by a sharp response and change in slope in the resistivity.  This change in slope is coincident with a gapping of the magnetic fluctuations and it was postulated that this temperature dependence in the gapped spin fluctuations was responsible for the ``metallic-like" behavior in Fe$_{1+x}$Te~\cite{Rodriguez13:88}.  As noted above, the resistivity from spin fluctuations can be calculated from $S(\vec{Q},\omega)$ and the electronic scattering from these fluctuations and the removal of low-energy decay channels was found to explain the change in resistivity.  While spin excitation gap magnitude, and also the ordering wavevector, is similar in Fe$_{1+x}$Te and La$_{2}$O$_{2}$Fe$_{2}$OSe$_{2}$, no such metal-``semimetal" transition has been reported in La$_{2}$O$_{2}$Fe$_{2}$OSe$_{2}$ despite the spin fluctuations showing a qualitatively similar temperature dependence across T$_{\mathrm{N}}$.

\begin{figure}[t]
\includegraphics[width=8.5cm] {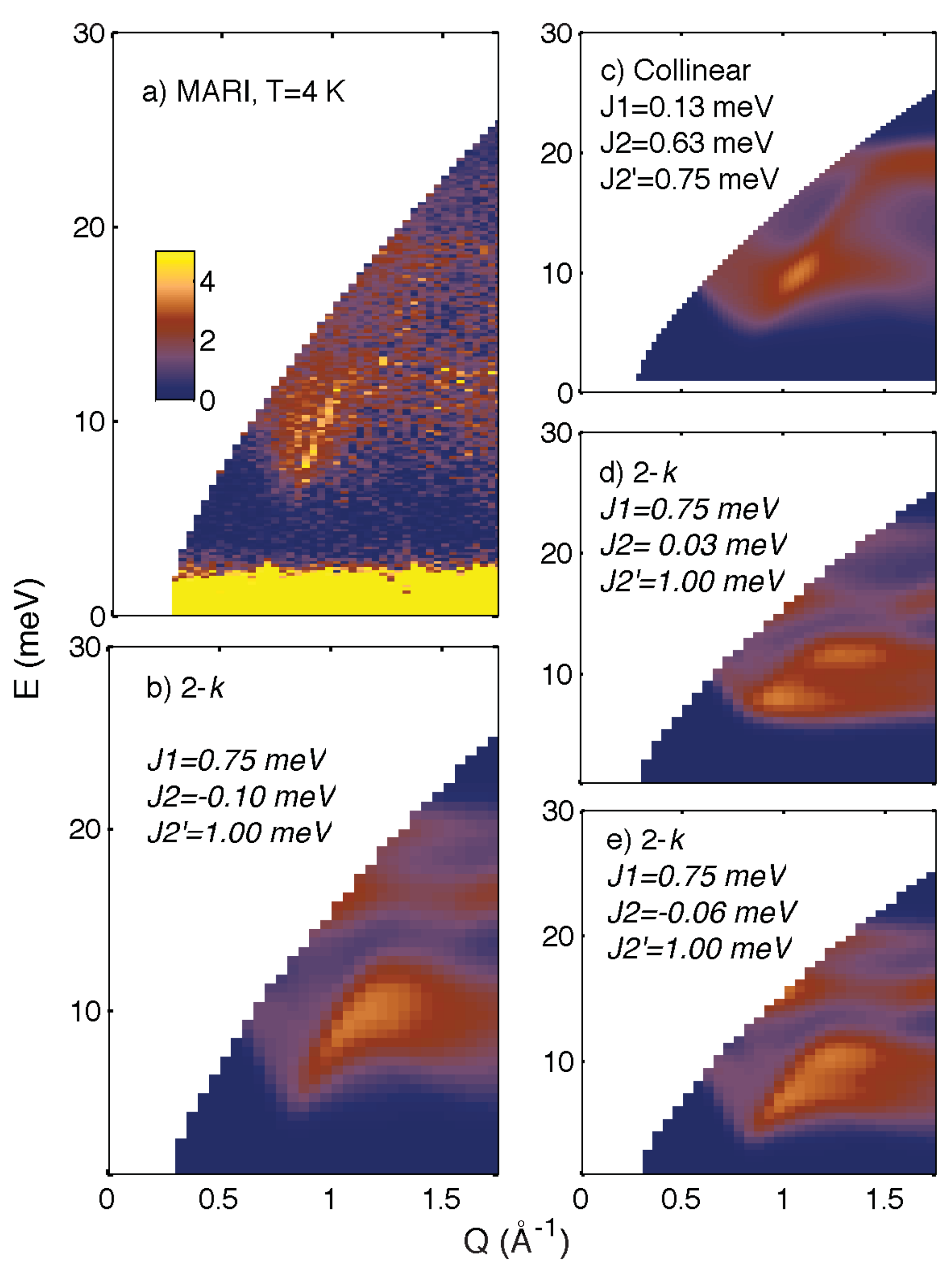}
\caption{\label{compare_model}   Low temperature neutron inelastic scattering for La$_{2}$O$_{2}$Fe$_{2}$OSe$_{2}$ compared against calculations based on a heuristic spin-wave model.  The results are taken from Ref. \onlinecite{McCabe10:150}.  $(a)$ shows a inelastic neutron scattering data compared against models supporting $(b)$ $2-k$ structure and $(c)$ collinear structure.  The effect of weak antiferromagnetic and ferromagnetic values on the spectra are shown in panels $(d)$ and $(e)$.}
\end{figure}

While the magnetic excitations are gapped in the magnetically ordered Neel state at low temperatures, they are gapless at high temperatures.  This may explain the origin of the observation of momentum broadened diffuse scattering in La$_{2}$O$_{2}$Co$_{2}$OSe$_{2}$ given neutron diffraction measurements are typically done in two-axis mode and hence energy integrating.  We emphasise, though, that no inelastic data has been reported for this particular Co compound and this observation is speculative with currently available data.  

The intensity distribution at the edge of the low temperature excitation gap is sensitive to the dimensionality of the magnetic interactions.  The dimensionality of the interactions is established in Fig. \ref{T_depend} $(d)$ where a first-moment sum rule analysis suggests that the interactions in La$_{2}$O$_{2}$Fe$_{2}$OSe$_{2}$  are two dimensional.  The figure shows the momentum integrated data compared against calculations based on the single-mode approximation for an isotropic dispersion in one-dimensional (1D) chain, 2D plane, or 3D structure.  The 2D model gives the best description consistent with the 2D-Ising critical properties discussed above.

Scans that probe larger energy transfers are shown in Figs. \ref{T_depend} $(f)-(h)$ where it is shown that the magnetic excitations extend up to energy transfers of $\sim$ 25 meV.  This small band width accounts for all of the expected spectral weight, confirmed by integrating the intensity in momentum and energy and comparing against the zeroth sum rule discussed above.  The total integral of both elastic and inelastic spectral weight was found to give an integral of 5.9 (4) which is close to the $S$=2 value of 6.   To give an estimate of the exchange coupling constants,  Ref. \onlinecite{McCabe10:150} compared the data against  calculations with a large single-ion anisotropy to fix the moment direction and considering only Heisenberg spin exchange.   This model is somewhat artificial as it uses anisotropy terms in the magnetic Hamiltonian to fix the moment direction to allow consistency with possible magnetic structures found from magnetic diffraction data.  As shown in Fig. \ref{compare_model}, the experimental spectrum can be reproduced reasonably well for the $2-k$ ground state with $J_{1}$=0.75 meV, $J_{2}$=-0.10 meV, and $J_{2'}$=1.0 meV.  Consistent models could be obtained for a collinear model, however these required an antfierromagnetic $J_{2}$ which is inconsistent with first principles calculations and also Goodenough rules.  Perhaps more conclusively from an experimental perspective,  the requirement of antiferromagnetic $J_{2}$ in the collinear model is surprising given the neutron spectroscopy data reviewed above on the chain compound Ce$_{2}$O$_{2}$FeSe$_{2}$.    Also, comparisons with magnetic high temperature susceptibility data find better agreement for the Weiss temperature with the parameters derived from $2-k$ model  than the corresponding parameters derived for the collinear model.  Therefore, through a combination of neutron diffraction and spectroscopy, Ref. \onlinecite{McCabe10:150} concluded that the magnetic structure of La$_{2}$O$_{2}$Fe$_{2}$OSe$_{2}$ is the $2-k$ structure.

It is interesting to note that while consistency is obtained in the sign of the exchange constants between Ce$_{2}$O$_{2}$FeSe$_{2}$ and La$_{2}$O$_{2}$Fe$_{2}$OSe$_{2}$, the  single mode analysis suggests that the ferromagnetic exchange in Ce$_{2}$O$_{2}$FeSe$_{2}$ is much larger than La$_{2}$O$_{2}$Fe$_{2}$OSe$_{2}$.  This can be attributed to the different local bond environment in both materials.  In Ce$_{2}$O$_{2}$FeSe$_{2}$, the Fe$^{2+}$ ion is in a local tetrahedral environment while La$_{2}$O$_{2}$Fe$_{2}$OSe$_{2}$ is more square planar or pseudo-octahedral.   In this context the Fe$^{2+}$ site La$_{2}$O$_{2}$Fe$_{2}$OSe$_{2}$ is quite different than iron based pnictide and chalcogenide systems where the iron is in a tetrahedral framework.  

The small exchange constants in La$_{2}$O$_{2}$Fe$_{2}$OSe$_{2}$ derived from this heuristic model and the bandwidth of the magnetic excitations are remarkable in the context of observations in the cuprates and also iron based pnictides.    Mott insulating La$_{2}$CuO$_{4}$~\cite{Coldea01:86} and YBa$_{2}$Cu$_{3}$O$_{6+x}$~\cite{Hayden96:54} both have bandwidths of over 300 meV and the parent phases of the pnictides have magnetic excitations that extend up to about $\sim$ 100 meV for BaFe$_{2}$As$_{2}$ and $\sim$ 150 meV in CaFe$_{2}$As$_{2}$~\cite{Zhao09:5,Dai15:87}.  The excitations in Fe$_{1+x}$Te chalcogenides extend up to $\sim$ 150-200 meV and the high energy excitations account for a large fraction of the total spectral weight~\cite{Stock14:90}.  Therefore, while it is tantalising to make a connection between the two dimensional oxyselenides to the cuprates and iron based superconductors owing to the qualitatively similar electronic phenomena, the magnetic excitations are very different with the oxyselenides displaying significantly smaller coupling.

\subsubsection{$Ln$$_{2}$O$_{2}$$M$$_{2}$OSe$_{2}$ for $M$=Mn$^{2+}$ or Co$^{2+}$}

At the time of writing this review, there has been no reports of neutron inelastic scattering data on Mn$^{2+}$ or Co$^{2+}$ analogues of the two dimensional oxyselenides discussed above.  Future work study the fluctuation spectrum in these materials will be useful in comparison to the work presented above in the context of the iron based oxyselenides.

\section{Electronic properties}

We now discuss the electronic properties of oxyselenides in terms of resistivity, optical measurements, x-ray spectroscopy, and calculations.  The ZrCuSiAs materials have been investigated in the context of thermoelectric properties while interest in compounds based on magnetic Fe$^{2+}$ have been pursued in the context of unconventional electronic properties in relation to superconductivity.

\subsection{ZrCuSiAs structures and related phases}

\begin{figure}[t]
\includegraphics[width=8.5cm] {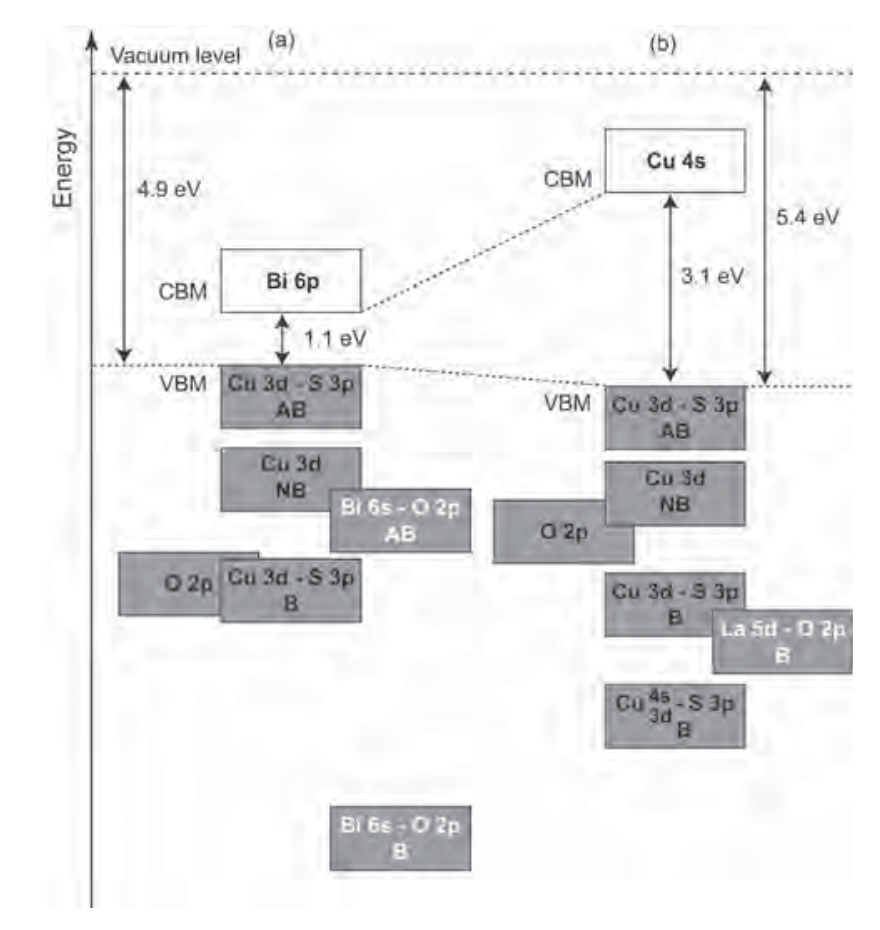}
\caption{\label{band_summary}  Schematic band structures of $(a)$ BiCuOS and $(b)$ LaCuOS reproduced from Ref. \onlinecite{Hiramatsu08:20}.}
\end{figure}

The optical transparency and p-type semiconducting properties of LaCuOQ materials have prompted several experimental and theoretical studies to understand their electronic structure.  Diffuse reflectance spectra measured for LaCuOSe indicate a band-gap of $\sim$2.8 eV,~\cite{Ueda02:91} although optical absorption measurements reveal sub-band gap absorptions~\cite{Hiramatsu10:132}.  LaCuO$Q$ ($Q$ = S, Se) phases show high conductivities and Seebeck and Hall measurements confirm that they behave as p-type semiconductors~\cite{Ueda02:91,Ueda03:15}.  Hole-doping (for example La$_{1-x}A_{x}$CuOSe; $A$ = Sr$^{2+}$, Mg$^{2+}$) can further enhance this conductivity~\cite{Ueda02:91,Hiramatsu07:91}. 

Energy band calculations show that the wide band gap arises primarily from the [Cu$_{2}$Se$_{2}$]$^{2+}$ layers with the valence band maximum composed of antibonding Cu 3d and Se 4p states, whilst Cu 4s states make up the conduction band minimum  (see Fig. \ref{band_summary})~\cite{Ueda04:69,Ueda04:16}. The connection of the band gap with Se states is further confirmed by investigations as a function of substituting S for Se which showed a large change in the band gap with doping (see Table \ref{band_gap} below for pure compound values).  These calculations indicate that the hole carriers are confined to the [Cu$_{2}$Se$_{2}$]$^{2+}$ layers by the insulating [La$_{2}$O$_{2}$]$^{2+}$ layers, giving significant two-dimensional character, consistent with features at the absorption edge of the material~\cite{Ueda04:69}.   More recently, density functional theory calculations have investigated the origin of the p-type semiconductivity and indicate that although this can be induced by aliovalent doping, Cu$^{+}$ vacancies are easily formed and are likely to be the dominant acceptor defect in samples~\cite{Hiramatsu10:207,Scanlon14:2}.   With the band gap dominated by the [Cu$_{2}$Se$_{2}$]$^{2+}$ layers, similar optical and electronic properties are observed for $Ln$CuOS ($Ln$ = La, Pr, Nd), with a slight decrease in band gap with decreasing $Ln^{3+}$ ionic radius~\cite{Ueda03:15}.

\begin{table}[ht]
\caption{Estimated band gaps from optical measurements taken from Ref. \onlinecite{Hiramatsu08:20}.}
\centering
\begin{tabular} {c c}
\hline\hline
Compound & $\Delta (eV)$  \\
\hline\hline
LaCuOS & 3.1  \\
BiCuOS & 1.1   \\
\hline
BiCuOSe & 0.8  \\
LaCuOSe & 2.8  \\
LaCuOTe & 2.4  \\
\hline \hline
\label{band_gap}
\end{tabular}
\end{table}

Bi$^{3+}$ ions can also be accommodated in the fluorite-like oxide layers and give a dramatic change in properties. BiCuOSe has a much higher electron conductivity than LaCuOSe and a smaller band gap ($\sim$0.8 eV with absorption in the near-infrared region) and similar behaviour is observed for other BiCuOQ phases~\cite{Hiramatsu08:20}.  Although electron conductivity was found to be different for BiCuOSe and LaCuOSe, they have similar hole conductivities,  implying that difference in the band structure can be attributed to the conduction band.  While the Bi$^{3+}$ 6s states are 2 - 5 eV below the Fermi energy, the 6p states form the bottom of the conduction band, deepening the conduction band and decreasing the band gap (Fig. \ref{band_summary})~\cite{Hiramatsu08:20,Zou13:1}.  The low thermal conductivity of BiCuOQ phases, coupled with their semiconducting behaviour makes them promising thermoelectric materials~\cite{Zhao14:7}.

Related cation-ordered La$_{2}$O$_{2}$CdSe$_{2}$ has an even larger band gap (~3.3 eV from diffuse reflectance measurements) and high electrical resistivity, and attempts to induce semiconducting behaviour by aliovalent doping were unsuccessful~\cite{Hiramatsu04:14}. Density functional theory calculations suggest that the valence band is predominantly composed of Se 4p states, similar to that of LaCuOSe, whilst the conduction band is composed of Cd 5s states and is much narrower than that of LaCuOSe. This narrower conduction band dispersion may account for the difficulty in doping La$_{2}$O$_{2}$CdSe$_{2}$~\cite{Hiramatsu04:108}. La$_{2}$O$_{2}$ZnSe$_{2}$ behaves similarly (with high electrical resistivity, a band gap of 3.4(2) eV from diffuse reflectance measurements, and difficulties with aliovalent doping) and density functional theory calculations indicate that the conduction band is mainly composed of La states, leading to a larger band gap than in LaCuOSe~\cite{Tuxworth13:52}. Unlike the $Ln$ = La systems, $Ln$ = Ce analogues often have much smaller band gaps and higher conductivities due to the Ce 4f and 5d bands near the band gap~\cite{Ainsworth15:54,Ueda03:15,Pitcher09:48}.

\begin{figure}[t]
\includegraphics[width=8.7cm] {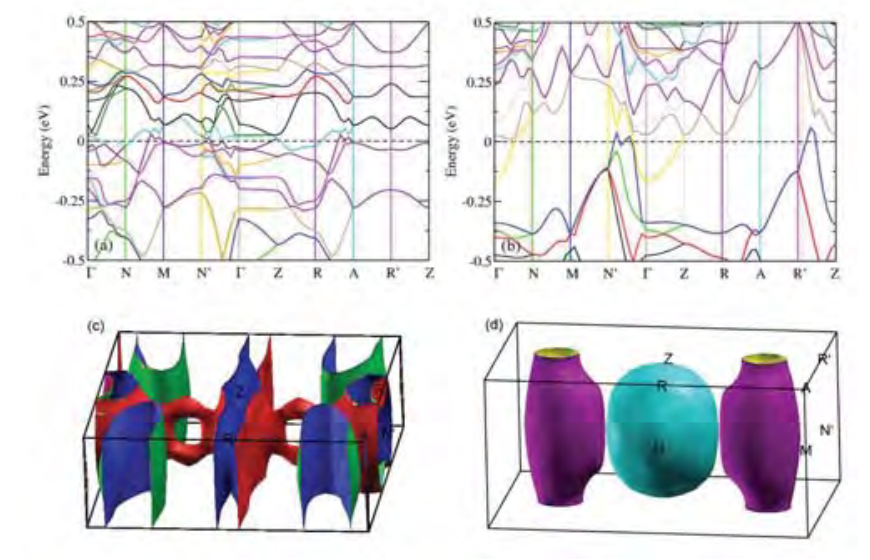}
\caption{\label{band_chain}  Electronic structure of $(a)$ Ce$_{2}$O$_{2}$FeSe$_{2}$ and $(b)$ (BaFe$_{2}$Se$_{3}$ in the nonmagnetic state.  The figure is taken from Ref. \onlinecite{Li14:9}.}
\end{figure}

The cation-ordered ZrCuSiAs-related phases $Ln_{2}$O$_{2}$FeSe$_{2}$ tend to have smaller band gaps and semiconducting behaviour. Polycrystalline samples of Ce$_{2}$O$_{2}$FeSe$_{2}$ are black and exhibit semiconducting behaviour (with electronic band gap $\sim$0.64 eV) and room temperature resistivity of $\sim$20 $\Omega$ cm~\cite{McCabe11:47}.  Electronic structure calculations were consistent with this and indicate itinerant Fe 3d states  and suggest that, despite the ``stripe" ordering, the electronic structure is far from being pseudo-one dimensional~\cite{Li14:9}.  This is illustrated in Fig. \ref{band_chain} which shows band structure presented in Ref. \onlinecite{Li14:9}.  The oxyselenide Ce$_{2}$O$_{2}$FeSe$_{2}$ is shown in Fig. \ref{band_chain} $(a)$ where it is seen that three bands cross the Fermi level.    Ref. \onlinecite{Li14:9} calculated the possible magnetic structures and found an ordered moment of 3.12 $\mu_{B}$ consistent with experimental work of 3.33 $\mu_{B}$~\cite{McCabe11:47}.

Density functional calculations suggest that, like the copper systems, the valence band and the conduction band are composed mainly of Fe 3d and Se 4p states~\cite{McCabe10:150}.  By contrast, the manganese analogues $Ln_{2}$O$_{2}$MnSe$_{2}$ generally have wider band gaps and more insulating behaviour: polycrystalline $Ln_{2}$O$_{2}$MnSe$_{2}$ (Ln = La, Pr, Nd) samples tend to be orange-brick red in colour and diffuse reflectance measurements on La$_{2}$O$_{2}$MnSe$_{2}$ indicate an optical band gap of 2.31 eV~\cite{Peschke15:641}.  Again, the Ce analogue Ce$_{2}$O$_{2}$MnSe$_{2}$ has slightly different properties, with polycrystalline samples being purple (single crystals orange~\cite{Wang15:27}), an activation energy for electronic conduction of 0.41(1) eV and room temperature conductivity of $\sim$ 9 $\times$ 10$^{-6}$ $\Omega^{-1}$ cm$^{-1}$, presumably due to the influence of Ce 4f states near E$_{F}$~\cite{Peschke15:641}.  The $\beta$- and monoclinic polymorphs of La$_{2}$O$_{2}$FeSe$_{2}$ allow us to consider the effect of the Fe coordination environment on the electronic structure.  $\beta$--La$_{2}$O$_{2}$FeSe$_{2}$ (with Fe(1)Se$_{4}$O$_{2}$ and Fe(2)Se$_{4}$ sites) is a black semiconductor with a band gap of $\sim$ 0.7 eV and room temperature resistivity $\sim$ 102 $\Omega$ cm~\cite{McCabe11:47}. The monoclinic polymorph of La$_{2}$O$_{2}$FeSe$_{2}$ with only FeSe$_{4}$O$_{2}$ sites is also a semiconductor but with a smaller band gap ($\sim$0.3 eV) and slightly lower room temperature resistivity~\cite{Nitsche14:640}.

\subsection{$Ln$$_{2}$O$_{2}$$M$$_{2}$OSe$_{2}$ structures and related phases}

Due to the connection with superconducting cuprates and iron based compounds, the $Ln_{2}$O$_{2}M_{2}$OSe$_{2}$ series of materials have been studied in depth.  These compounds have typically smaller band gaps than the ZrCuSiAs compounds discussed above.  

The oxyselenides are all, nearly universally, semiconductors and sometimes described as ``bad-metals" with the resistivity increasing with decreasing temperature.   This occurs, for example in La$_{2}$O$_{2}$Fe$_{2}$OSe$_{2}$,  even though magnetic fluctuations are gapped as described above.  This differs from the case of Fe$_{1+x}$Te where gapped magnetic excitations were found to coincide with a transition from a ``semi metallic/bad metal" state to a metallic resistivity.  The connection between low-energy spin fluctuations and the resistivity using the formula outlined above in the experimental section was made in Ref. \onlinecite{Rodriguez13:88}.  A summary of the activation energies extracted by fitting the resistivity to $\rho=\rho_{\circ}e^{E_{a}/k_{B}T}$ is summarized in Table \ref{activation_energy} for a series of oxyselenides.

\begin{table}[ht]
\caption{Activation energies extracted from resistivity data ($\rho=\rho_{\circ}e^{E_{a}/k_{B}T}$).}
\centering
\begin{tabular} {c c c}
\hline\hline
Compound & $E_{a}$ & Reference \\
\hline\hline
La$_{2}$O$_{2}$Fe$_{2}$OSe$_{2}$ & 0.19 &  \onlinecite{Zhu10:104} \\
Ce$_{2}$O$_{2}$Fe$_{2}$OSe$_{2}$ & 0.26 &  \onlinecite{Ni11:83} \\
Pr$_{2}$O$_{2}$Fe$_{2}$OSe$_{2}$ & 0.15 &  \onlinecite{Ni11:83} \\
Nd$_{2}$O$_{2}$Fe$_{2}$OSe$_{2}$ & 0.15 &  \onlinecite{Ni11:83} \\
Sm$_{2}$O$_{2}$Fe$_{2}$OSe$_{2}$ & 0.18 &  \onlinecite{Ni11:83} \\
\hline\
La$_{2}$O$_{2}$Fe$_{2}$OS$_{2}$ & 0.24 &  \onlinecite{Zhu10:104} \\
\hline\
Ce$_{2}$O$_{2}$FeSe$_{2}$  & 0.32 & \onlinecite{McCabe11:47}\\
\hline\
BaFe$_{2}$Se$_{2}$O  & 0.29 & \onlinecite{Lei12:86}\\
\hline\
La$_{2}$O$_{2}$Mn$_{2}$OSe$_{2}$  & 0.24 & \onlinecite{Free11:23}\\
\hline\
La$_{2}$O$_{2}$Co$_{2}$OSe$_{2}$  & 0.35 & \onlinecite{Free11:23}\\
\hline \hline
\label{activation_energy}
\end{tabular}
\end{table}

\begin{figure}[t]
\includegraphics[width=8.5cm] {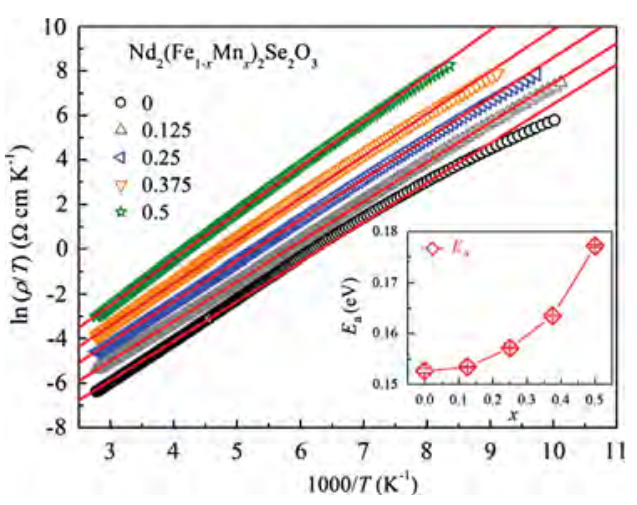}
\caption{\label{act_energy}  The resistivity and activation energies for Nd$_{2}$O$_{2}$(Fe$_{1-x}$Mn$_{x}$)$_{2}$OSe$_{2}$ taken from Ref. \onlinecite{Liu15:618}.}
\end{figure}

The data on these oxyselenides is difficult to interpret, but it generally seems to follow a trend that for decreasing rare earth radius the activation energy decreases.  This might imply a correlation between lattice constant and cell volume and activation energy.   This trend is confirmed in Fig. \ref{act_energy} which plots the resistivity  fit to a ``small polaron hopping" model $\rho(T)=AT\exp(E_{a}/k_{B}T)$.   With increasing Mn$^{2+}$ doping the lattice constants also increase and therefore this plot confirms the trend which was suggested from the data in Table \ref{activation_energy} that the activation energy scales with the lattice constant.  We note that the trend of decreasing cell volume corresponding to lower activation energies is reflected in comparing data on Sr$_{2}$OBi$_{2}$Se$_{3}$ (E$_{g}$=0.0092(1) eV) and Ba$_{2}$OBi$_{2}$Se$_{3}$ (E$_{g}$=0.11(1) eV) ~\cite{Panella16:28} and indeed may be a general feature across the oxyselenides.

\begin{figure}[t]
\includegraphics[width=8.5cm] {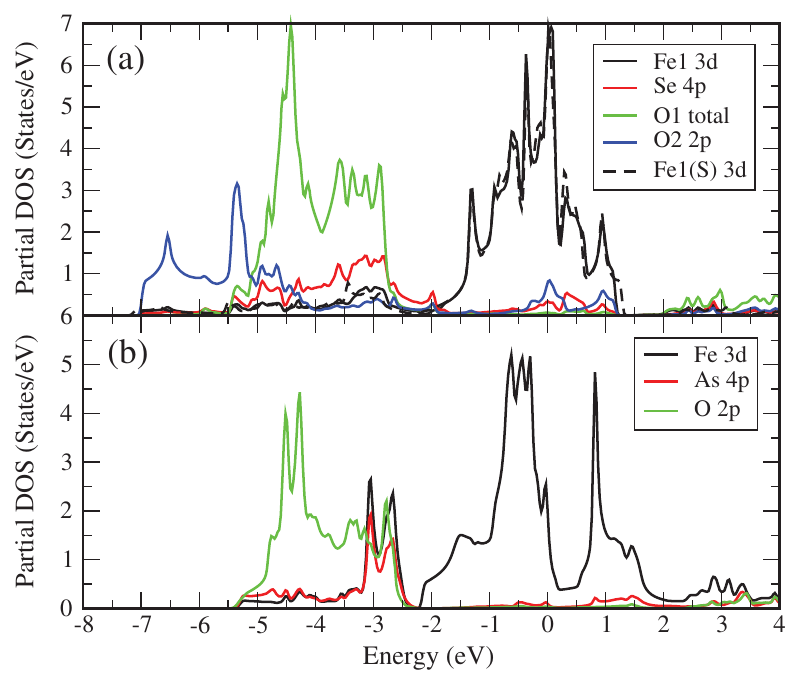}
\caption{\label{dos}   The partial density of states for paramagnetic La$_{2}$O$_{2}$Fe$_{2}$OSe$_{2}$  $(a)$ and LaOFeAs $(b)$ taken from Ref. \onlinecite{Zhu10:104}.  The curve fits are to the ``small polaron hopping" model described in the main text.  An increase of activation energy with increasing lattice constant is observed.}
\end{figure}

Electronic structure calculations have been performed for a number of  $Ln_{2}$O$_{2}M_{2}$O(Se,S)$_{2}$ oxyselenides.  In particular for La$_{2}$O$_{2}$Fe$_{2}$O(Se,S)$_{2}$ the electronic structure was calculated using density functional theory and is reported in Ref. \onlinecite{Zhu10:104}.  Fig. \ref{dos} shows the projected density of states for both compounds were it is shown that the $3d$ electrons on the iron site contribute strongly to the density of states at the Fermi energy.  The other point noted in the calculation is that the $d$-electron density of states for iron is mostly confined between -2 and 1.2 eV.  This represents a considerable narrowing of the iron $d$-electron band compared with pnictides such as the ZrCuSiAs-related as LaFeAsO (Fig. \ref{dos} panel $(b)$) where the band occurs between -2.2 and 2 eV and also FeTe or FeSe based chalcogenides~\cite{Subedi08:78}.  The narrower iron $3d$ electronic bands point to enhanced correlations.  This is substantiated by resistivity data showing that the behavior is insulating in contrast to iron based pnictides and chalcogenides which are typically metallic.  Resistivity data  reported in Ref. \onlinecite{Zhu10:104} find activation energy gaps of 0.19 and 0.24 eV for La$_{2}$O$_{2}$Fe$_{2}$O(Se,S)$_{2}$ respectively.  Based on the electronic behaviour combined with the antiferromagnetic ordering at low temperatures, these materials were classified as Mott insulators~\cite{Zhu10:104}.  

\begin{figure}[t]
\includegraphics[width=8.7cm] {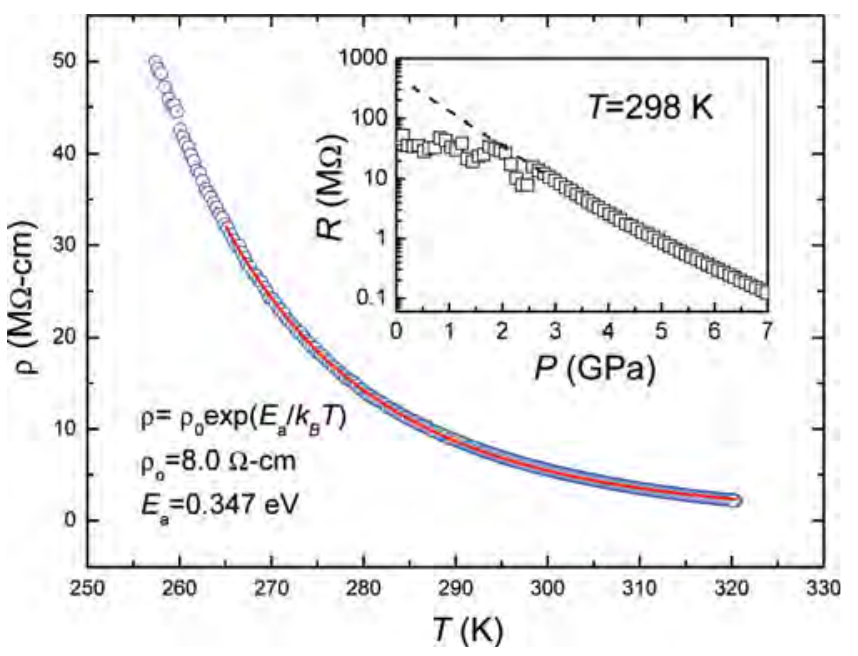}
\caption{\label{rho_pressure}  Resistivity as a function of temperature and pressure for La$_{2}$O$_{2}$Co$_{2}$OSe$_{2}$.  The figure is taken from Ref. \onlinecite{Wang10:132}.}
\end{figure}

A similar picture for the electronic structure is found for  La$_{2}$O$_{2}$Co$_{2}$O(Se,S)$_{2}$ and reported in Ref. \onlinecite{Wu10:82}.  Density functional calculations find an even narrower electronic bandwidth than for the iron analogue with Co $3d$ bandwidth of $\sim$ 2.8 eV~\cite{Wu10:82}.  Resistivity data was reported in Ref. \onlinecite{Wang10:132} with an activation energy of 0.35 eV which led to the conclusion that the 3d electrons associated with Co$^{2+}$ were localized.  In the context of the discussion in relation to Mott insulating behavior, Ref. \onlinecite{Wang10:132} investigated the pressure dependence of the resistivity where a sharp (several orders of magnitude) drop in resistivity was reported.  Based on this, and a comparison to first principle calculations, the authors concluded that La$_{2}$O$_{2}$Co$_{2}$OSe$_{2}$ was a ``marginal" Mott insulator and the pressure dependence (Fig. \ref{rho_pressure}) suggests that  La$_{2}$O$_{2}$Co$_{2}$OSe$_{2}$ is proximate to a metallic state.

While the first principle calculations above have pointed towards Mott insulating behaviour where insulating, or semiconducting, properties are the result of electron correlations, recent x-ray inelastic scattering data has come to a slightly different conclusion~\cite{Freelon15:92}.    By combining resonant inelastic x-ray spectroscopy with first principle calculations, Ref. \onlinecite{Freelon15:92} suggested that the electronic properties of La$_{2}$O$_{2}$Fe$_{2}$OSe$_{2}$ are more reminiscent of a Kondo insulator where a gap opens due to hybridisation of orbitals.  This was established through density functional calculations of the orbitally resolved self-energies.  It was proposed that La$_{2}$O$_{2}$Fe$_{2}$OSe$_{2}$ was a ``Mott-Kondo" insulator.  

The suggestion of a combination of electronic correlations (termed Mott insulators) and orbital effects (Kondo insulators) mimics recent proposals for a new type of metallic state term ``Hunds metals"~\cite{Georges13:4}.  The idea of Hunds metals has evolved from a proposal based on local density approximation calculations in LaOFeAs where it was noted that the splitting of the crystal fields due to a tetragonal distortion are comparable to the overall crystal splitting between the $|e\rangle$ and $|t\rangle$ states.  In this case, it was noted that even a small Hunds coupling would result in a spin transition from S=2 to S=1~\cite{Haule09:11}.  Such a framework could explain the low ordered moments in the pnictides (for example $gS$=0.5 $\pm$ 0.05 $\mu_{B}$ in FeAs from neutron diffraction) where much larger moments are clearly expected in the case of weak Hunds coupling or in the intermediate crystal field description.  The model also provides a means of explaining a strongly correlated metal and has been applied to LaO$_{1-x}$F$_{x}$FeAs~\cite{Haule08:100}.  The tuning from a Hunds metal to a Mott insulator has been proposed to be sensitive to the Fe-Fe distance~\cite{Yin10:105,Yin12:86} which is interesting in the context of the resistivity measurements under pressure noted above.

\subsection{Superconductivity and the Oxyselenides}

\begin{figure}[t]
\includegraphics[width=8.7cm] {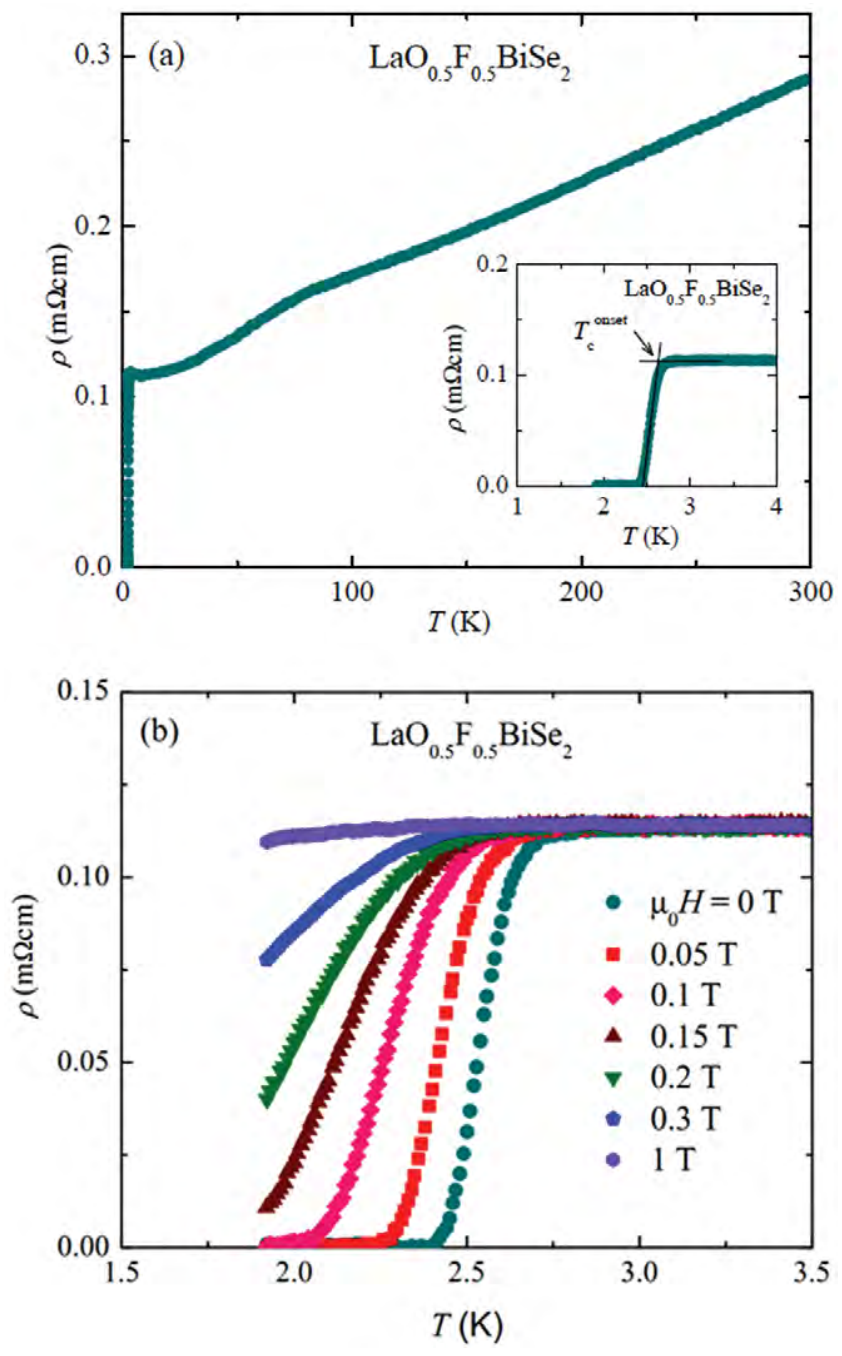}
\caption{\label{superconduct}  The resistivity of LaO$_{0.5}$F$_{0.5}$BiSe$_{2}$ as a function of temperature and magnetic field taken from Ref. \onlinecite{Krzton14:26}.}
\end{figure}

Iron oxychalcogenides FeQ (Q = S, Se, Te) have been shown to display unconventional superconductivity with properties tuned by intercalation chemistry between the antifluorite-like FeQ layers~\cite{Rodriguez11:2,Taylor13:87,Lynn15:92,Borg16:93,Vivanvo16:xx,Dagotto13:85}. To the best of our knowledge, the only oxychalcogenide superconductors are of the LnOBiS$_{2}$ family~\cite{Tanaka14:219} including LaO$_{0.5}$F$_{0.5}$BiSe$_{2}$ with superconducting T$_{c}$ = 2.6 K (Fig. \ref{superconduct})~\cite{Mizoguchi11:133,Krzton14:26}, LaO$_{0.5}$F$_{0.5}$BiSeS with a T$_{c}$ = 3.8 K~\cite{Wang15:27}, and related rare earth substituted compounds~\cite{Demura15:84,Demura13:82}.

LaO$_{0.5}$F$_{0.5}$BiSe$_{2}$ displays metallic conductivity (rather than semiconducting or bad metallic conductivity described above for other oxyselenides). The low superconducting transition temperatures have been attributed to the significant distortion of the BiSe$_{2}$ layers~\cite{Tanaka14:219}, with transition temperatures suppressed further with pressure~\cite{Kotegawa12:9}.   Electronic studies combining photoemission spectroscopy~\cite{Nagira14:83,Ye14:90,Zeng14:90} and photoelectron spectroscopy~\cite{Saini14:90} on these and related oxyselenides give good agreement with band structure calculations, suggesting that correlation effects may not be important.  However, it has been suggested that LaO$_{0.54}$F$_{0.46}$BiS$_{2}$ is close to a topology change in the Fermi surface~\cite{Terashima14:90}. This conclusion and their more conductive nature clearly set these compounds apart from the other oxyselenides discussed above. These $Ln$OBiQ$_{2}$-related materials represent a new development in the field of oxyselenide research.

\section{Conclusion}

There are several concepts that underlie the magnetic and electronic properties of the oxyselenides. The first is dimensionality, usually as a result of the anion-ordering, giving layered crystal structures. This influences the electronic structures, highlighted by the band narrowing in La$_{2}$O$_{2}$Fe$_{2}$O$Q_{2}$ predicted by Zhu \textit{et al.}~\cite{Zhu10:104} and consistent with the small values for magnetic exchange interactions observed experimentally~\cite{McCabe14:90}, and also by the confinement of holes in LaCuOSe materials confined to the [Cu$_{2}$Se$_{2}$]$^{2-}$ layers by the insulating [La$_{2}$O$_{2}$]$^{2+}$ layers~\cite{Ueda04:69}. This dimensionality also influences the magnetic ordering with magnetic stacking faults and longer magnetic correlation lengths within layers found for several $Ln_{2}$O$_{2}M_{2}$OSe$_{2}$ materials~\cite{Zhao13:87,McCabe14:90,Ni10:82,McCabe14:90}.  

The second key concept is connectivity which influences the magnetic and electronic structures: the magnetic frustration resulting from the tetrahedral arrangement of magnetic $Ln^{3+}$ ions in $Ln_{4}$O$_{4}$Se$_{3}$ materials discussed in Section 3a is a good example of this, as well as the strong interplay between Cr$^{3+}$ and $Ln^{3+}$ magnetism in $Ln$CrOS$_{2}$ materials with Cr$^{3+}$ and $Ln^{3+}$ coordination polyhedra linked via oxide anions~\cite{Takano99:85,Takano02:122,Winterberger87:70}. The Ce$_{2}$O$_{2}$FeSe$_{2}$ oxyselenide with one-dimensional chains of FeSe$_{4}$ tetrahedra~\cite{McCabe11:47,McCabe14:90} is surprising in this respect: in contrast to its one-dimensional connectivity, its electronic structure is far from one-dimensional~\cite{Li14:9}. The preparation of several polymorphs of $Ln_{2}$O$_{2}$FeSe$_{2}$ ($Ln$ = La, Ce) built up from FeSe$_{4}$ tetrahedra, FeSe$_{4}$O$_{2}$ pseudo-octahedra and from combinations of these~\cite{Nitsche14:640} coordination environments will provide an ideal means to investigate the role of the coordinating anion on the electronic structure (band widths) and magnetic structures.

The final concept to highlight is the local environment and particularly crystal field effects. While results from magnetic neutron diffraction are consistent with weak/intermediate crystal fields for transition metal sites, recent suggestions imply that coupling between electronic and orbital properties may occur.~\cite{Chen09:80}  Crystal field effects have been shown to have a role in the magnetic and structural behaviour of other mixed-anion systems~\cite{Kimber10:82,Wildman15:54,Kimber08:78} and may also be relevant to understanding the low-temperature behaviour of Pr$_{2}$O$_{2}$$M_{2}$OSe$_{2}$ ($M$ = Mn, Fe)~\cite{Ni11:83,Free11:23}. Oxyselenides provide a diverse series of materials in which novel magnetic and electronic phenomena can be studied. An example of this is the orthogonal, $2-k$ magnetic structure adopted by La$_{2}$O$_{2}$Fe$_{2}$OSe$_{2}$ and Sr$_{2}$F$_{2}$Fe$_{2}$OS$_{2}$, which, to the best of our knowledge, is unique among magnetically ordered systems~\cite{Zhao13:87,McCabe14:90} This structure results from coupling between orbital and electronic properties and from competition between anisotropy and competing exchange interactions, which will continue to challenge theory.

While a number of studies on oxyselenides have been performed, the lack of large single crystals is hampering efforts to fully understand the magnetism of these systems and synthetic efforts to produce sufficiently large crystals would be enhance the field.  The lack of unconventional superconductivity among magnetic oxyselenides (in contrast to the selenides and to oxypnictides) is interesting, especially given the strong electron correlation effects. The connections between cuprates and iron-based superconductors will remain a point of future study.



%

\end{document}